\documentclass[12pt,letterpaper]{article}
\pdfoutput=1
\usepackage[affil-it]{authblk}
\usepackage{amsmath}
\usepackage{amssymb}
\usepackage{graphicx}
\usepackage{natbib}
\usepackage{bm}
\usepackage{url}
\usepackage{amsthm}
\usepackage{thmtools}
\usepackage[margin=2.54cm]{geometry}
\usepackage{setspace}
\usepackage{paralist}
\usepackage{lineno}
\usepackage{setspace}
\usepackage{appendix}
\usepackage{siunitx}
\usepackage{xr}
\usepackage{fancyhdr}
\usepackage{lscape}

\fancypagestyle{firststyle}
{
   \fancyhf{}
   \fancyhead[C]{Running head: stochastic community dynamics}
}

\usepackage{etoolbox}
\usepackage{lmodern}

\makeatletter
\patchcmd{\@maketitle}{\LARGE \@title}{\fontsize{16}{19.2}\selectfont\@title}{}{}
\makeatother

\begin{document}

\raggedright

\title{Among-site variability in the stochastic dynamics of East African coral reefs}
\author{Katherine A. Allen \\School of Environmental Sciences, University of Liverpool, Liverpool, L69 3GP, UK \footnote{k.a.allen@liverpool.ac.uk. Current address: Institute of Integrative Biology, University of Liverpool, Liverpool L69 7ZB, UK} \and
John F. Bruno \\Department of Biology, University of North Carolina at Chapel Hill, Chapel Hill, North Carolina 27599-3300, USA \footnote{jbruno@unc.edu} \and
Fiona Chong \\School of Environmental Sciences, University of Liverpool, Liverpool, L69 3GP, UK \footnote{f.chong@student.liverpool.ac.uk} \and
Damian Clancy \\School of Mathematical and Computer Sciences, Actuarial Mathematics and Statistics, Heriot-Watt University, Edinburgh, UK \footnote{d.clancy@hw.ac.uk} \and
Tim R. McClanahan\\Wildlife Conservation Society, 2300 Southern Boulevard, Bronx, NY 10460, USA \footnote{tmcclanahan@wcs.org} \and
Matthew Spencer\\School of Environmental Sciences, University of Liverpool, Liverpool, L69 3GP, UK \footnote{m.spencer@liverpool.ac.uk} \and
Kamila \.Zychaluk\\Department of Mathematical Sciences, University of Liverpool, Liverpool, L69 7ZL, UK \footnote{kamila.zychaluk@liverpool.ac.uk}
}
\date{}

\maketitle
\thispagestyle{firststyle}

\section*{Abstract}
Coral reefs are dynamic systems whose composition is highly influenced by unpredictable biotic and abiotic factors. Understanding the spatial scale at which long-term predictions of reef composition can be made will be crucial for guiding conservation efforts. Using a 22-year time series of benthic composition data from 20 reefs on the Kenyan and Tanzanian coast, we studied the long-term behaviour of Bayesian vector autoregressive state-space models for reef dynamics, incorporating among-site variability. We estimate that if there were no among-site variability, the total long-term variability would be approximately one third of its current value. Thus among-site variability contributes more to long-term variability in reef composition than does temporal variability. Individual sites are more predictable than previously thought, and predictions based on current snapshots are informative about long-term properties. Our approach allowed us to identify a subset of possible climate refugia sites with high conservation value, where the long-term probability of coral cover $\leq 0.1$ was very low. Analytical results show that this probability is most strongly influenced by among-site variability and by interactions among benthic components within sites. These findings suggest that conservation initiatives might be successful at the site scale as well as the regional scale.

\section*{Keywords}

vector autoregressive model, state-space model, stochastic dynamics, community composition, spatial variability, temporal variability, coral reef, Bayesian statistics

\clearpage

\section*{Introduction}
``Probabilistic language based on stochastic models of population growth'' has been proposed as a standard way to evaluate conservation and management strategies \citep{Ginzburg_1728}. For example, a stochastic population model can be used to estimate the probability of abundance falling below some critical level. Such population viability analyses are widely used, and may be reasonably accurate if sufficient data are available \citep{Brook_3196}. In principle, the same approach could be used for communities, provided that a sufficiently simple model of community dynamics can be found.

A good candidate for such a model is the vector autoregressive model of order 1 or VAR(1) \citep{Lutkepohl93, Ives03}. This is a discrete-time model for the vector of log abundances of a set of species or groups, which includes environmental stochasticity and may include environmental explanatory variables. It makes the simplifying assumptions that inter- and intraspecific interactions can be represented by a linear approximation on the log scale, and that future abundances are conditionally independent of past abundances, given current abundances. Where possible, it is desirable to use a state-space form of the VAR(1) model, which also includes measurement error \citep{Lindegren09, Mutshinda09}. 

\citet{Hampton13} review applications of VAR(1) models in community ecology, which include studying the stability of freshwater plankton systems \citep{Ives03}, designing adaptive management strategies for the Baltic Sea cod fishery \citep{Lindegren09}, and estimating the contributions of environmental stochasticity and species interactions to temporal fluctuations in abundance of moths, fish, crustaceans, birds and rodents \citep{Mutshinda09}. Recently, VAR(1) models have been applied to the dynamics of the benthic composition of coral reefs \citep{Cooper15, Gross15}, using a log-ratio transformation \citep{Egozcue03} rather than a log transformation, to deal with the constraint that proportional cover of space-filling benthic groups sums to 1.

Coral reefs are dynamic systems influenced by both deterministic factors such as interactions between macroalgae and hard corals \citep{MHE07}, and stochastic factors such as temperature fluctuations \citep{Baker08} and storms \citep{Con97}. In general, high coral cover is considered a desirable state for a coral reef, and there is some evidence that coral cover of at least 0.1 is important for long-term maintenance of reef function \citep{Kennedy13, Perry13, Roff15}. Thus, coral cover of 0.1 might be an appropriate threshold against which to evaluate reef conservation strategies, and VAR(1) models can be used to estimate the probability of coral cover falling to or below this threshold \citep{Cooper15}.

There is evidence for systematic differences in reef dynamics among locations. For example, on the Great Barrier Reef, coral cover has declined more strongly at southern and central than at northern sites \citep{Death12}, and in the U.S. Virgin Islands, VAR(1) models showed that sites differed in their sensitivity to disturbance and speed of recovery \citep{Gross15}. Some sites in a region may therefore represent coral refugia, where reefs are either protected from or able to adapt to changes in environmental conditions \citep{MAM+07}. Although it may be possible to associate differences in dynamics among sites with differences in environmental variables, it is also possible to treat among-site differences as another random component of a VAR(1) model. This will allow estimation of the relative importance of among-site variability and within-site temporal variability, which is important for the design of conservation strategies. If within-site temporal variability dominates, it will not be possible to identify good sites to conserve based on current status, while if among-site variability dominates, even a ``snapshot'' sample at one time point may be enough to identify good sites. Thus, for example, the reliability of among-site patterns from surveys at one time point, such as the relationship between benthic composition and human impacts on remote Pacific atolls \citep{Sandin08}, depends on among-site variability dominating within-site temporal variability. Furthermore, since among-site variability will affect the probability of undesirable community composition (such as coral cover $\leq 0.1$), conservation strategies that explicitly address among-site variability may be effective.

Here, we develop a state-space VAR(1) model for regional dynamics of East African coral reefs, including random site effects and measurement error, and use it to answer four key questions about spatial and temporal variability. How important is among-site variability in the dynamics of benthic composition, relative to within-site temporal variability? How much variability is there among sites in the probability of low ($\leq 0.1$) coral cover? What is the most effective way (in terms of altering model parameters) to reduce the probability of low coral cover in the region? How informative is a single snapshot in time about the long-term properties of a site?

\section*{Methods}

\subsection*{Data collection}

Surveys of 20 spatially distinct reefs in Kenya and Tanzania (supporting information, Table \ref{tab:reefs}, Figure \ref{fig:map}) were conducted annually during the period 1991-2013 (generally in November or December prior to 1998, but January or February from 1998 onwards). Those in the north were typically fringing reefs, \SIrange{100}{2000}{\metre} from the shore, while those in the south were typically smaller and more isolated patch reefs, further from the shore \citep{McClanahan01c}. We categorized reefs as either fished or unfished, although there was substantial heterogeneity within these categories, because some fished reefs were community management areas with reduced harvesting intensity \citep{Cinner15}, and some unfished reefs had only recently been designated as reserves. Of the 20 reefs, 10 were divided into two sites separated by \SIrange{20}{100}{\metre}, while the remaining 10 reefs comprised only one site. The selection of sites represents available data rather than a random sample from all the locations at which coral reefs are present in the geographical area (and all of the longest time series are from Kenyan fringing reefs). Thus, when we refer below to `a randomly-chosen site' we strictly mean `a site drawn at random from the population for which data could have been available.'

Each of the 30 sites was visited at least twice (data from sites visited once were omitted), with a maximum of 20 visits. A version of line-intercept sampling \citep{Kaiser83,McClanahan01b} was used to estimate reef composition. In total, 2665 linear transects were sampled across all sites and years, with between 5 and 18 transects (median 9) at each site in a single year. Transects were randomly placed between two points $\SI{10}{\metre}$ apart, but as the transect line was draped over the contours of the substrate, the measured lengths varied between $\SI{10}{\metre}$ and $\SI{15}{\metre}$. Cover of benthic taxa was recorded as the sum of draped lengths of intersections of patches of each taxon with the line, divided by the total draped length of the line. Intersections with length less than $\SI{3}{\centi\metre}$ were not recorded. Taxa were identified to species or genus level, but for this study cover was grouped into three broad categories: hard coral, macroalgae and other (algal turf, calcareous and coralline algae, soft corals and sponges). Sand and seagrass were recorded, but excluded from our analysis, which focussed on hard substrate. The dynamics of a subset of these data were analyzed using different methods in \citet{Zychaluk12}.

\subsection*{Data processing}

The three cover values form a three-part composition, a set of three positive numbers whose sum is 1 \citep[Definition 2.1, p. 26]{Aitchison86}. Standard multivariate statistical techniques are not appropriate for untransformed compositional data, due to the absence of an interpretable covariance structure and the difficulties with parametric modelling \citep[chapter 3]{Aitchison86}. To avoid these difficulties, the proportional cover data were transformed to orthogonal, unconstrained, isometric log-ratio (ilr) coordinates \citep{Egozcue03}. The transformed data at site $i$, transect $j$, time $t$ were represented by the vector $\mathbf y_{i,j,t} = [y_{1,i,j,t}, y_{2,i,j,t}]^T$, in which the first coordinate $y_{1,i,j,t}$ was proportional to the natural log of the ratio of algae to coral, and the second coordinate $y_{2,i,j,t}$ was proportional to the natural log of the ratio of other to the geometric mean of algae and coral (supporting information, section \ref{sec:datatransform}). The $T$ denotes transpose: throughout, we work with column vectors.

\subsection*{The model}

The true value $\mathbf x_{i,t} = [x_{1,i,t},x_{2,i,t}]^T$ of the isometric log-ratio transformation of cover of hard corals, macroalgae and other at site $i$ at time $t$ was modelled by a vector autoregressive process of order 1 (i.e. a process in which the cover in a given year depends only on cover in the previous year), an approach used in other recent models of coral reef dynamics \citep{Cooper15, Gross15}. Unlike previous models, we include a random term representing among-site variation, and explicit treatment of measurement error (making this a state-space model). The full model is
\begin{equation}
\begin{aligned}
\mathbf x_{i,t+1} &= \mathbf a + \bm \alpha_i + \mathbf B \mathbf x_{i,t} + \bm \varepsilon_{i,t}, \\
\bm \alpha_i &\sim \mathcal N(\mathbf 0, \mathbf Z),\\
\bm \varepsilon_{i,t} &\sim \mathcal N(\mathbf 0, \bm \Sigma)\\
\mathbf y_{i,j,t} &\sim t_2(\mathbf x_{i,t}, \mathbf H, \nu). 
\end{aligned}
\label{eq:model}
\end{equation}

The column vector $\mathbf a$ represents the among-site mean proportional changes in $\mathbf x_{i,t}$ evaluated at $\mathbf x_{i,t} = \mathbf 0$. The column vector $\bm \alpha_i$ represents the amount by which these proportional changes for the $i$th site differ from the among-site mean, and is assumed to be drawn from a multivariate normal distribution with mean vector $\mathbf 0$ and $2 \times 2$ covariance matrix $\mathbf Z$. The $2 \times 2$ matrix $\mathbf B$ represents the effects of $\mathbf x_{i,t}$ on the proportional changes, and can be thought of as summarizing intra- and inter-component interactions such as competition. The column vector $\bm \varepsilon_{i,t}$ represents random temporal variation, and is assumed to be drawn from a multivariate normal distribution with mean vector $\mathbf 0$ and covariance matrix $\bm \Sigma$. We assume that there is no temporal or spatial autocorrelation in $\bm \varepsilon$, and that $\bm \varepsilon$ is independent of the among-site variation $\bm \alpha$. 

The observed transformed compositions $\mathbf y_{i,j,t}$ vary around the corresponding true compositions $\mathbf x_{i,t}$ due to both small-scale spatial variation in true composition among transects within a site, and measurement error in estimating composition from a transect. We cannot easily separate these sources of variation because transects were located at different positions in each year, and there were no repeat measurements within transects. Observed log-ratio transformed cover $\mathbf y_{i,j,t}$ in the $j$th transect of site $i$ at time $t$ was assumed to be drawn from a bivariate $t$ distribution (denoted by $t_2$) with location vector equal to the corresponding $\mathbf x_{i,t}$, and unknown scale matrix $\mathbf H$ and degrees of freedom $\nu$ \citep{Lange89}. The bivariate $t$ distribution can be interpreted as a mixture of bivariate normal distributions whose covariance matrices are the same up to a scalar multiple \citep{Lange89}, and therefore allows a simple form of among-site or temporal variation in the distribution of measurement error or small-scale spatial variation, whose importance increases as the degrees of freedom decrease. Preliminary analyses suggested that it was important to allow this variation, because the model in Equation \ref{eq:model} fitted the data much better than a model with a bivariate normal distribution for $\mathbf y_{i,j,t}$ (supporting information, section \ref{sec:measurement}).

We make the important simplifying assumptions that $\mathbf B$ is the same for all sites, and that the causes of among-site and temporal variation are not of interest. A separate $\mathbf B$ for each site, or even a hierarchical model for $\mathbf B$, would be difficult to estimate from the amount of data we have. It might be possible to explain some of the random temporal variation using temporally-varying environmental covariates such as sea surface temperature, and some of the among-site variation using temporally constant covariates such as management strategies \citep{Cooper15}. However, it is not necessary to do so in order to answer the questions listed at the end of the introduction, and keeping the model as simple as possible is important because parameter estimation is quite difficult. Furthermore, some of the relevant environmental variables may be associated with management strategies, making it difficult to separate the effects of environmental variation and management. For example, although some water quality variables were not strongly associated with protection status \citep{CarreiroSilva12}, unfished reefs were designated as protected areas due to their relatively good condition and are generally found in deeper lagoons with lower and more stable water temperatures than fished reefs (T. R. McClanahan, personal observation).

To understand the features of dynamics common to all sites, we plotted the back-transformations from ilr coordinates to the simplex of the overall intercept parameter $\mathbf a$ and the columns $\mathbf a_1$ and $\mathbf a_2$ of a matrix $\mathbf A$, which is related to $\mathbf B$ and describes the effects of current reef composition on the change in reef composition from year to year \citep{Cooper15}. We plotted $\mathbf A$ rather than $\mathbf B$ because it leads to a simpler visualization of effects (supporting information, section \ref{sec:visualize}). For example, a point lying to the left of the line representing equal proportions of coral and algae (the 1:1 coral-algae isoproportion line) corresponds to a parameter tending to increase coral relative to algae.

\subsection*{Parameter estimation}

We estimated all model parameters and checked model performance using Bayesian methods implemented in the Stan programming language \citep{stan-software:2015}, as described in the supporting information (section \ref{sec:estimation}). Stan uses the No-U-Turn Sampler, a version of Hamiltonian Monte Carlo, which can converge much faster than random-walk Metropolis sampling when parameters are correlated \citep{Hoffman14}. For most results, we report posterior means and $95\%$ highest posterior density (HPD) intervals \citep{Hyndman_4081}, calculated in R \citep{RCore15}.

\subsection*{Long-term behaviour}
In the long term, the true transformed composition $\mathbf x^*$ of a randomly-chosen site will converge to a stationary distribution, provided that all the eigenvalues of $\mathbf B$ lie inside the unit circle in the complex plane \citep[e.g.][p. 10]{Lutkepohl93}. If the eigenvalues of $\mathbf B$ are complex, the system will oscillate as it approaches the stationary distribution. Details of long-term behaviour are in the supporting information, section \ref{sec:longterm}.

This stationary distribution is the multivariate normal vector
\begin{equation}
\mathbf x^* \sim \mathcal N(\bm \mu^*, \bm \Sigma^* + \mathbf Z^*),
\label{eq:stationary}
\end{equation}
whose stationary mean $\bm \mu^*$ depends on $\mathbf B$ and $\mathbf a$, and whose stationary covariance is the sum of the stationary within-site covariance $\bm \Sigma^*$ (which depends on $\mathbf B$ and $\bm \Sigma$) and the stationary among-site covariance $\mathbf Z^*$ (which depends on $\mathbf B$ and $\mathbf Z$). 

For a fixed site $i$, the value of $\bm \alpha_i$ is fixed and the stationary distribution is given by
\begin{equation}
\mathbf x_i^* \sim \mathcal N(\bm \mu_i^*, \bm \Sigma^*),
\label{eq:stationaryi}
\end{equation}
whose stationary mean $\bm \mu_i^*$ depends on $\mathbf B$, $\mathbf a$ and $\bm \alpha_i$, and whose stationary covariance matrix is $\bm \Sigma^*$. Note that $\mathbf B$, which describes intra- and inter-component interactions on an annual time scale, affects all the parameters of both stationary distributions, and therefore affects both within- and among-site variability in the long term. Also, the back-transformation of the stationary mean $\bm \mu^*$ of the transformed composition, rather than the arithmetic mean vector of the untransformed composition, is the appropriate measure of the centre of the stationary distribution \citep{Aitchison89}.

\subsection*{How important is among-site variability?}
The covariance matrix of the stationary distribution for a randomly-chosen site (Equation \ref{eq:stationary}) contains contributions from both among- and within-site variability. To quantify the contributions from these two sources, we calculated
\begin{equation}
\rho = \left(\frac{|\bm \Sigma^*|}{|\bm \Sigma^* + \mathbf Z^*|} \right)^{1/2},
\label{eq:vrat}
\end{equation}
(supporting information, section \ref{sec:amongreef}), which is the ratio of volumes of two unit ellipsoids of concentration \citep{Kenward79}, the numerator corresponding to the stationary distribution in the absence of among-site variation (or for a fixed site, as in Equation \ref{eq:stationaryi}), and the denominator to the full stationary distribution of transformed reef composition in the region. The volume of each ellipsoid of concentration is a measure of the dispersion of the corresponding distribution. Thus $\rho$ provides an indication of how much of the total variability would remain if all among-site variability was removed. A similar statistic was used by \citet{Ives03} to measure the contribution of species interactions to stationary variability.

\subsection*{How much variability is there among sites in the probability of low coral cover?}

For a given coral cover threshold $\kappa$, we define $q_{\kappa,i}$ as the long-term probability that site $i$ has coral cover less than or equal to $\kappa$. This can be interpreted either as the proportion of time for which the site will have coral cover less than or equal to $\kappa$ in the long term, or as the probability that the site will have coral cover less than or equal to $\kappa$ at a random time, in the long term. We set $\kappa = 0.1$, which has been suggested as a threshold for a positive net carbonate budget, based on simulation models and data from Caribbean reefs \citep{Kennedy13, Perry13, Roff15}. We calculated $q_{0.1,i}$ for each site numerically (supporting information, section \ref{sec:plowintegral}). In order to determine whether differences in $q_{0.1,i}$ were related to current coral cover, we plotted $q_{0.1,i}$ against the corresponding sample mean coral cover for each site, over all transects and years. In order to determine whether differences in $q_{0.1, i}$ had obvious explanations, we distinguished between fished and unfished reefs, and patch and fringing reefs. In order to determine whether there was strong spatial pattern in the probability of low coral cover, we calculated spline correlograms \citep{Bjornstad01} for a sample from the posterior distribution of $q_{0.1,i}$ (supporting information, section \ref{sec:plowcorr}).

\subsection*{What is the most effective way to reduce the probability of low coral cover?}

For a given coral cover threshold $\kappa$, we define $q_\kappa$ as the long-term probability that a randomly-chosen site has coral cover less than or equal to $\kappa$. This is equal to the expected long-term probability that coral cover is less than or equal to $\kappa$ over the region, and can be calculated numerically (supporting information, section \ref{sec:plowintegral}). To find the most effective way to reduce $q_\kappa$, we calculated its derivatives with respect to each model parameter. As above, we concentrated on $\kappa = 0.1$. However, we also compared results from $\kappa = 0.05$ and $\kappa = 0.20$. The probability $q_\kappa$ is a function of 12 parameters: all four elements of $\mathbf B$; both elements of $\mathbf a$; elements $\sigma_{11}$, $\sigma_{21}$ and $\sigma_{22}$ of $\bm \Sigma$; and elements $\zeta_{11}$, $\zeta_{21}$ and $\zeta_{22}$ of $\mathbf Z$. The negative of the gradient vector of derivatives of $q_\kappa$ with respect to these parameters describes the direction of movement through parameter space in which the probability of low coral cover will be reduced most rapidly, and the elements of this vector with the largest magnitudes correspond to the parameters to which $q_\kappa$ is most sensitive. To understand why $q_\kappa$ responds to each model parameter, note that $q_\kappa$ depends on the parameters $\bm \mu^*$, $\bm \Sigma^*$ and $\mathbf Z^*$ of the stationary distribution (Equation \ref{eq:stationary}), which are in turn affected by the model parameters. We therefore used the chain rule for matrix derivatives \citep[p.108]{Magnus07} to break down the derivatives into effects of $\bm \mu^*$, $\bm \Sigma^*$ and $\mathbf Z^*$ on $q_\kappa$, and effects of model parameters on $\bm \mu^*$, $\bm \Sigma^*$ and $\mathbf Z^*$ (supporting information, section \ref{sec:plow}).

\subsection*{How informative is a snapshot about long-term site properties?}

In a stochastic system, how much can a ``snapshot'' survey at a single point in time tell us about the long-term behaviour of the system? For example, are differences among sites that appear to be in good and bad condition likely to be maintained in the long term? To make this question more precise, suppose that we draw a site at random from the region, and at one point in time, draw the true state of the site at random from the stationary distribution for the site. This scenario matches Diamond's definition of ``natural snapshot experiments'' as ``comparisons of communities assumed to have reached a quasi-steady state'' \citep{Diamond86}. For simplicity, we assume that we can estimate the true state accurately (for example, by taking a large number of transects). To quantify how informative this is about the long term properties of the site, we computed the correlation coefficients between corresponding components of the true state at a given site at a given time and of stationary mean for that site (supporting information, section \ref{sec:snapshot}). If these correlations are high, then a snapshot will be informative about long term properties.

\section*{Results}
\subsection*{Overall dynamics}

At all sites, the model appeared to provide a good description of observed dynamics, although sometimes with high uncertainty. The back-transformed posterior mean true states from the model (e.g. Figure \ref{fig:obstrueback}, grey lines) closely tracked the centres of the distributions of cover estimates from individual transects, although there was substantial among-transect variability at a given site in a given year (e.g. Figure \ref{fig:obstrueback}, circles). Figure \ref{fig:obstrueback} shows two examples, and time series for all sites are plotted in the supporting information, Figures \ref{fig:Bongoyo1} to \ref{fig:Watamu1}. There were also substantial differences in patterns of temporal change among sites. For example, Kanamai1 (Figure \ref{fig:obstrueback}a-c), a fished site, had consistently low algal cover and no dramatic changes in cover of any component. In contrast, Mombasa1 (Figure \ref{fig:obstrueback}d-f), an unfished site, had a sudden decrease in coral cover in 1998, and algal cover was high from 2007 onwards. As a result, Mombasa1 was unusual in that the current estimate of true algal cover was well above the stationary mean estimate (Figure \ref{fig:obstrueback}e: black circle at end of time series). For most other sites, current estimated true cover was close to the stationary mean (supporting information, Figures \ref{fig:Bongoyo1} to \ref{fig:Watamu1}, black circles at ends of time series). The uncertainty in true states (Figure \ref{fig:obstrueback}, grey polygons represent $95\%$ highest posterior density (HPD) credible intervals) was higher during intervals with missing observations (e.g. 2008 in Figure \ref{fig:obstrueback}). In general, uncertainty in true states (grey polygons) and stationary means (black bars at end of time series) was highest for sites with few observations (e.g. Bongoyo1, Figure \ref{fig:Bongoyo1}). 

The overall intercept parameter $\mathbf a$ (Figure \ref{fig:parms}, green), which describes the dynamics of reef composition at the origin (where each component is equally abundant) was consistent with the observed low macroalgal cover in the region (e.g. Figure \ref{fig:obstrueback}b, e). The back-transformation of $\mathbf a$ lay close to the coral-other edge of the ternary plot, and slightly above the 1:1 coral-other isoproportion line. It therefore represented a strong year-to-year decrease in algae, and a slight increase in other relative to coral, at the origin. 

Current reef composition acts on year-to-year change in composition (through matrix $\mathbf A$) so as to maintain fairly stable reef composition. The first column $\mathbf a_1$ of $\mathbf A$, which represents the effects of the transformed ratio of algae to coral on year-to-year change in composition, lay (when back-transformed) to the left of the 1:1 coral-algae isoproportion line, above the 1:1 other-algae isoproportion line, and below the 1:1 coral-other isoproportion line (Figure \ref{fig:parms}, orange). Thus, increases in algae relative to coral resulted in decreases in algae relative to coral and other, and increases in coral relative to other, in the following year. The second column $\mathbf a_2$ of $\mathbf A$, which represents the effects of the transformed ratio of other to algae and coral on year-to-year change in composition, lay (when back-transformed) on the 1:1 coral-algae isoproportion line, below the 1:1 other-algae isoproportion line, and below the 1:1 coral-other isoproportion line (Figure \ref{fig:parms}, blue). Thus, increases in other relative to algae and coral resulted in little change in the ratio of coral to algae, but decreases in other relative to both coral and algae. Consistent with the above interpretation of year-to-year dynamics, every set of parameters in the Monte Carlo sample led to a stationary distribution, since both eigenvalues of $\mathbf B$ lay inside the unit circle in the complex plane (supporting information, section \ref{sec:dynamics}). The magnitudes of these eigenvalues were smaller than those for a similar model for the Great Barrier Reef \citep{Cooper15}, indicating more rapid approach to the stationary distribution. There was some evidence for complex eigenvalues of $\mathbf B$, leading to rapidly-decaying oscillations in both components of transformed reef composition on approach to this distribution. This contrasts with the Great Barrier Reef, where there was no evidence for oscillations \citep{Cooper15}.

\subsection*{How important is among-site variability?}

There was substantial among-site variability in the locations of stationary means (Figure \ref{fig:amongreef}, dispersion of points). Stationary mean algal cover was always low, but there was a wide range of stationary mean coral cover. Although our primary focus is not on the causes of among-site variability, there was a tendency for most of the reefs with highest stationary mean coral cover to be patch reefs (Figure \ref{fig:amongreef}, circles). The stationary means did not clearly separate by management (Figure \ref{fig:amongreef}, open symbols fished, filled symbols unfished). The long-term temporal variability around the stationary means was also substantial (Figure \ref{fig:amongreef}, green lines), as was the uncertainty in the values of the stationary means (Figure \ref{fig:amongreef}, grey dashed lines). The $\rho$ statistic (Equation \ref{eq:vrat}), which quantifies the posterior mean contribution of within-site variability to the total stationary variability in reef composition in the region, was 0.29 ($95\%$ HPD interval $[0.20, 0.39]$), or approximately one third. Thus, while within-site temporal variability around the stationary mean was not negligible, among-site variability in the stationary mean was more important in the long term.

For all three components of variability (within-site, among-site, and measurement error/small-scale spatial variability), variation in algal cover was larger than variation in coral or other. This can be seen in the shapes of the back-transformed unit ellipsoids of concentration (Figure \ref{fig:covellipsesback}: within-site, green; among-site, orange; measurement error and small-scale spatial variability, blue) which were all elongated to some extent along the 1:1 coral-other isoproportion line. This was similar to, but less extreme than, the pattern observed in the Great Barrier Reef \citep{Cooper15}. The among-site ellipsoid almost entirely enclosed the within-site ellipsoid, consistent with the estimate above that among-site variability was more important than within-site variability in the long term. The large estimated measurement error/small-scale spatial variability component was consistent with the substantial observed variability in cover among transects at any given site and time (Figure \ref{fig:obstrueback}, circles and supporting information, Figures \ref{fig:Bongoyo1} to \ref{fig:Watamu1}, circles). The low estimated degrees of freedom $\nu$ for the bivariate $t$ distribution of measurement error/small-scale spatial variability (posterior mean 2.99, $95\%$ HPD interval $[2.64, 3.35]$) suggested that some aspect of the process leading to variation in measured composition among transects at a given site was varying substantially over space or time, although we cannot determine the mechanism.

\subsection*{How much variability is there among sites in the probability of low coral cover?}

There was also substantial among-site variability in the probability of low coral cover. For a randomly-chosen site, the posterior mean probability of coral cover less than or equal to 0.1 ($q_{0.1}$) in the long term was 0.12 ($95\%$ credible interval $[0.04, 0.21]$). The corresponding site-specific probabilities $q_{0.1,i}$ varied from $8 \times 10^{-5}$ to 0.52 but were low for most sites, with a strong negative relationship between probability of low coral cover and observed mean coral cover (Figure \ref{fig:problowcoralobs}). There was no clear distinction between fished and unfished reefs (Figure \ref{fig:problowcoralobs}, open symbols fished, filled symbols unfished). However, probability of low coral cover appeared to be systematically lower on patch reefs, which were mainly in Tanzania (Figures \ref{fig:problowcoralobs} and \ref{fig:map}, circles: median of posterior means $2 \times 10^{-3}$, first quartile $4 \times 10^{-4}$, third quartile 0.04) than on fringing reefs (Figures \ref{fig:problowcoralobs} and \ref{fig:map}, triangles: median of posterior means 0.08, first quartile 0.04, third quartile 0.11). One site (Ras Iwatine) had a much higher probability of low coral cover than all others, and is relatively polluted compared to other sites in this study, due to high levels of nutrient effluent from a large hotel (T.R. McClanahan, personal observation). 

There was little evidence for strong spatial autocorrelation in the probability of low coral cover, because the $95\%$ envelope for the spline correlogram included zero for all distances other than \SIrange{261}{322}{\kilo \metre} (supporting information, Figure \ref{fig:spline}). The general lack of strong spatial autocorrelation reflects the substantial variation in probability of coral cover less than or equal to 0.1 ($q_{0.1,i}$) among nearby sites, while the possibility of negative spatial autocorrelation at scales of around $\SI{300}{\kilo\metre}$ may reflect the generally low values of $q_{0.1,i}$ for Tanzanian patch reefs, separated from sites in the north of the study area with generally higher $q_{0.1,i}$ by approximately $\SI{300}{\kilo\metre}$ (Figure \ref{fig:map}).

\subsection*{What is the most effective way to reduce the probability of low coral cover?}

Both among-site variability and internal dynamics, particularly of other relative to algae and coral (component 2), were important in determining the probability $q_{0.1}$ of coral cover $\leq 0.1$ in the region. Figure \ref{fig:derivs} shows the direction in parameter space along which the probability of low coral cover will reduce most rapidly (the estimated gradient vector of $q_{0.1}$  with respect to all the model parameters). The four parameters to which $q_{0.1}$ was most sensitive were (in descending order: Figure \ref{fig:derivs}) $\zeta_{21}$ (among-site covariance between transformed components 1 and 2), $b_{22}$ (effect of component 2 on next year's component 2), $\zeta_{22}$ (among-site variance of component 2), and $b_{12}$ (effect of component 2 on next year's component 1). Although there was substantial variability among Monte Carlo iterations in the values of these derivatives, the rank order of magnitudes was fairly consistent (supporting information, Figure \ref{fig:rankderivs}). All four most important parameters had positive effects on $q_{0.1}$ (Figure \ref{fig:derivs}), so reducing these parameters will reduce $q_{0.1}$. The effects of within-site temporal variability on the probability of low coral cover were relatively unimportant (Figure \ref{fig:derivs}, derivatives of $q_{0.1}$ with respect to $\sigma_{11}$, $\sigma_{21}$ and $\sigma_{22}$ all had posterior means close to zero). The signs of the effects of each parameter on $q_{0.1}$, and results for coral cover thresholds 0.05 and 0.1, are discussed further in the supporting information (sections \ref{sec:derivsigns} and \ref{sec:thresholds}).

\subsection*{How informative is a snapshot about long-term site properties?}

For both components of transformed composition, a snapshot of reef composition at a single time on a randomly-chosen site will be informative about the stationary mean (correlations between true value at a given time and stationary mean: component 1 posterior mean 0.84, 95$\%$ HPD interval $[0.75, 0.91]$; component 2 posterior mean 0.82, 95$\%$ HPD interval $[0.73, 0.90]$). This is consistent with the negative relationship between long-term probability of coral cover $\leq 0.1$ and observed mean coral cover (Figure \ref{fig:problowcoralobs}). Thus, while long-term monitoring of East African coral reefs is important for other reasons, it should be possible to identify those with high conservation value (in terms of benthic composition) from a single survey.

\section*{Discussion}

In the long term, among-site variability dominates within-site temporal variability in East African coral reefs. In consequence, the long-term probability of coral cover $\leq 0.1$ varied substantially among sites. This suggests that it is in principle possible to make reliable decisions about the conservation value of individual sites based on a survey of multiple sites at one point in time, and to design conservation strategies at the site level. This was not the only possible outcome: if within-site temporal variability dominated among-site variability, among-site differences would be neither important nor predictable in the long term. Given the large positive effect of among-site variability on the long-term probability of coral cover $\leq 0.1$, reducing among-site variability in compositional dynamics may be an effective conservation strategy.

The dominance of among-site variability has important implications for conservation. There was clear evidence for the existence of a stationary distribution of long-term reef composition in East Africa. The overall shape of this distribution (Figure \ref{fig:amongreef}) was similar to that estimated by \citet{Zychaluk12} for a subset of the same data, using a different modelling approach. However, our new analysis shows that this distribution is generated by a combination of spatial and temporal processes, with substantial long-term differences among sites. Thus, the distribution in \citet{Zychaluk12} may be a good approximation to the long-term distribution for a randomly-chosen site, but there will be much less variability over time in the distribution for any fixed site. In consequence, the sites having the highest long-term conservation value can be identified even from single-survey snapshots, and conservation strategies at the site scale may be possible. Furthermore, in cases where among-site variability in dynamics is dominant, it will be misleading to generalize from observations of a few sites to regional patterns \citep{Bruno09}.

In our study, the sites with the highest long-term conservation value are those with very low long-term probabilities of coral cover $\leq 0.1$ (Figure \ref{fig:problowcoralobs}), a threshold chosen based on evidence that coral cover $\leq 0.1$ is detrimental to reef persistence \citep{Kennedy13, Perry13, Roff15}. Many of these sites are Tanzanian patch reefs, which may have maintained high coral cover despite disturbance because of local hydrography \citep{MAM+07}, and are priority sites for conservation, with high alpha and beta diversity \citep{Ateweberhan16}. In the light of these observations, we experimented with a model in which reef type was included as an explanatory variable. Although the estimated effects of reef type were consistent with lower long-term probabilities of coral cover $\leq 0.1$, including reef type did not improve the expected predictive accuracy of the model (F. Chong, unpublished results), probably because only 482 out of 2665 transects were from patch reefs, and all but one patch reefs had only very short time series (supporting information, Table \ref{tab:reefs}). Furthermore, the absence of strong spatial autocorrelation in long-term probabilities of coral cover $\leq 0.1$ suggests that it will be necessary to consider conservation value at small spatial scales, rather than simply to identify subregions with high conservation value. Similarly, \citet{Vercelloni14} found that trajectories of coral cover on the Great Barrier Reef were consistent at the scale of $\si{\kilo\metre^2}$, but not at larger spatial scales. They argued that it would therefore be appropriate to focus management actions at the $\si{\kilo\metre^2}$ scale. Also, it may be easier to persuade local communities to accept management at such scales than at larger scales \citep{McClanahan16}.

A key result is that if we want to minimize the long-term probability $q_{0.1}$ that a randomly-chosen site has coral cover $\leq 0.1$, we should minimize among-reef variability in dynamics, other things being equal. This is because the centre of the stationary distribution lies outside the set of compositions with coral cover $\leq 0.1$ (Supporting Information, Section \ref{sec:derivsigns}). Conversely, if the centre lay inside this set, then (other things being equal) maximizing among-site variability would minimize $q_{0.1}$. This result is very general, applying to any model of community composition which has a stationary distribution, for which increasing among-site variability increases stationary variability, and for any conservation objective based on a composition threshold.

Conservation strategies that might minimize among-site variability include distributing a fixed amount of human activity such as coastal development or fishing evenly, rather than concentrating it in a few locations. On the other hand, many conservation strategies will affect both the mean dynamics and the among-site variability in dynamics. For example, protecting the sites that are already in the best condition will tend to increase among-site variability, while moving the centre of the stationary distribution away from the set of compositions with coral cover $\leq 0.1$. 

Minimizing among-site variability in dynamics may conflict with other proposed conservation strategies. It has been suggested that increased beta diversity is associated with lower temporal variability in metacommunities, for at least some taxa, and that regions of high beta diversity may therefore be priority regions for conservation \citep{Mellin14}. It is likely that increased beta diversity will also be associated with increased among-site variability in dynamics, because different species are likely to have different population-dynamic characteristics. Hence, it may not always be possible to manage for both low among-site variability in dynamics and high beta diversity. It is not yet clear which of these objectives is more important in general.

Our analyses were based on the long-term consequences of current environmental conditions, and may therefore not be relevant if environmental conditions change. For example, if changes in climate or local human activity altered the vector $\mathbf a$ so as to transpose the centre of the stationary distribution into the set with coral cover $\leq 0.1$, then maximizing among-site variability would become the best strategy. Since declining coral cover trends have been observed at the regional level \citep[e.g.][]{Cote05, Death12}, such a shift in the best strategy may occur. It is therefore better to view a stationary distribution under current conditions as a ``speedometer'' that tells us about the long-term outcome if these conditions were maintained, rather than as a prediction \citep[p. 30]{Cas01}.

In conclusion, our analysis extends the broadly-applicable vector autoregressive approach to community dynamics \citep[reviewed by][]{Hampton13} by quantifying random among-site variability in dynamics. This gives a new perspective on the long-term behaviour of the set of communities in a region, as a set of stationary distributions with random but persistent differences. The extent of these differences relative to temporal variability determines how predictable the behaviour of individual sites will be. Since these differences may be associated with differences in conservation value, probabilistic risk assessment based on this approach can be used to suggest conservation strategies at both site and regional scales. At site scales, our approach can be used to identify potential coral refugia, while at regional scales, it can identify the parameters with most influence on conservation objectives.

\section*{Acknowledgments}

This work was funded by NERC grant NE/K00297X/1 awarded to MS. 

\clearpage

\section*{Figure legends}

\begin{figure}[h]
\includegraphics[angle=90, origin=c, height=17cm]{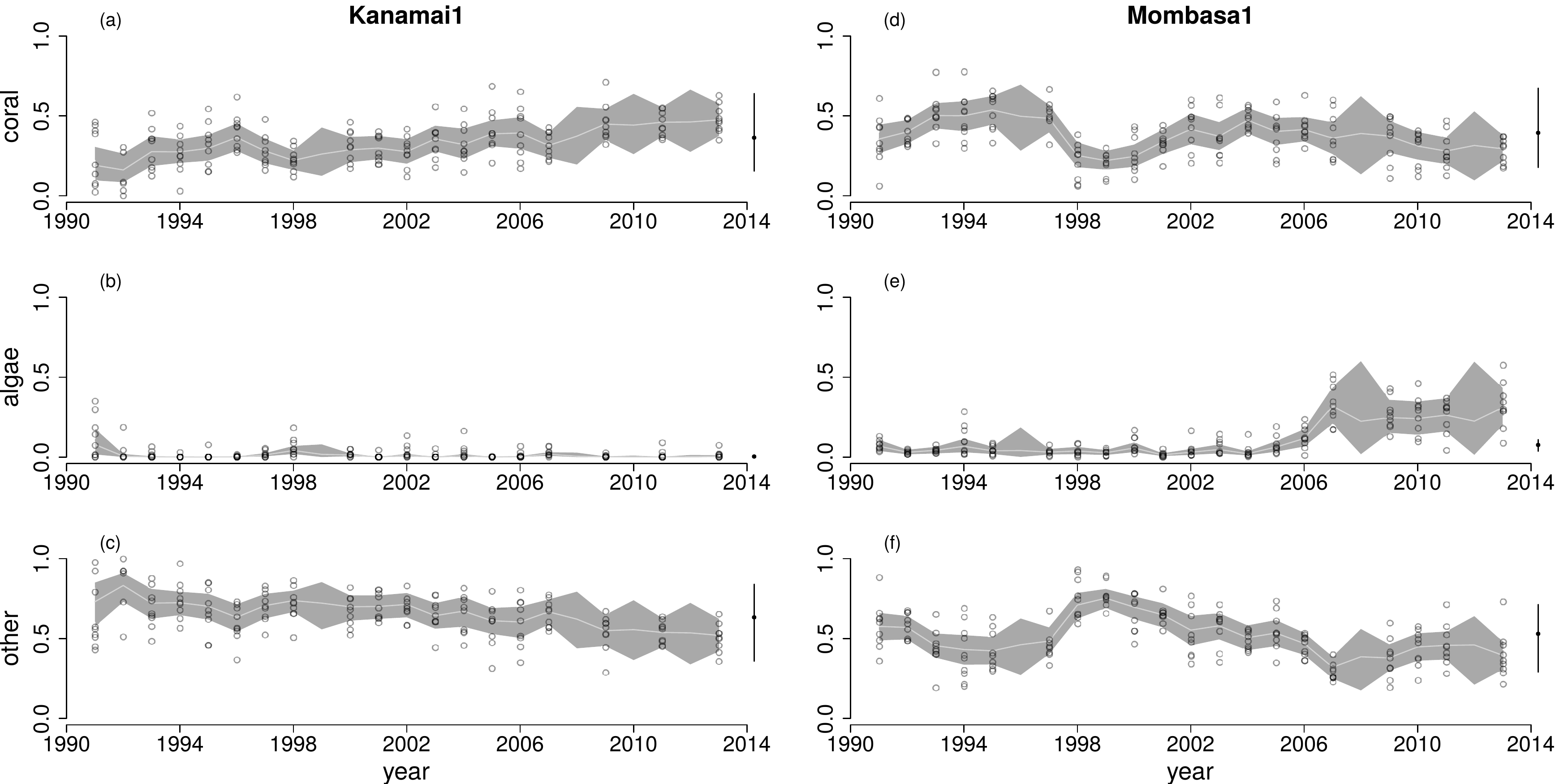}
\caption{}
\label{fig:obstrueback}
\end{figure}

Figure \ref{fig:obstrueback}. Time series of cover of hard corals, macroalgae and other at two of the 30 sites surveyed: Kanamai1 (fished, a-c) and Mombasa1 (unfished, d-f). Circles are observations from individual transects. Grey lines join back-transformed posterior mean true states from Equation \ref{eq:model}, and the shaded region is a 95$\%$ highest posterior density interval. The back-transformed stationary mean composition for the site is the black dot after the time series and the bar is a 95$\%$ highest posterior density interval.

\begin{figure}[h]
\includegraphics[height=18cm]{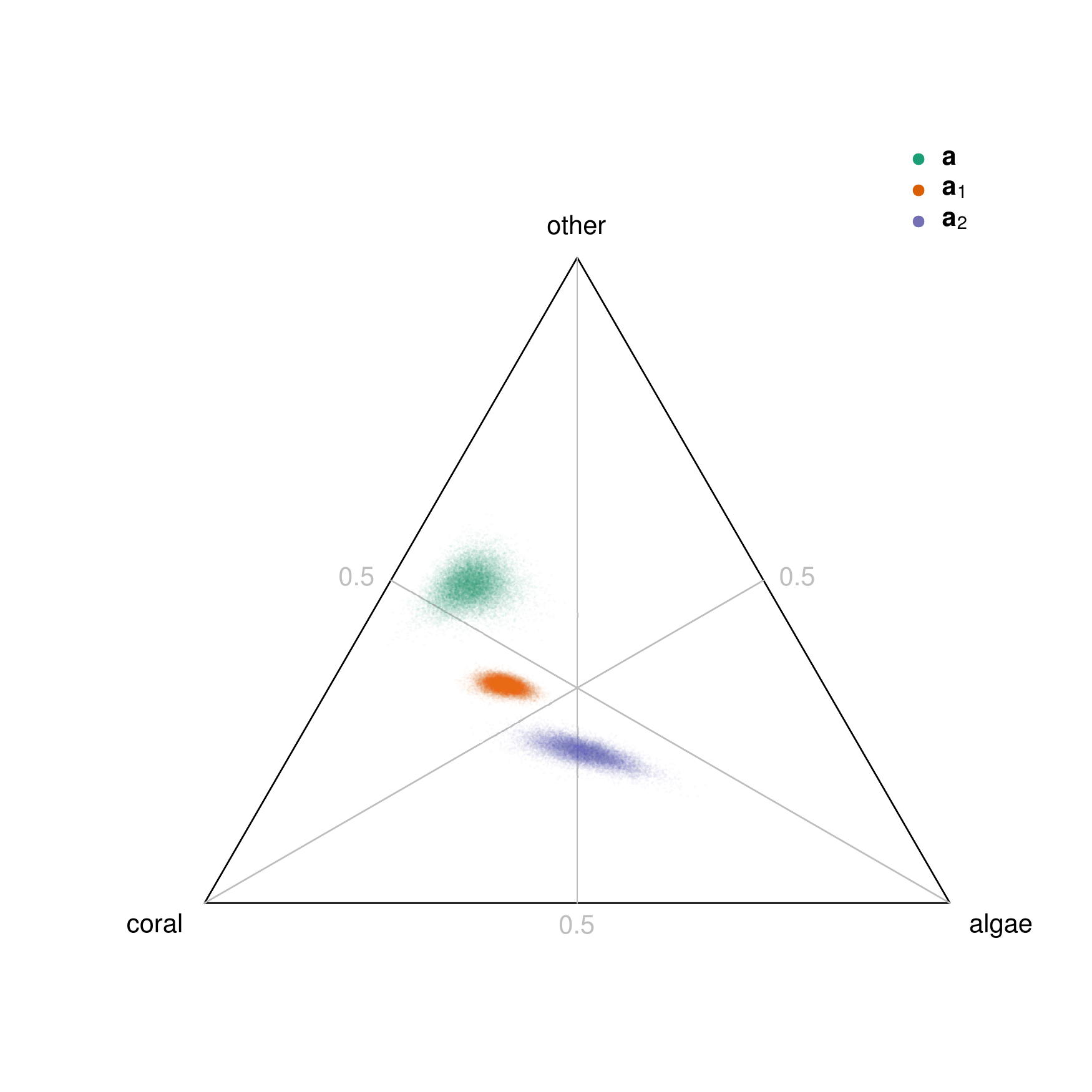}
\caption{}
\label{fig:parms}
\end{figure}

Figure \ref{fig:parms}. Posterior distributions of the back-transformed overall intercept $\mathbf a$ (green), effect $\mathbf a_1$ of component 1 (proportional to log(algae/coral)) on year-to-year change (orange), and effect $\mathbf a_2$ of component 2 (proportional to log(other/geometric mean(algae,coral)) on year-to-year change (blue).

\begin{figure}[h]
\includegraphics[height=18cm]{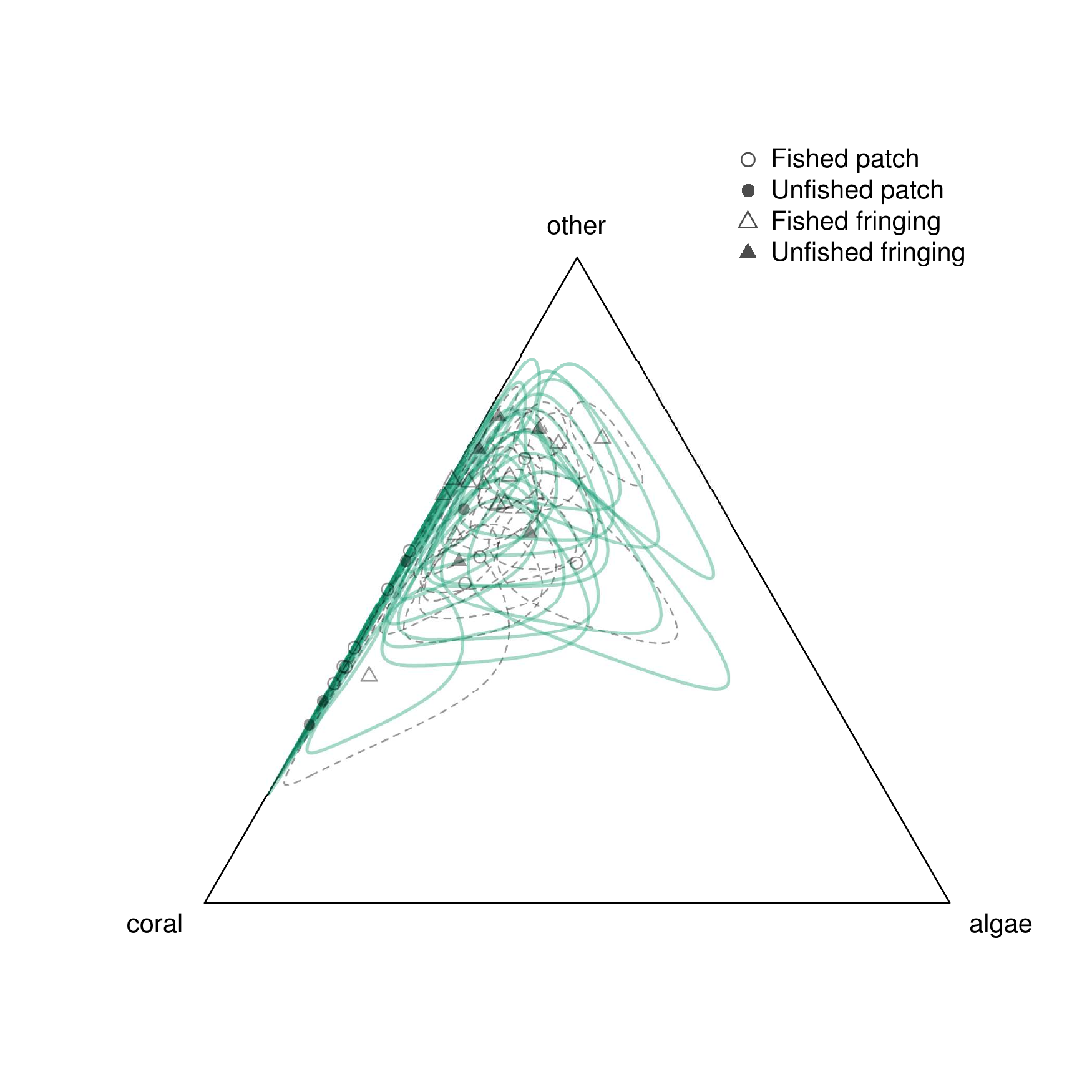}
\caption{}
\label{fig:amongreef}
\end{figure}

Figure \ref{fig:amongreef}. Stationary among- and within-site variation in benthic composition. Grey points: back-transformed stationary means for each site (open circles fished patch, filled circles unfished patch, open triangles fished fringing, filled triangles unfished fringing, posterior means of of stationary means). Grey dashed curves: back-transformed unit ellipsoids of concentration representing uncertainty in stationary means (calculated using sample covariance matrices from Monte Carlo iterations). Green solid curves: back-transformed unit ellipsoids of concentration representing within-site stationary variation (calculated using posterior mean within-site covariance matrix).

\begin{figure}[h]
\includegraphics[height=18cm]{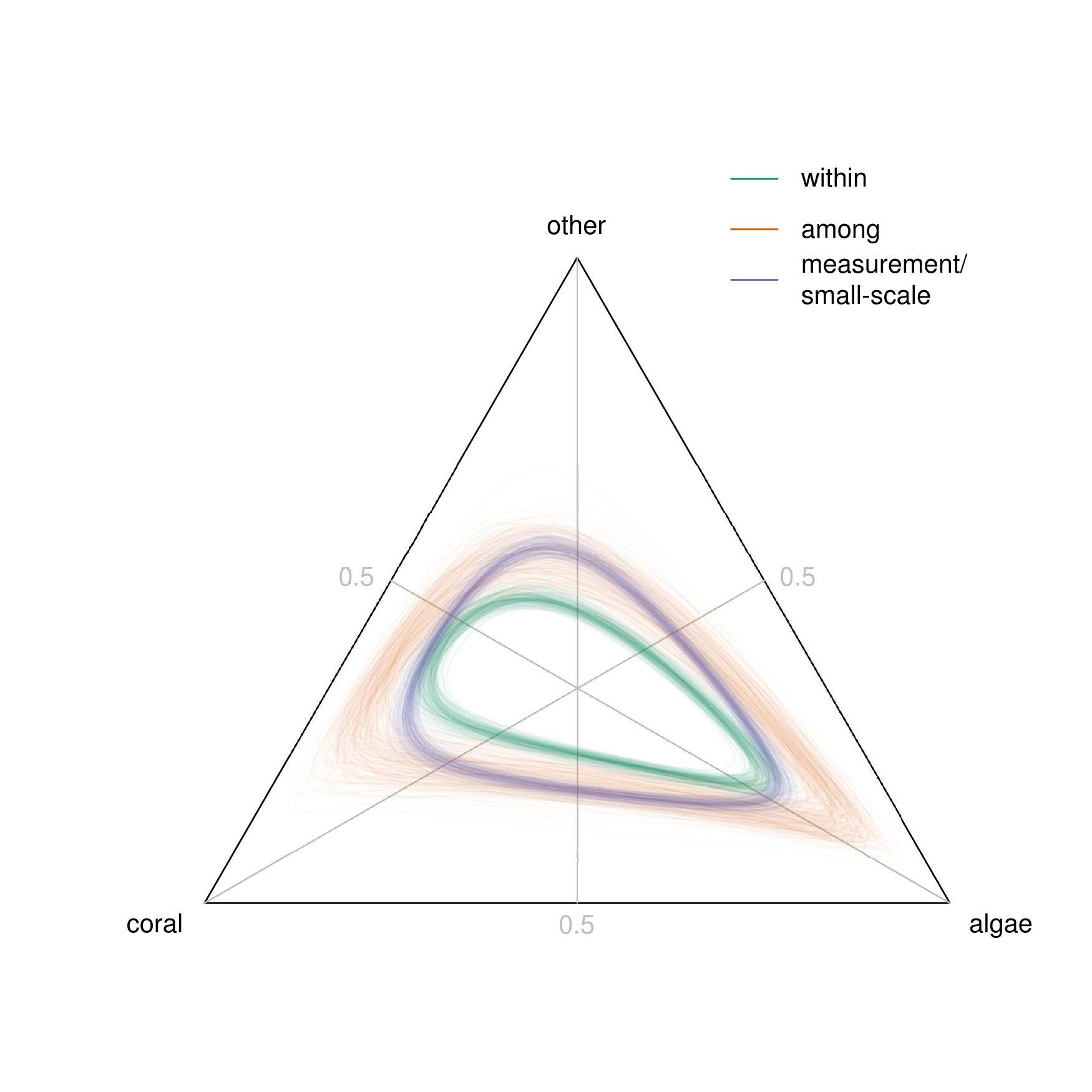}
\caption{}
\label{fig:covellipsesback}
\end{figure}

Figure \ref{fig:covellipsesback}. Back-transformed unit ellipsoids of concentration for stationary within-site covariance $\bm \Sigma^*$ (green), stationary among-site covariance $\mathbf Z^*$ (orange), and measurement error/small-scale spatial variation $\nu \mathbf H /(\nu-2)$ (blue). In each case, 200 ellipsoids drawn from the posterior distribution are plotted, centred on the origin.

\begin{figure}[h]
\includegraphics[angle=0, origin=c,height=17cm]{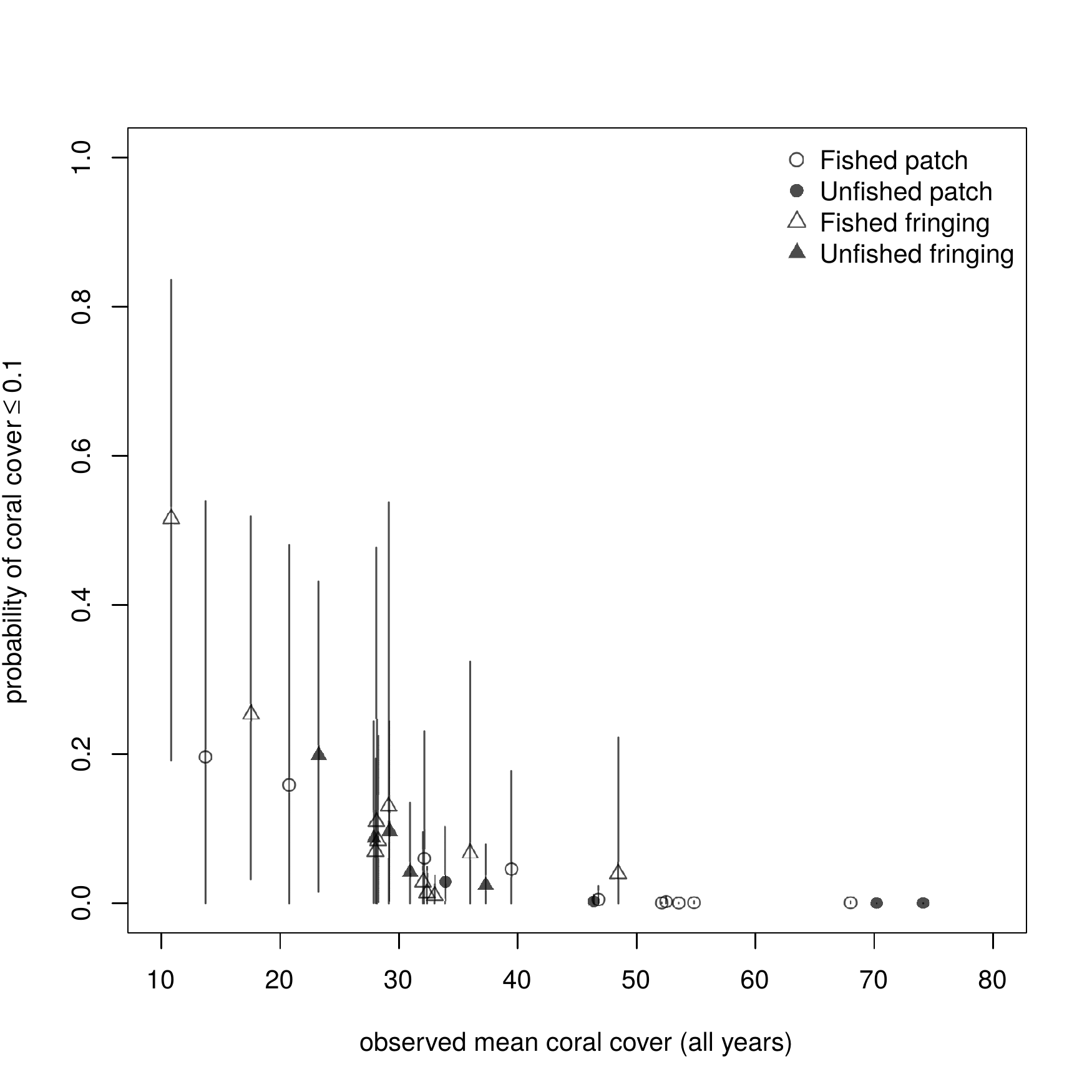}
\caption{}
\label{fig:problowcoralobs}
\end{figure}

Figure \ref{fig:problowcoralobs}. Long-term probability of coral cover less than or equal to 0.1 at each site against mean observed coral cover across all years. Circles are patch reefs and triangles are fringing reefs. Open symbols are fished reefs and shaded symbols are unfished. Vertical lines are 95$\%$ highest posterior density intervals.

\begin{figure}[h]
\includegraphics[height=18cm]{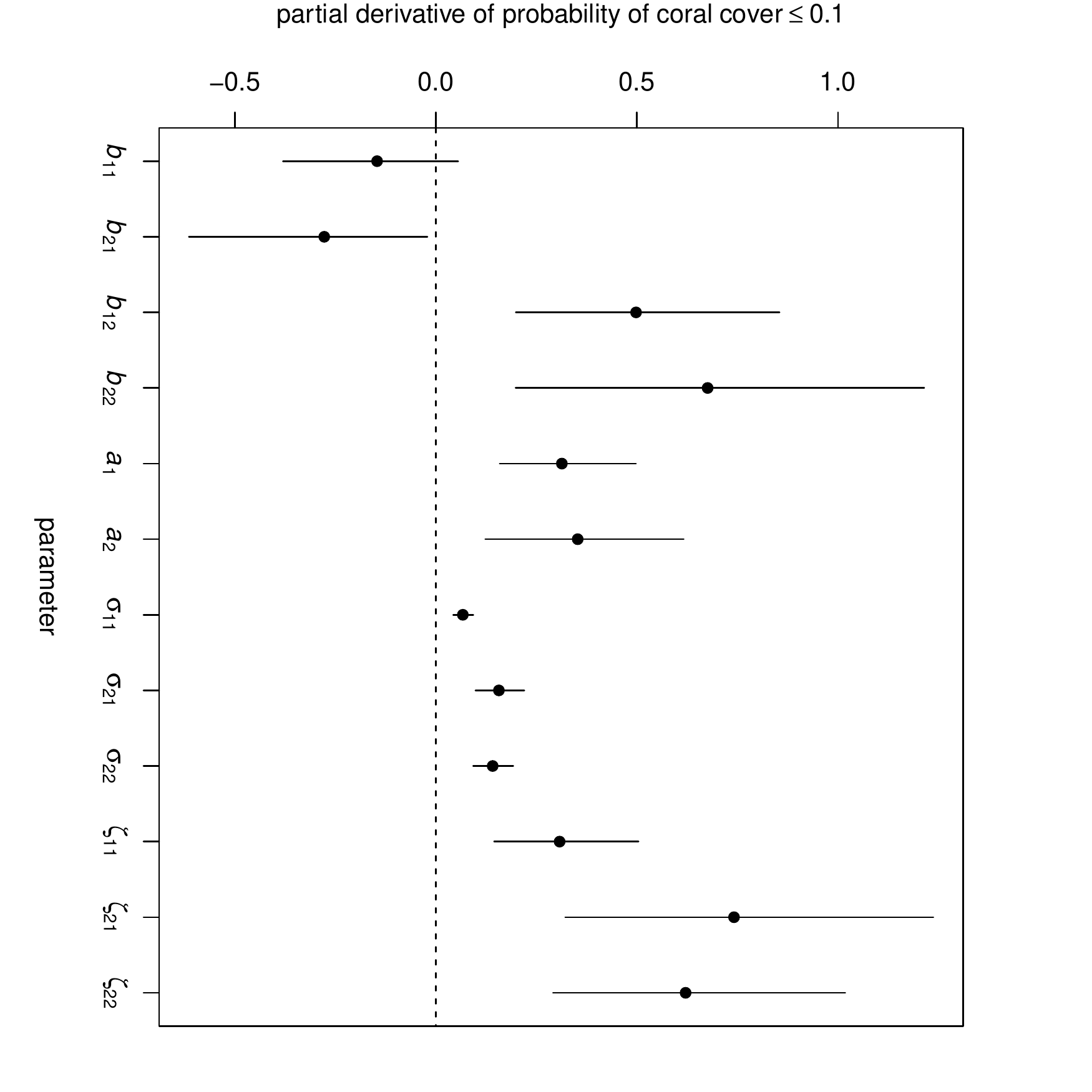}
\caption{}
\label{fig:derivs}
\end{figure}

Figure \ref{fig:derivs}. Elements of the gradient vector of partial derivatives of the long-term probability of coral cover less than or equal to 0.1 with respect to elements of the $\mathbf B$ matrix (effects of transformed composition in a given year on transformed composition in the following year), the $\mathbf a$ vector (overall intercept, representing among-site mean proportional changes in transformed composition at the origin), the covariance matrix of random temporal variation $\bm \Sigma$, and the covariance matrix of among-site variability $\mathbf Z$. For each parameter, the dot is the posterior mean and the bar is a 95$\%$ highest posterior density credible interval. For the covariance matrices, the elements $\sigma_{12}$ and $\zeta_{12}$ are not shown, because they are constrained to be equal to $\sigma_{21}$ and $\zeta_{21}$ respectively. The horizontal dashed line is at zero, the no-effect value.

\clearpage

\renewcommand{\thesection}{A\arabic{section}}
\renewcommand{\theequation}{A.\arabic{equation}}
\renewcommand{\thefigure}{A\arabic{figure}}
\renewcommand{\thetable}{A\arabic{table}}

\section{Data transformation}
\label{sec:datatransform}
\begin{figure}[h]
\includegraphics[height=20cm]{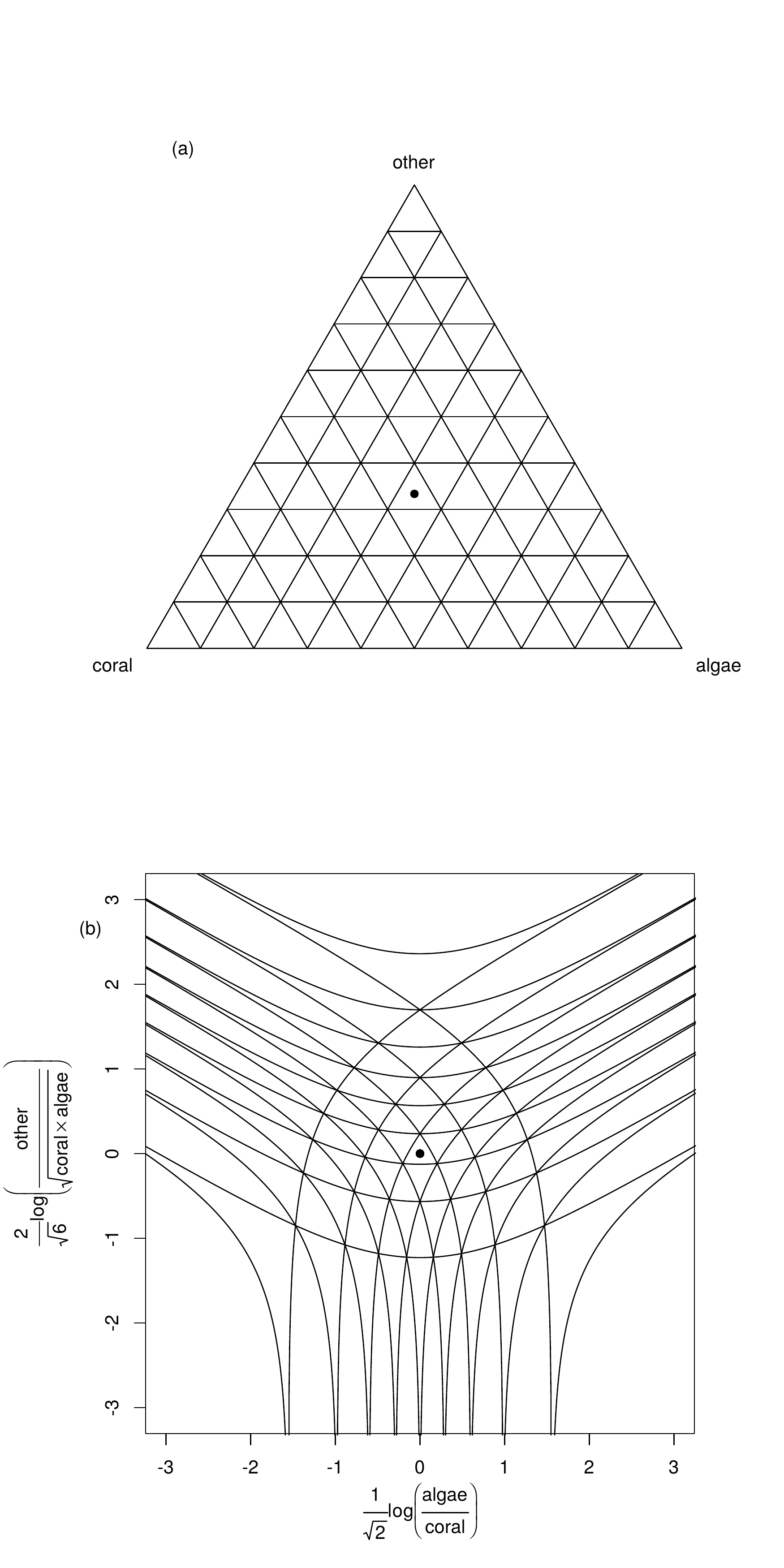}
\caption{The ilr transformation given by Equation \ref{eq:ilr}. (a) The open 2-simplex $\mathbb S^3$, in which three-part compositions lie. The dot represents the composition with equal relative abundances of coral, algae and other. Lines are contours of constant relative abundance of one part. (b) The ilr-transformed composition in $\mathbb R^2$, with dot and contours as in (a).}
\label{fig:ilrintuition}
\end{figure}

Proportional cover data were transformed to isometric log-ratio (ilr) coordinates \citep{Egozcue03}. Let $\mathbf z_{i,j,t} = [z_{1,i,j,t},z_{2,i,j,t},z_{3,i,j,t}]^T$ denote a vector of observed proportional cover of coral ($z_{1,i,j,t}$), algae ($z_{2,i,j,t}$) and other ($z_{3,i,j,t}$) at site $i$, transect $j$, at time $t$ (the $T$ denotes transpose). Then the ilr transformation for our data is given by
\begin{equation}
\begin{aligned}
\mathrm{ilr} \colon \mathbb S^3 &\to \mathbb R^2,\\
\mathbf z_{i,j,t} = [z_{1,i,j,t},z_{2,i,j,t},z_{3,i,j,t}]^T &\mapsto \left[\frac{1}{\sqrt{2}} \log \left( \frac{z_{2,i,j,t}} {z_{1,i,j,t}} \right),  \frac{2}{\sqrt{6}} \log \left( \frac{z_{3,i,j,t}}{\sqrt{z_{1,i,j,t} z_{2,i,j,t}}} \right) \right]^T,\\
\end{aligned}
\label{eq:ilr}
\end{equation}
where $\mathbb S^3$ denotes the open 2-simplex in which three-part compositions lie. The first element of the transformed composition is proportional to the natural log of the ratio of algae to coral, and the second element is proportional to the natural log of the ratio of other to the geometric mean of algae and coral. The transformation can be thought of as stretching out the open 2-simplex (Figure \ref{fig:ilrintuition}(a)) so that it covers the whole of the real plane (Figure \ref{fig:ilrintuition}(b)).

As the domain of the transformation is the open simplex, which does not include compositions with zero parts, any observed zeros were replaced by half the smallest non-zero value recorded (0.0008) before transformation, and the other components rescaled accordingly. This is the simple replacement strategy described in \citet{MartinFernandez03}, although more sophisticated approaches are possible. We denote the resulting transformed observations by $\mathbf y_{i,j,t} = [y_{1,i,j,t},y_{2,i,j,t}]^T$.

\section{The model}

For convenience, we reproduce the full model equations here:
\begin{equation}
\begin{aligned}
\mathbf x_{i,t+1} &= \mathbf a + \bm \alpha_i + \mathbf B \mathbf x_{i,t} + \bm \varepsilon_{i,t}, \\
\bm \alpha_i &\sim \mathcal N(\mathbf 0, \mathbf Z),\\
\bm \varepsilon_{i,t} &\sim \mathcal N(\mathbf 0, \bm \Sigma)\\
\mathbf y_{i,j,t} &\sim t_2(\mathbf x_{i,t}, \mathbf H, \nu), 
\end{aligned}
\label{eq:modelapp}
\end{equation}
where $\mathbf x_{i,t}$ is the true transformed composition at site $i$, time $t$, $\mathbf a$ is a vector of among-site mean proportional changes evaluated at $\mathbf x_{i,t} = \mathbf 0$, $\bm \alpha_i$ represents the amount by which these proportional changes for the $i$th site differ from the among-site mean, the $2 \times 2$ matrix $\mathbf B$ represents the effects of $\mathbf x_{i,t}$ on the proportional changes, $\bm \varepsilon_{i,t}$ represents random temporal variation,

\begin{equation*}
\mathbf Z = \begin{bmatrix} \zeta_{11} & \zeta_{12} \\
\zeta_{21} & \zeta_{22}
\end{bmatrix}
\end{equation*}
is the covariance matrix of the among-site term $\bm \alpha_i$ (note that throughout, a diagonal element such as $\zeta_{ii}$ of a covariance matrix represent the variance of the $i$th variable), 

\begin{equation*}
\mathbf \Sigma = \begin{bmatrix} \sigma_{11} & \sigma_{12} \\
\sigma_{21} & \sigma_{22}
\end{bmatrix}
\end{equation*}
is the covariance matrix of the temporal variation, $\mathbf y_{i,j,t}$ is the observed log-ratio transformed cover in the $j$th transect of site $i$ at time $t$,  

\begin{equation*}
\mathbf H = \begin{bmatrix} \eta_{11} & \eta_{12} \\
\eta_{21} & \eta_{22}
\end{bmatrix}
\end{equation*}
is the scale matrix of the bivariate $t$ distribution of the $\mathbf y_{i,j,t}$, and $\nu$ is the corresponding degrees of freedom.

\section{Describing measurement error and small-scale temporal variability}
\label{sec:measurement}
We initially considered using a bivariate normal distribution to describe the variability of observed transformed composition $\mathbf y_{i,j,t}$ around true composition $\mathbf x_{i,t}$, but preliminary analyses showed that a heavier-tailed distribution was needed. We therefore used the bivariate $t$ distribution with location vector $\mathbf x_{i,t}$, scale matrix $\mathbf H$ and degrees of freedom $\nu$, which for $\nu > 2$ has covariance matrix $\nu \mathbf H/(\nu-2)$ \citep{Lange89}. Support for the choice of the $t$ over the normal distribution was provided by expected predictive accuracy based on leave-one-out cross-validation \citep{Vehtari15}, which was much higher for the bivariate $t$ model than for the bivariate normal model (difference in leave-one-out cross-validation score 527, standard error 48). 

\section{Visualizing model parameters}
\label{sec:visualize}
The effects of reef composition on short-term dynamics are most easily visualized by the back transformation from ilr coordinates to the simplex of the columns of the matrix $\mathbf A = \mathbf B - \mathbf I_2$, where $\mathbf I_k$ denotes the $k \times k$ identity matrix. The matrix $\mathbf A$ describes effects of transformed reef composition on year-to-year changes in transformed reef composition \citep{Cooper15}. This is a better visualization than the back transformation of $\mathbf B$, because in the random walk case (where there are no interesting composition effects), $\mathbf A = \mathbf 0_2$ (the $2 \times 2$ matrix of zeros), and each column of the back-transformation of $\mathbf A$ represents a point at the origin of the simplex. In contrast, in the random walk case, each column of the back transformation of $\mathbf B = \mathbf I_2$ represents a point at a different location in the simplex. The first column $\mathbf a_1$ of $\mathbf A$ represents the effect of a unit increase in the first component of reef composition (proportional to log(algae/coral)) on year-to-year change in reef composition. For example, if the back-transformation of $\mathbf a_1$ lies to the left of the centre of the simplex (the origin, with equal proportions of coral, algae and other), but on the line of equal relative abundances of coral and other (the 1:1 coral-other isoproportion line), it indicates that high algal cover relative to coral tends to result in a decrease in algae relative to coral in the following year. Similarly, the second column $\mathbf a_2$ of $\mathbf A$ represents the effect of a unit increase in the second component of reef composition (proportional to log(other/geometric mean(algae,coral))) on year-to-year change in reef composition.

\section{Parameter estimation}
\label{sec:estimation}

Code for all analyses is available at \url{https://www.liverpool.ac.uk/~matts/kenya.zip}.

\subsection{Priors}

For $\mathbf Z$ and $\bm \Sigma$, our priors were based on data from the Great Barrier Reef \citep{Cooper15}. We inspected the sample covariance matrices for ilr-transformed year-to-year changes in composition, and among-site variation in mean composition, on 55 sites in the Great Barrier Reef, where observation error is thought to be fairly small \citep{Cooper15}. We chose inverse Wishart priors \citep[p. 574]{GCSR03} with 4 degrees of freedom (the smallest value for which the prior mean exists, giving a fairly uninformative prior). We chose identity scale matrices, because ellipses of unit Mahalanobis distance around the origin for the mean of this prior almost enclosed corresponding ellipses for the sample covariance matrices of both year-to-year changes and among-site mean composition, and strong correlations among transformed components are neither assumed nor ruled out. Thus, this seems a plausible prior for $\bm \Sigma$ and $\mathbf Z$. In the absence of strong prior information, we used the same prior for $\mathbf H$. 

For the degrees of freedom of measurement error, $\nu$, we assumed a $U(2,30)$ distribution. The lower bound was dictated by the requirement that $\nu > 2$ for the covariance to exist, and the upper bound was chosen to be large enough that the resulting measurement error distribution was able to approach a multivariate normal if necessary. In practice, the posterior distribution of $\nu$ did not pile up against either of these bounds, indicating that the precise choice of prior was unlikely to matter.

We chose vague priors for the other parameters. We assumed independent $\mathcal N(0,10)$ priors on each element of $\mathbf x_{i,0}$ for each site $i$ (where the subscript 0 denotes the first time point at which the site was observed). For each element of $\mathbf a$ and $\mathbf B$, we assumed independent $\mathcal N(0,100)$ priors. 

\subsection{Monte Carlo simulation}
\begin{figure}[h]
  \includegraphics[height=17cm]{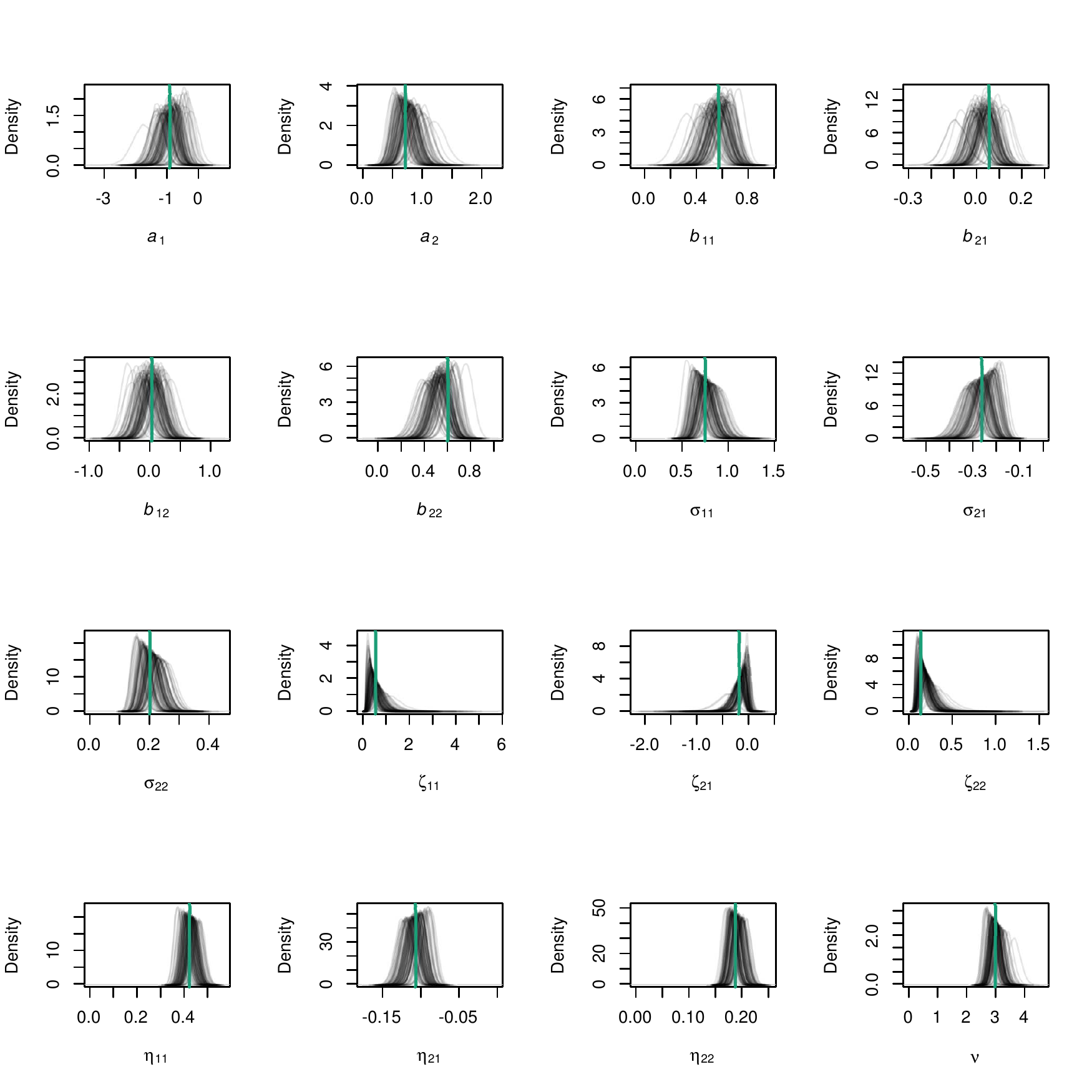}
\caption{Posterior distributions of parameters estimated from simulated data. Thick green vertical lines: parameter values used to generate simulated data (posterior means from real data). Black lines: kernel density estimates of posterior distributions from 100 simulated data sets, each with the same number of sites, number and spacing of observation times, and numbers of transects at each observation time, as the real data. Number of simulated data sets in which true value was within 95$\%$ HPD interval: 89 ($a_1$), 95 ($a_2$), 97 ($b_{11}$), 91 ($b_{21}$), 95 ($b_{12}$), 90 ($b_{22}$), 99 ($\sigma_{11}$), 96 ($\sigma_{21}$), 93 ($\sigma_{22}$), 96 ($\zeta_{11}$), 93 ($\zeta_{21}$), 98 ($\zeta_{22}$), 93 ($\eta_{11}$), 93 ($\eta_{21}$), 96 ($\eta_{22}$), 93 ($\nu$).}
\label{fig:simdata}
\end{figure}

We ran four Monte Carlo chains in parallel for 5000 iterations each, after a 5000-iteration warmup period. This took approximately two hours on a 64-bit Ubuntu 12.04 system with 4 3.2 GHz Intel Xeon cores and 16 GiB RAM. The potential scale reduction statistic, which takes the value 1 if all chains have converged to a common distribution, was 1.00 to two decimal places for all parameters, consistent with satisfactory convergence \citep[pp. 414-415]{stan-manual:2015}. Effective sample sizes, which measure the size of the sample from the posterior distribution after accounting for autocorrelation in the Monte Carlo chains \citep[pp. 417-419]{stan-manual:2015}, were at least 2839 for all parameters (most were much larger, with first quartile 12430 and median 17490). Inspection of trace plots did not reveal any obvious problems with sampling. In addition, we evaluated the model's performance in estimating known parameters. We generated 100 simulated data sets with identical structure to the real data, using posterior mean estimates for each parameter. We sampled the $\bm \alpha_i$, $\bm \varepsilon_{i,t}$ and $\mathbf y_{i,j,t}$ from distributions defined by Equation \ref{eq:modelapp}, and set the initial true transformed compositions at a given site to the sample means from all years and transects on that site in the real data. The estimates were reasonably close to the true values, and lay within the 95$\%$ HPD intervals in 89-99 out of 100 cases (Figure \ref{fig:simdata}). Thus, while estimating state-space models from ecological time series data can be challenging \citep{AugerMethe15}, performance appears adequate in this case, perhaps because we have many replicate transects from which to estimate measurement error and small-scale spatial variability, and most parameters are estimated using data across many sites.

\subsection{Model checking}
\begin{figure}[h]
\includegraphics[height=20cm]{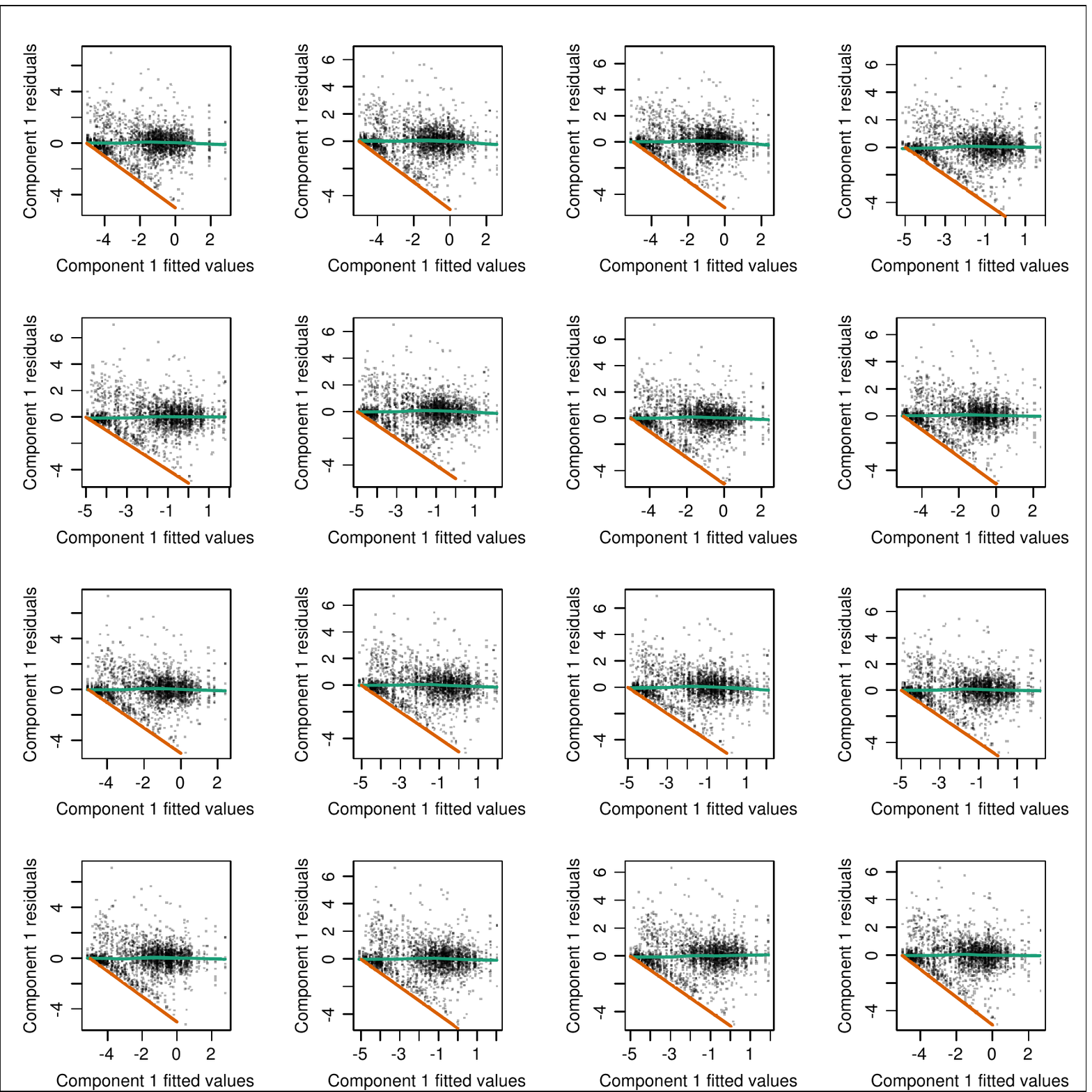}
\caption{Fitted values against Bayesian residuals for component 1. Each panel is a single randomly-chosen Monte Carlo iteration. Dots represent Bayesian residuals against fitted values for individual transects. The green line is a loess smoother. The orange line is the minimum possible value for component 1 residuals.} 
\label{fig:resids1}
\end{figure}

\begin{figure}[h]
\includegraphics[height=20cm]{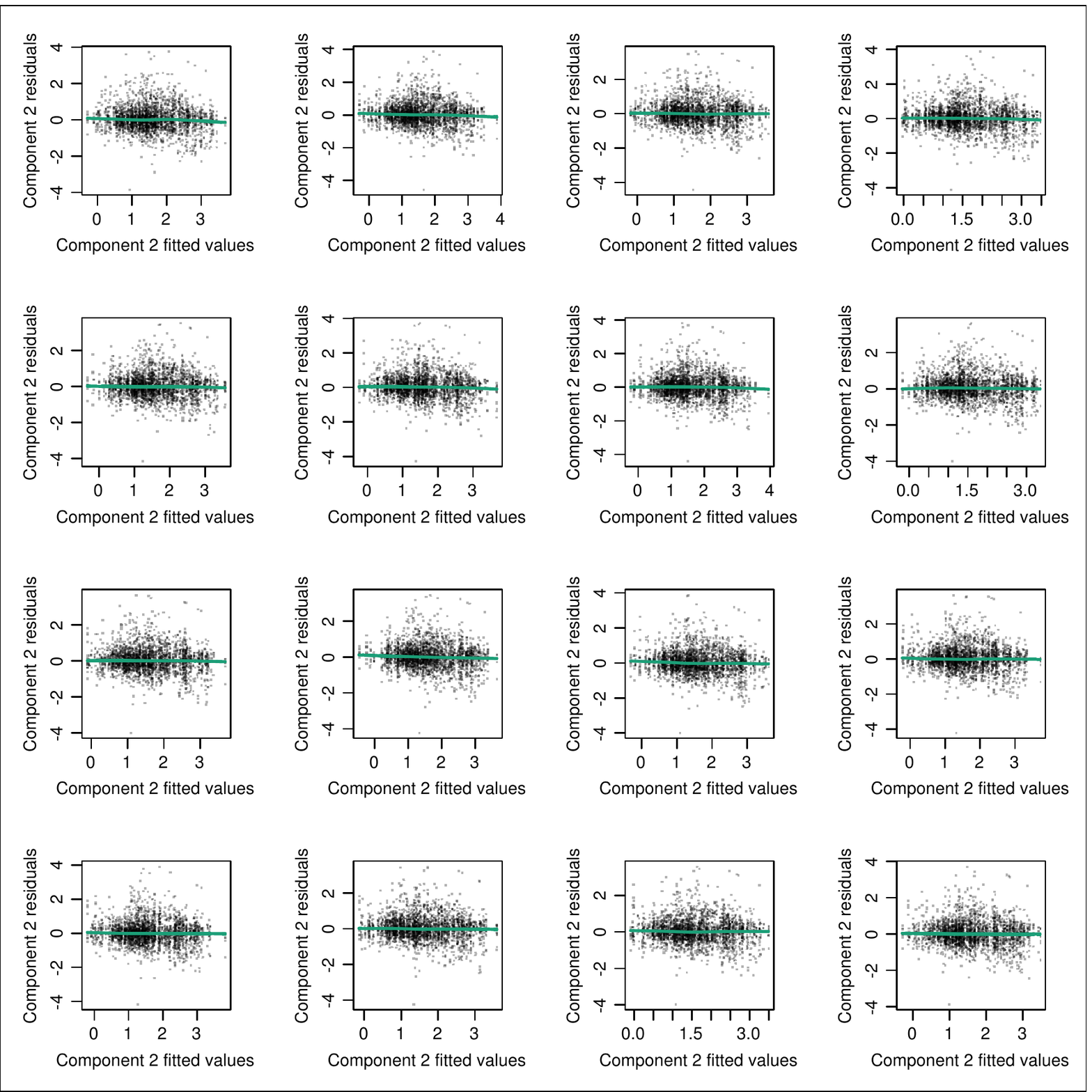}
\caption{Fitted values against residuals for component 2. Each panel is a single randomly-chosen Monte Carlo iteration. Dots represent Bayesian residuals against fitted values for individual transects. The green line is a loess smoother.}
\label{fig:resids2} 
\end{figure}

We examined plots of Bayesian residuals \citep[p. 170]{GCSR03} against predicted values of the two components of transformed reef composition. For the $k$th Monte Carlo iteration, the Bayesian residual for the $j$th transect on the $i$th site at time $t$ is $\mathbf y_{i,j,t} - \mathbf x_{i,t}|\bm \theta_k$, where $\bm \theta_k$ denotes the estimated parameters in the $k$th iteration. If the model is performing well, there should be no obvious relationship between residuals and fitted values. We checked 16 randomly-chosen iterations, which did not reveal any major cause for concern (Figures \ref{fig:resids1}, \ref{fig:resids2}). However, no residuals for component 1 fell below an obvious diagonal line (Figure \ref{fig:resids1}), which results from the treatment of observed zeros. Given the simple replacement strategy for zeros described in Section \ref{sec:datatransform} and the definition of component 1 of the transformed composition in Equation \ref{eq:ilr}, 

\begin{equation*}
\begin{aligned}
y_{1,i,j,t} &= \frac{1}{\sqrt{2}} \log \left( \frac{z_{2,i,j,t}} {z_{1,i,j,t}} \right)\\
&\geq \frac{1}{\sqrt{2}} \log \left(\frac{0.0008}{0.9984}  \right)
&= -5.0216.
\end{aligned}
\end{equation*}

Thus the Bayesian residual for component 1 is constrained by

\begin{equation*}
y_{1,i,j,t} - x_{1,i,t}|\bm \theta_k \geq -5.0216 - x_{1,i,t}|\bm \theta_k,
\end{equation*}
the orange line on Figure \ref{fig:resids1}. Thus the assumption of a multivariate $t$ distribution for individual transect deviations from true values (Equation \ref{eq:modelapp}) cannot hold exactly. It might in future be worth attempting to develop a more mechanistic model of the process generating observed zeros, but we do not attempt this here because the majority of data are unaffected. Although a similar constraint exists on component 2, it did not appear to be important in practice, because there is no obvious diagonal line of residuals on Figure \ref{fig:resids2}.

Inspection of quantile-quantile plots and histograms of estimated skewness and kurtosis for 16 iterations did not indicate any major problems with the assumptions of multivariate normal distributions with zero mean, covariance matrices $\mathbf Z$ and $\bm \Sigma$ respectively for $\bm \alpha$ and $\bm \varepsilon$, and a multivariate $t$ distribution with zero location vector, scale matrix $\mathbf H$, for Bayesian residuals. Quantile-quantile plots used the natural log of a squared Mahalanobis-like distance/2 against natural log of quantiles of $\chi^2(2)$ for multivariate normal distributions, or against natural log of quantiles of $F(2,\nu)$ for multivariate $t$ distributions \citep[modified from][]{Lange89}.  We did not transform to asymptotically standard normal deviates because the degrees of freedom for the $t$ distribution were small. We found it helpful to log transform both axes, particularly for the multivariate $t$ distribution, for which some observations may have very large squared Mahalanobis-like distance. We obtained the $p$-values for several tests of multivariate normality of $\bm \alpha$ and $\bm \varepsilon$: Royston's $H$ \citep{Royston82}, Henze-Zirkler's test \citep{Henze90}, and Mardia's skewness and kurtosis \citep{Mardia70} using the MVN package in R \citep{Korkmaz14}. There were more small $p$-values than expected (the distribution of $p$-values should be approximately uniform in the interval (0,1) if the data are normal) but that often is the case for very large samples, and does not indicate a major cause for concern.

\section{Long-term behaviour}
\label{sec:longterm}
Iterating Equation \ref{eq:modelapp} from a fixed initial transformed composition $\mathbf x_{i,0}$,
\begin{equation}
\mathbf x_{i,t} = \sum_{j=0}^{t-1} \mathbf B^j \mathbf a + \sum_{j=0}^{t-1} \mathbf B^j \bm \alpha_i + \mathbf B^t \mathbf x_0 + \sum_{j=0}^{t-1} \mathbf B^j \bm \varepsilon_{i,t-1-j}
\label{eq:iterate}
\end{equation}

If all the eigenvalues of $\mathbf B$ lie inside the unit circle in the complex plane, the system will converge to a stationary distribution as $t \rightarrow \infty$ \citep[e.g.][p. 10]{Lutkepohl93}. If the eigenvalues of $\mathbf B$ are complex, they will form a complex conjugate pair $\lambda = re^{\pm i \theta}$ (where $r$ is the magnitude and $\theta$ is the argument), and there will be oscillations with period $2 \pi/\theta$, whose amplitudes will change by a factor of $r$ each year \citep[e.g.][p. 355]{Otto07}.

The first term in Equation \ref{eq:iterate} is deterministic, and converges to 
\begin{equation}
\bm \mu^* = (\mathbf I_2 - \mathbf B)^{-1} \mathbf a
\label{eq:stationarymean}
\end{equation}
 \citep[e.g.][p. 10]{Lutkepohl93}, which represents the among-site mean of stationary mean transformed composition. The third term is also deterministic, and converges to $\mathbf 0$, so that initial conditions are forgotten.

 The second term, representing among-site variation, has mean vector $\mathbf 0$ by definition, and the covariance matrix of its limit is

\begin{align}
\mathbf Z^* &= \mathrm V \left[ (\mathbf I_2 - \mathbf B)^{-1} \bm \alpha_i \right] \notag \\
&= (\mathbf I_2 - \mathbf B)^{-1} \mathrm V \left[ \bm \alpha_i \right] \left((\mathbf I_2 - \mathbf B)^{-1} \right) ^T \notag\\
&= (\mathbf I_2 - \mathbf B)^{-1} \mathbf Z \left((\mathbf I_2 - \mathbf B)^{-1} \right) ^T, \label{eq:Zetastar}
\end{align}
since $(\mathbf I_2 - \mathbf B)^{-1}$ is a constant matrix and $\bm \alpha_i$ is a random vector. The covariance matrix $\mathbf Z^*$ represents the among-site variation in stationary mean transformed composition.

The fourth term represents the long-term effects of temporal variability. It has mean vector $\mathbf 0$ by definition, and it can be shown that it has covariance matrix
\begin{equation}
\bm \Sigma^* = \mathrm{vec}^{-1} \left((\mathbf I_4 - \mathbf B \otimes \mathbf B)^{-1} \mathrm{vec} \left( \bm \Sigma \right) \right) \label{eq:Sigmastar}
\end{equation}
\citep[e.g.][p. 22]{Lutkepohl93}, where the $\mathrm{vec}$ operator stacks the columns of a matrix, $\mathrm{vec}^{-1}$ unstacks them, and $\otimes$ is the Kronecker product. The covariance matrix $\bm \Sigma^*$ can be interpreted as the stationary covariance of transformed reef composition, conditional on the value of $\bm \alpha_i$. Since among-site variation and temporal variation were assumed independent, the unconditional stationary covariance is $\bm \Sigma^* + \mathbf Z^*$. Both the conditional and unconditional stationary distributions are multivariate normal, since both $\bm \varepsilon_{i,t}$ and $\bm \alpha_i$ were assumed multivariate normal. Thus the stationary distribution for a randomly-chosen site is the multivariate normal vector
\begin{equation}
\mathbf x^* \sim \mathcal N(\bm \mu^*, \bm \Sigma^* + \mathbf Z^*).
\label{eq:stationaryapp}
\end{equation}

To find the long-term behaviour for a given site $i$, we condition on the value of $\bm \alpha_i$. Thus Equation \ref{eq:stationarymean} is replaced by

\begin{equation*}
\bm \mu_i^* = (\mathbf I_2 - \mathbf B)^{-1} (\mathbf a + \bm \alpha_i),
\end{equation*}
and the stationary distribution is

\begin{equation*}
\mathbf x_i^* \sim \mathcal N(\bm \mu_i^*, \bm \Sigma^*).
\end{equation*}

\section{How important is among-site variability?}
\label{sec:amongreef}

From Equation \ref{eq:stationaryapp}, the covariance matrix $\bm \Sigma^* + \mathbf Z^*$ of the stationary distribution for a randomly-chosen site contains contributions from both among- and within-site variability. To quantify the contributions from these two sources, we will use a statistic based on a ratio of generalized variances. 

The generalized variance of a multivariate distribution is defined as the determinant of the covariance matrix \citep[section 3.4]{Wilks32, Johnson07}. In the specific case of a multivariate normal distribution, the generalized variance may be interpreted in terms of {\em ellipsoids of concentration}, defined as follows. Suppose a random vector $\mathbf W$ is distributed according to a $p$-dimensional normal distribution with mean vector $\bm \mu$ and covariance matrix $\mathbf V$. Then for any constant $k \ge 0$, the set $E_k = \left\{ \mathbf w : \left( \mathbf w - \bm \mu \right)^T \mathbf V^{-1} \left( \mathbf w- \bm \mu \right) = k \right\}$ consists of points $\mathbf w$ of constant probability density. In $p=2$ dimensions, $E_k$ is an ellipse, and may be referred to as a probability density contour. In $p>2$ dimensions $E_k$ is known as an ellipsoid of concentration of $\mathbf V$ about $\bm \mu$ \citep{Kenward79}. Taking $k=1$, the set $E_1$ is known as the unit ellipsoid of concentration. The volume within the unit ellipsoid~$E_1$ may be used as a measure of the dispersion of the distribution, and is equal to $S_p \sqrt{\left| \mathbf V \right|}$, where $S_p$ is the volume of the $p$-dimensional sphere of radius~1.

In the light of the above interpretation, we chose to measure the contribution of within-site variability to total variability using the quantity
\begin{equation}
\rho = \left(\frac{|\bm \Sigma^*|}{|\bm \Sigma^* + \mathbf Z^*|} \right)^{1/2},
\label{eq:vratapp}
\end{equation}
which is the ratio of volumes of two unit ellipsoids of concentration, the numerator corresponding to the stationary distribution in the absence of among-site variation, and the denominator to the full stationary distribution of transformed reef composition in the region. This ratio is undefined if $\bm \Sigma^* + \mathbf Z^*$ is not of full rank, but this does not occur in our application. From Minkowski's theorem \citep[][section 13.5]{Mirsky55} it follows that $| \bm \Sigma^*| + | \mathbf Z^*| \leq | \bm \Sigma^* + \mathbf Z^*|$, so that $0 \leq \rho \leq 1$. However, in general $| \bm \Sigma^*| + | \mathbf Z^*| \neq | \bm \Sigma^* + \mathbf Z^*|$, so that $\rho$ cannot be simply interpreted as the proportion of total variability explained by within-site variation. Nevertheless, $\rho$ provides an indication of how much of the total variability would remain if all among-site variability was removed. Furthermore, $\rho^2$ is analogous to Wilks' Lambda \citep{Wilks32, Kenward79}, a likelihood-ratio test statistic often used in multivariate analysis of variance.

\section{Probability of low coral cover}
\label{sec:plowintegral}
For a given site $i$, the long-term probability $q_{\kappa,i}$ of coral cover less than or equal to $\kappa$ is the integral of the multivariate normal stationary density for the site over the shaded area in Figure \ref{fig:derivintuition} (for $\kappa=0.1$). This can be written as
\begin{equation}
q_{\kappa,i} = 1- \int_{-\infty}^u P(X_2 \leq \gamma | X_1 = x_1) f_{X_1}(x_1) \, \mathrm{d} x_1,
\label{eq:integralplow}
\end{equation}
where, using Equations \ref{eq:ilr} and the constraint that the untransformed components of benthic composition must sum to 1,

\begin{equation*}
u = \frac{1}{\sqrt{2}}\log \left(\frac{1}{\kappa} -1 \right)
\end{equation*}
is the largest value of the first ilr component $x_1$ for which it is possible to have coral cover less than or equal to $\kappa$,

\begin{equation*}
\gamma = \frac{2}{\sqrt{6}} \log \left(\frac{1 - \kappa \left( 1 + e^{\sqrt{2} x_1}\right)}{\kappa \sqrt{e^{\sqrt{2} x_1}}} \right)
\end{equation*}
is the value of the second ilr component $x_2$ for which coral cover is equal to $\kappa$, given the value of $x_1$, $P(X_2 \leq \gamma | X_1 = x_1)$ is the conditional marginal cumulative distribution of $x_2$, given the value of $x_1$, and $f_{X_1}(x_1)$ is the unconditional marginal density of the first ilr component $x_1$.

Since 

\begin{equation*}
\mathbf X = [X_1,X_2]^T \sim \mathcal N(\mathbf \mu_i^*, \bm \Sigma_i^*),
\end{equation*}
the unconditional marginal distribution of $x_1$ is
\begin{equation}
\mathcal N(\mu_{1,i}^*, \sqrt{\sigma_{11,i}^*}),
\label{eq:x1margin}
\end{equation}
 and the conditional marginal distribution of $x_2$ given $x_1$ is
\begin{equation}
\mathcal N\left(\mu_{2,i}^* + \frac{\sigma_{21,i}^*}{\sigma_{11,i}^*} (x_1 - \mu_{i,1}^*), \sigma_{22,i}^* - \frac{(\sigma_{21,i}^*)^2}{\sigma_{11,i}^*} \right)
\label{eq:x2condmargin}
\end{equation}
\citep[p. 579]{GCSR03}. Then the integral in Equation \ref{eq:integralplow} can be approximated numerically using the \verb+integrate()+ function in R \citep{RCore15}, which is based on routines in \citet{Piessens83}. The same approach can be used for $q_\kappa$ for a randomly-chosen site, replacing the elements of $\bm \mu_i^*$ and $\bm \Sigma_i^*$ in Equations \ref{eq:x1margin} and \ref{eq:x2condmargin} with the corresponding elements of $\bm \mu^*$ and $\bm \Sigma^*$.

\section{Spline correlograms for spatial pattern in probability of low coral cover}
\label{sec:plowcorr}

We calculated a spline correlogram \citep{Bjornstad01} for each set of $q_{0.1,i}$ in the 20000 Monte Carlo iterations, using the \verb+spline.correlog()+ function in the R package \verb+ncf+ version 1.15. We constructed a 95$\%$ highest-density envelope \citep{Hyndman_4081} for the resulting set of correlograms using the R package \verb+hdrcde+ version 3.1.

\section{What is the most effective way to reduce the probability of low coral cover?}
\label{sec:plow}
For a given threshold $\kappa$, we can calculate (by numerical integration) the probability $q_\kappa = P(\text{coral cover} \leq \kappa)$, for a composition drawn from the stationary distribution on a site chosen at random from the region. The probability $q_\kappa$ is a function of 12 parameters: all four elements of $\mathbf B$; both elements of $\mathbf a$; elements $\sigma_{11}$, $\sigma_{21}$ and $\sigma_{22}$ of $\bm \Sigma$; and elements $\zeta_{11}$, $\zeta_{21}$ and $\zeta_{22}$ of $\mathbf Z$. Note that because $\bm \Sigma$ and $\mathbf Z$ are covariance matrices, they must be symmetric, and so $\sigma_{12}$ and $\zeta_{12}$ are not free parameters. These 12 parameters can be thought of as the coordinates of a point in $\mathbb R^{12}$. The steepest reduction in $q_\kappa$ as we move through $\mathbb R^{12}$ is achieved by moving in the direction of $- \bm \nabla q_\kappa$, where $\bm \nabla q_\kappa$ is the gradient vector $[ \partial q_\kappa/\partial b_{11}, \ldots , \partial q_\kappa / \partial \zeta_{22}]^T$ \citep[p. 355]{RHB02}.

To understand the effects of each parameter, note that the probability $q_\kappa$ depends on these parameters only through $\bm \mu^*$, $\bm \Sigma^*$ and $\mathbf Z^*$. Thus, for any parameter matrix $\bm \Theta$, using the chain rule for matrix derivatives,

\begin{equation*}
D q_\kappa(\bm \Theta) = D q_\kappa(\bm \mu^*) D \bm \mu^*(\bm \Theta) + D q_\kappa(\bm \Sigma^*) D \bm \Sigma^*(\bm \Theta) + D q_\kappa (\mathbf Z^*) D \mathbf Z^*(\bm \Theta),  
\end{equation*}
where $D \mathbf E(\mathbf X)$ denotes the matrix derivative of $\mathbf E$ with respect to $\mathbf X$ \citep[p. 108]{Magnus07}. This allows us to break up the effects of a parameter into its effects via the stationary mean and stationary within- and among-site covariances. In each term, the first factor ($D q_\kappa(\bm \mu^*)$, $D q_\kappa(\bm \Sigma^*)$ or $D \bm \Sigma^*(\bm \Theta)$) can only be found numerically. The non-zero second factors are

\begin{align}
D \bm \mu^*(\mathbf B) &= (\mathbf a^T \otimes \mathbf I_2)\left[ \left( (\mathbf I_2 - \mathbf B)^{-1} \right)^T \otimes (\mathbf I_2 - \mathbf B)^{-1}\right], \label{eq:DmustarB}\\
D \bm \Sigma^*(\mathbf B) &= \mathbf F \left[ (\mathrm{vec} \bm \Sigma)^T \otimes \mathbf I_4 \right] \left[ \left( (\mathbf I_4 - \mathbf B \otimes \mathbf B)^{-1} \right)^T \otimes (\mathbf I_4 - \mathbf B \otimes \mathbf B)^{-1}\right] \nonumber \\ &(\mathbf I_2 \otimes \mathbf K_4 \otimes \mathbf I_2)  (\mathbf I_4 \otimes \mathrm{vec} \mathbf B + \mathrm{vec} \mathbf B \otimes \mathbf I_4),\nonumber\\
D \mathbf Z^*(\mathbf B) &= \mathbf F \left[(\mathrm{vec} \mathbf Z)^T \otimes \mathbf I_4 \right] (\mathbf I_2 \otimes \mathbf K_4 \otimes \mathbf I_2) \left[ \mathbf I_4 \otimes \mathrm{vec} (\mathbf I_2 - \mathbf B)^{-1} + \mathrm{vec} (\mathbf I_2 - \mathbf B)^{-1} \otimes \mathbf I_4 \right] \nonumber \\ &\left[ \left( (\mathbf I_2 - \mathbf B)^{-1} \right)^T \otimes (\mathbf I_2 - \mathbf B)^{-1} \right],\nonumber\\
D \bm \mu^*(\mathbf a) &= (\mathbf I_2 - \mathbf B)^{-1},\nonumber\\
D \bm \Sigma^*(\bm \Sigma) &= \mathbf F (\mathbf I_4 - \mathbf B \otimes \mathbf B)^{-1} \mathbf G,\nonumber\\
D \mathbf Z^*(\mathbf Z) &= \mathbf F \left[(\mathbf I_2 - \mathbf B)^{-1} \otimes (\mathbf I_2 - \mathbf B)^{-1} \right] \mathbf G, \nonumber
\end{align}
where $\mathbf K_4$ is the $4 \times 4$ commutation matrix \citep[p. 54]{Magnus07},

\begin{equation*}
\mathbf F = \begin{bmatrix}1 & 0 & 0 & 0 \\0 & 1 & 0 & 0 \\ 0 & 0 & 0 & 1 \end{bmatrix},
\end{equation*}
and

\begin{equation*}
\mathbf G = \begin{bmatrix} 1 & 0 & 0 \\0 & 1 & 0 \\ 0 & 1 & 0 \\ 0 & 0 & 1 \end{bmatrix}.
\end{equation*}

\section{How informative is a snapshot about long-term site properties?}
\label{sec:snapshot}
Denote the true state of a randomly-chosen site at a given time by $\mathbf x$, and the corresponding stationary mean for that site by $\bm \mu^*$. Under the model of Equation \ref{eq:modelapp}, $\bm \mu^*$ has covariance matrix $\mathbf Z^*$ (Equation \ref{eq:Zetastar}). Write the true state as $\mathbf x = \bm \mu^* + \bm \Delta$, where $\bm \Delta$ is the deviation from the stationary mean, which has covariance matrix $\bm \Sigma^*$ (Equation \ref{eq:Sigmastar}). The correlation $\rho_k$ between the $k$th component $x_k$ of $\mathbf x$ and the corresponding component $\mu_k^*$ of $\bm \mu^*$ is an obvious way to measure how informative the snapshot will be for this component. This is

\begin{equation*}
\begin{aligned}
\rho_k &= \frac{\text{cov}(\mu^*_k + \Delta_k, \mu^*_k)}{\sqrt{V[\mu^*_k + \Delta_k] V[\mu^*_k]}}\\
&= \frac{V[\mu^*_k] + \text{cov}(\mu^*_k,\Delta_k)}{\sqrt{V[\mu^*_k + \Delta_k] V[\mu^*_k]}}\\
&= \frac{V[\mu^*_k]}{\sqrt{(V[\mu^*_k] + V[\Delta_k]) V[\mu^*_k]}} \quad \text{(because $\bm \alpha$ and $\bm \varepsilon$ assumed independent)}\\
&= \left( \frac{\zeta^*_{kk}}{\zeta^*_{kk} + \sigma^*_{kk}} \right)^{1/2},
\end{aligned}
\end{equation*}
where $\zeta^*_{kk}$ is the $k$th diagonal element of $\mathbf Z^*$, and $\sigma^*_{kk}$ is the $k$th diagonal element of $\bm \Sigma^*$. If $\rho_k$ is far from zero, a snapshot will be a reliable guide to the long-term value of the $k$th component of transformed reef composition. On the other hand, if $\rho_k$ is close to zero, a snapshot will be unreliable. Thus $\rho_k$ measures the extent to which conservation and management decisions could be based on observations at a single time point. We computed both $\rho_1$ which tells us how much we could learn about the log of the ratio of algae to coral and $\rho_2$, which tells us how much we could learn about the log of the ratio of other to the geometric mean of coral and algae.

\section{Dynamics}
\label{sec:dynamics}
\begin{figure}[h]
\includegraphics[height=18cm]{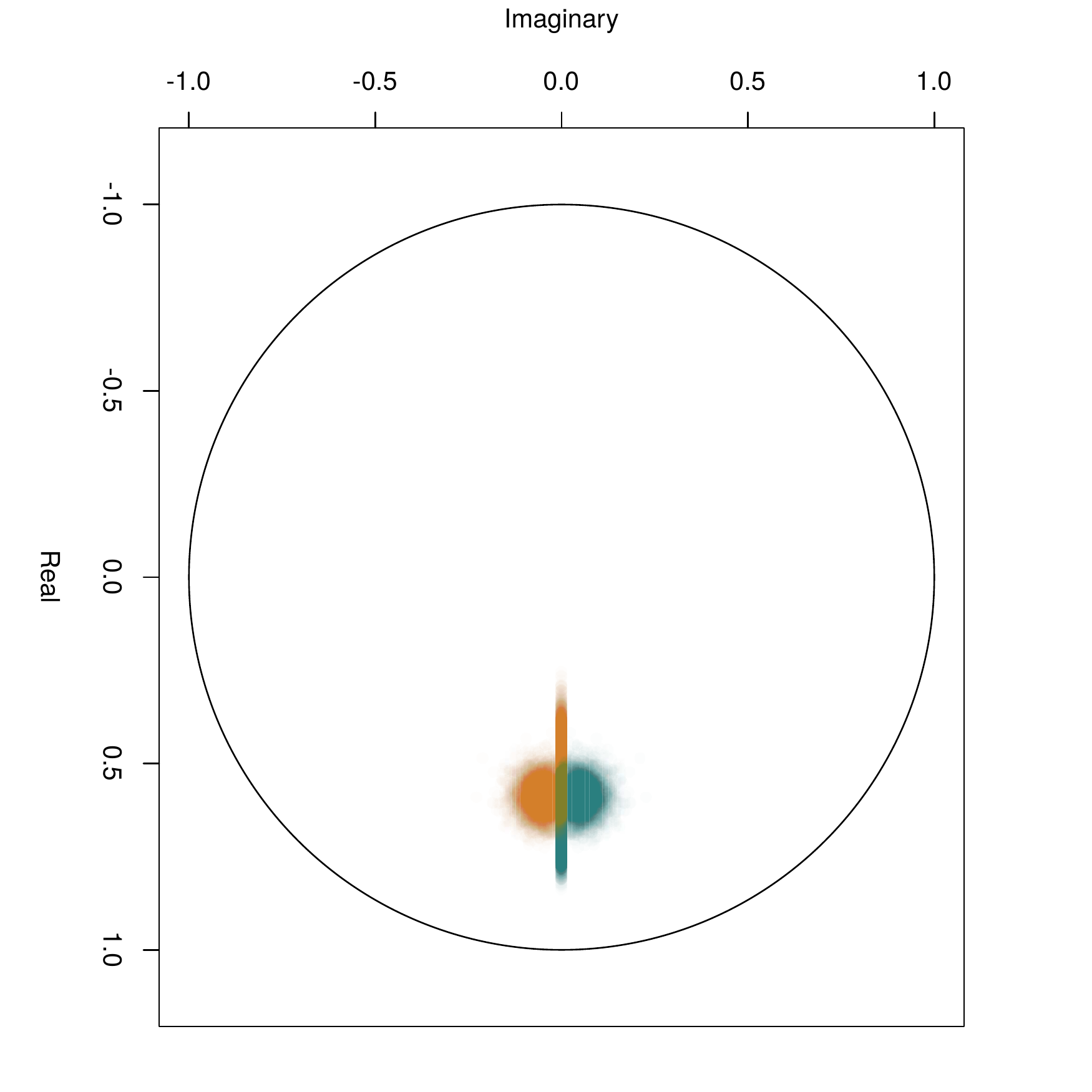}
\caption{Distribution of the two eigenvalues of $\mathbf B$ in the complex plane. Each Monte Carlo sample gives a pair of eigenvalues, represented by two points: $\lambda_1$ (green), posterior mean magnitude 0.64, $95\%$ HPD interval $[0.53, 0.75]$; $\lambda_2$ (orange), posterior mean magnitude 0.53, $95\%$ HPD interval $[0.41, 0.66]$)}
\label{fig:eigen}
\end{figure}

 Consistent with the patterns suggesting negative feedbacks that will tend to maintain fairly stable reef composition, every set of sampled parameters led to a stationary distribution (Figure \ref{fig:eigen}: all sampled eigenvalues of $\mathbf B$ fell inside the unit circle in the complex plane, with maximum magnitude 0.84). In $27\%$ of iterations, there was evidence for oscillations on the approach to the stationary distribution, because the eigenvalues were complex. In such cases, the oscillations had a long period (posterior mean 113 years, $95\%$ HPD interval $[21, 284]$ years), but their amplitude more than halved within three years because the magnitudes of the eigenvalues involved were small (original posterior mean magnitude of complex eigenvalues 0.59, $95\%$ credible interval $[0.51, 0.67]$, cubed posterior mean magnitude 0.21, $95\%$ HPD interval $[0.13, 0.30]$). The distribution of eigenvalues was very different from that of the Great Barrier Reef \citep[Appendix A.10]{Cooper15}, where the largest eigenvalue lay close to the point beyond which the stationary distribution would not exist (bootstrap mean magnitude 0.95), and there was no evidence for oscillations (no bootstrap replicates had complex eigenvalues). However, a different estimation method was used in \citet{Cooper15}, so the eigenvalues may not be directly comparable.

\section{Probability of low coral cover: signs of derivatives}
\label{sec:derivsigns}
\begin{figure}[h]
\includegraphics[width=18cm]{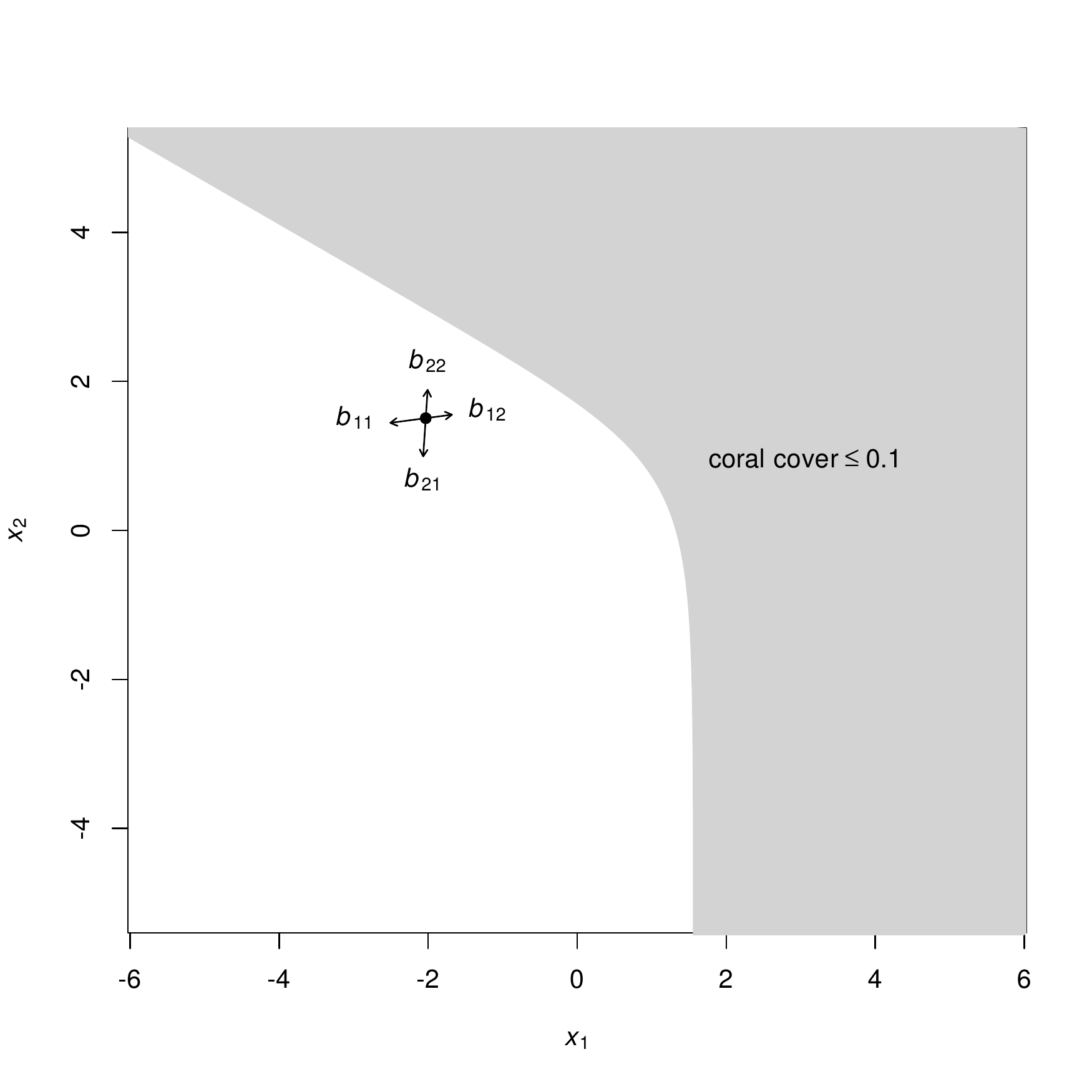}
\caption{Effects of the elements of $\mathbf B$ on the location of the stationary mean $\bm \mu^*$. Axes: the two components of isometric logratio transformed benthic composition (Equation \ref{eq:ilr}). Component $x_1$ is proportional to the log of the ratio of algae to coral. Component $x_2$ is proportional to the log of the ratio of other to the geometric mean of algae and coral. Black dot: point estimate of stationary mean $\bm \mu^*$, calculated from Equation \ref{eq:stationarymean} using posterior means of $\mathbf a$ and $\mathbf B$. Arrows: directions of derivatives of $\bm \mu^*$ with respect to each element of $\mathbf B$ (Equation \ref{eq:DmustarB}). Shaded region: coral cover $\leq 0.1$.}
\label{fig:derivintuition}
\end{figure}

Here, we explain the signs of the derivatives of the probability of low coral cover with respect to each parameter. We concentrate on coral cover threshold 0.1. 
The overall stationary mean $\bm \mu^*$ lies in the region where coral cover is greater than 0.1 for all iterations (Figure \ref{fig:derivintuition}, black circle, shows a point estimate for $\bm \mu^*$, based on the stationary means of $\mathbf a$ and $\mathbf B$). The shaded region of Figure \ref{fig:derivintuition} has coral cover $\leq 0.1$. Because of the shape of the boundary of the shaded region, either increasing $\mu_1^*$ (increasing the ratio of algae to coral) or increasing $\mu_2^*$ (increasing the ratio of other to the geometric mean of coral and algae) will move the stationary mean closer to this region. Also, since the stationary mean lies outside the region of interest, increasing the variability in the stationary distribution by increasing the elements of $\bm \Sigma^*$ or $\mathbf Z^*$ will increase the probability of falling in the region of interest. Hence the derivatives of $q_{0.1}$ with respect to $\bm \mu^*$, $\bm \Sigma^*$, $\mathbf Z^*$ contain only positive elements.

It is then intuitively obvious that the derivatives of $q_{0.1}$ with respect to $\bm \Sigma$ and $\mathbf Z$ will contain only positive elements. Increasing the amount of year-to-year temporal variability or among-site variability will increase the variability in the stationary distribution, and hence the long-term probability of coral cover less than or equal to 0.1.

The signs of the derivatives of $q_{0.1}$ with respect to $\mathbf a$ are also easy to understand. The components $a_1$, $a_2$ represent the rates of increase of $x_1$ and $x_2$ respectively, so we would expect that increasing either of them will increase the corresponding component of the stationary mean. Thus the derivatives of $\bm \mu^*$ with respect to $\mathbf a$ will be positive, and from Figure \ref{fig:derivintuition}, increasing either component of $\bm \mu^*$ will increase the probability of coral cover $\leq 0.1$.

The derivatives of $q_{0.1}$ with respect to $\mathbf B$ are a little harder to understand. They are (predominantly) negative with respect to $b_{11}$ and $b_{21}$, but positive with respect to $b_{12}$ and $b_{22}$. Since $\mathbf B$ affects both the stationary mean (Equation \ref{eq:stationarymean}) and the stationary covariance, which is the sum of $\bm \Sigma^*$ (Equation \ref{eq:Sigmastar}) and $\mathbf Z^*$ (Equation \ref{eq:Zetastar}), all of these effects could be important. However, in 93$\%$ of iterations,

\begin{equation*}
| Dq_{0.1}(\bm \mu*) D \bm \mu^*(\mathbf B) | \succ | Dq_{0.1}(\bm \Sigma^*) D \bm \Sigma^*(\mathbf B) + Dq_{0.1}(\mathbf Z^*) D \mathbf Z^*(\mathbf B) |,
\end{equation*}
where $\succ$ is an elementwise inequality, and $| \mathbf D|$ indicates the elementwise magnitude, such that for two matrices $\mathbf D$ and $\mathbf E$ with the same dimensions, $| \mathbf D | \succ |\mathbf E|$ if and only if the magnitude of every $d_{ij}$ is greater than the magnitude of the corresponding $e_{ij}$. In other words, in almost all iterations, the sign of the effect of $\mathbf B$ on $q_{0.1}$ via $\bm \mu^*$ determines the sign of the overall effect of $\mathbf B$ on $q_{0.1}$. We therefore concentrate on understanding how $\mathbf B$ affects $\bm \mu^*$.

To understand the signs of the effects of $b_{11}$ and $b_{22}$ on $\bm \mu^*$, consider the one-dimensional deterministic analogue

\begin{equation*}
x_{t+1} = a + b x_t.
\end{equation*}
Iterating this gives

\begin{equation*}
x_t = a \left( 1 + b + b^2 + \ldots + b^{t-1} \right) + b^t x_0.
\end{equation*}
For $0 < b < 1$, the term $b^t x_0 \rightarrow \ 0$ as $t \rightarrow \infty$. Then the derivative of $x_\infty$ with respect to $b$ has the same sign as $a$. In our system, $a_1 < 0$ and $a_2 > 0$, so we expect the signs of derivatives of $\bm \mu^*$ with respect to $b_{11}$ to be negative, and the signs of derivatives of $\bm \mu^*$ with respect to $b_{22}$ to be positive.

To understand the signs of the effects of $b_{12}$ and $b_{21}$ on $\bm \mu^*$, recall that $b_{12}$ is the effect of component 2 (which typically takes positive values) on component 1, and $b_{21}$ is the effect of component 1 (which typically takes negative values) on component 2. If, as in our system, $b_{12}$ and $b_{21}$ are both positive, and the system is linear, we would expect that the signs of their effects on $\bm \mu^*$ will be the same as the signs of components 2 and 1 respectively.

Then, by the graphical argument above (Figure \ref{fig:derivintuition}), we expect the signs of the derivatives of $q_{0.1}$ with respect to $b_{11}$, $b_{21}$, $b_{12}$ and $b_{22}$ to be $-$,$-$,$+$,$+$ respectively.

\section{Probability of low coral cover: rank order and other thresholds}
\label{sec:thresholds}
\begin{figure}[h]
\includegraphics[height=18cm]{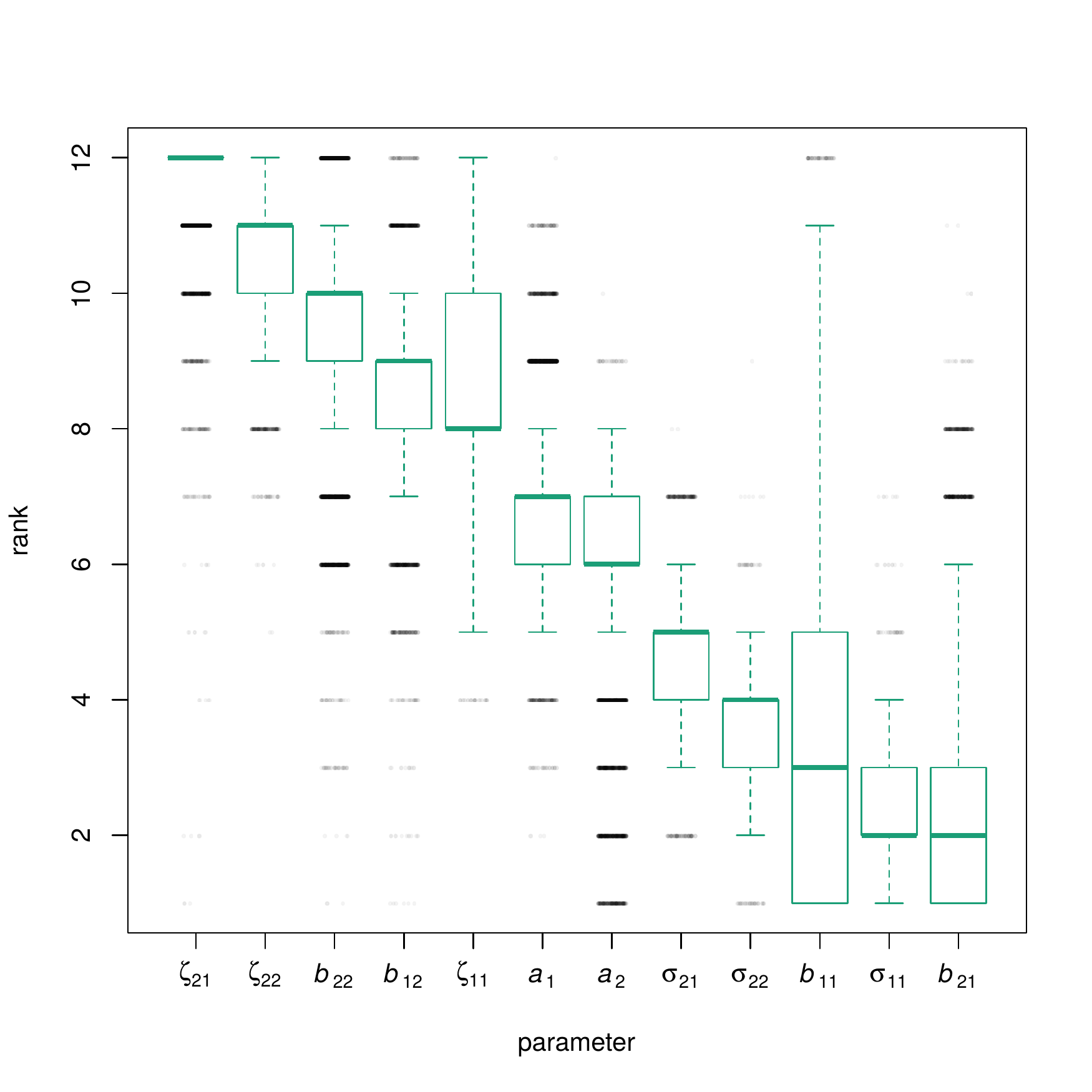}
\caption{Ranks of partial derivatives of the long-term probability of coral cover less than or equal to 0.05 with respect to elements of the $\mathbf B$ matrix, the $\mathbf a$ vector, the covariance matrix of random temporal variation $\bm \Sigma$, and the covariance matrix of among-site variability $\mathbf Z$. Parameters are ranked in descending order of median rank (higher ranks indicate larger magnitudes of partial derivative). Outliers are indicated as jittered black dots. For the covariance matrices, the elements $\sigma_{12}$ and $\zeta_{12}$ are not shown, because they are constrained to be equal to $\sigma_{21}$ and $\zeta_{21}$ respectively.}
\label{fig:rankderivs0.05}
\end{figure}

\begin{figure}[h]
\includegraphics[height=18cm]{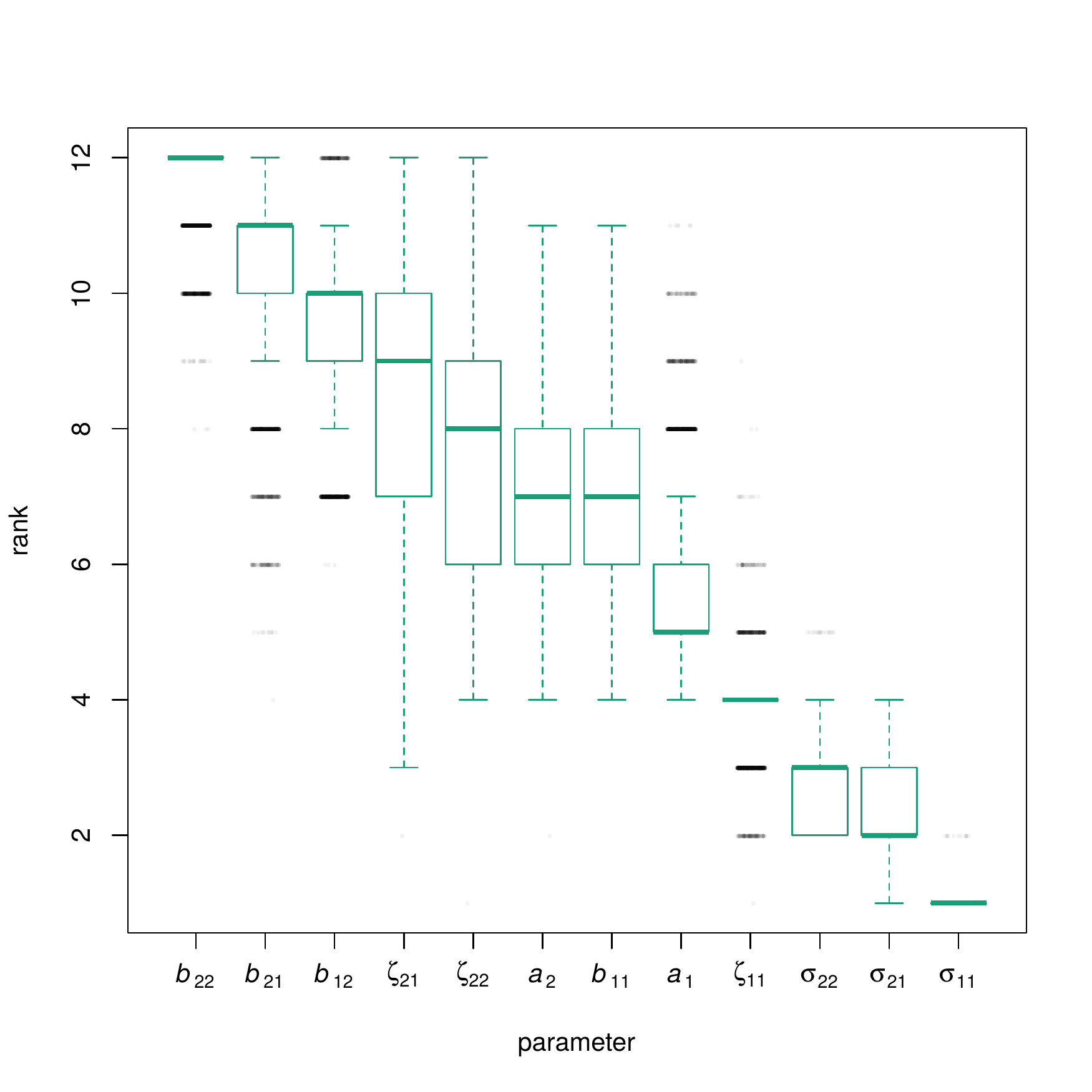}
\caption{Ranks of partial derivatives of the long-term probability of coral cover less than or equal to 0.2 with respect to elements of the $\mathbf B$ matrix, the $\mathbf a$ vector, the covariance matrix of random temporal variation $\bm \Sigma$, and the covariance matrix of among-site variability $\mathbf Z$. Parameters are ranked in descending order of median rank (higher ranks indicate larger magnitudes of partial derivative). Outliers are indicated as jittered black dots. For the covariance matrices, the elements $\sigma_{12}$ and $\zeta_{12}$ are not shown, because they are constrained to be equal to $\sigma_{21}$ and $\zeta_{21}$ respectively.}
\label{fig:rankderivs0.2}
\end{figure}

For threshold 0.05, the signs of the effects of $b_{11}$ and $b_{21}$ were not clearly negative. The four most important parameters were (in descending order: Figure \ref{fig:rankderivs0.05}) $\zeta_{21}$, $\zeta_{22}$, $b_{22}$ and $b_{12}$ (the same four as for threshold 0.1, but in a different order). For threshold 0.2, the signs were as for threshold 0.1, but the four most important parameters were (in descending order) $b_{22}$, $b_{21}$, $b_{12}$ and $\zeta_{21}$ (with $\zeta_{22}$ now in fifth place: Figure \ref{fig:rankderivs0.2}). Thus, while the details depend to some extent on the threshold, the overall conclusion that both internal dynamics and among-site variability are the most important factors affecting the probability of low coral cover is robust.

\begin{figure}[h]
\includegraphics[height=18cm]{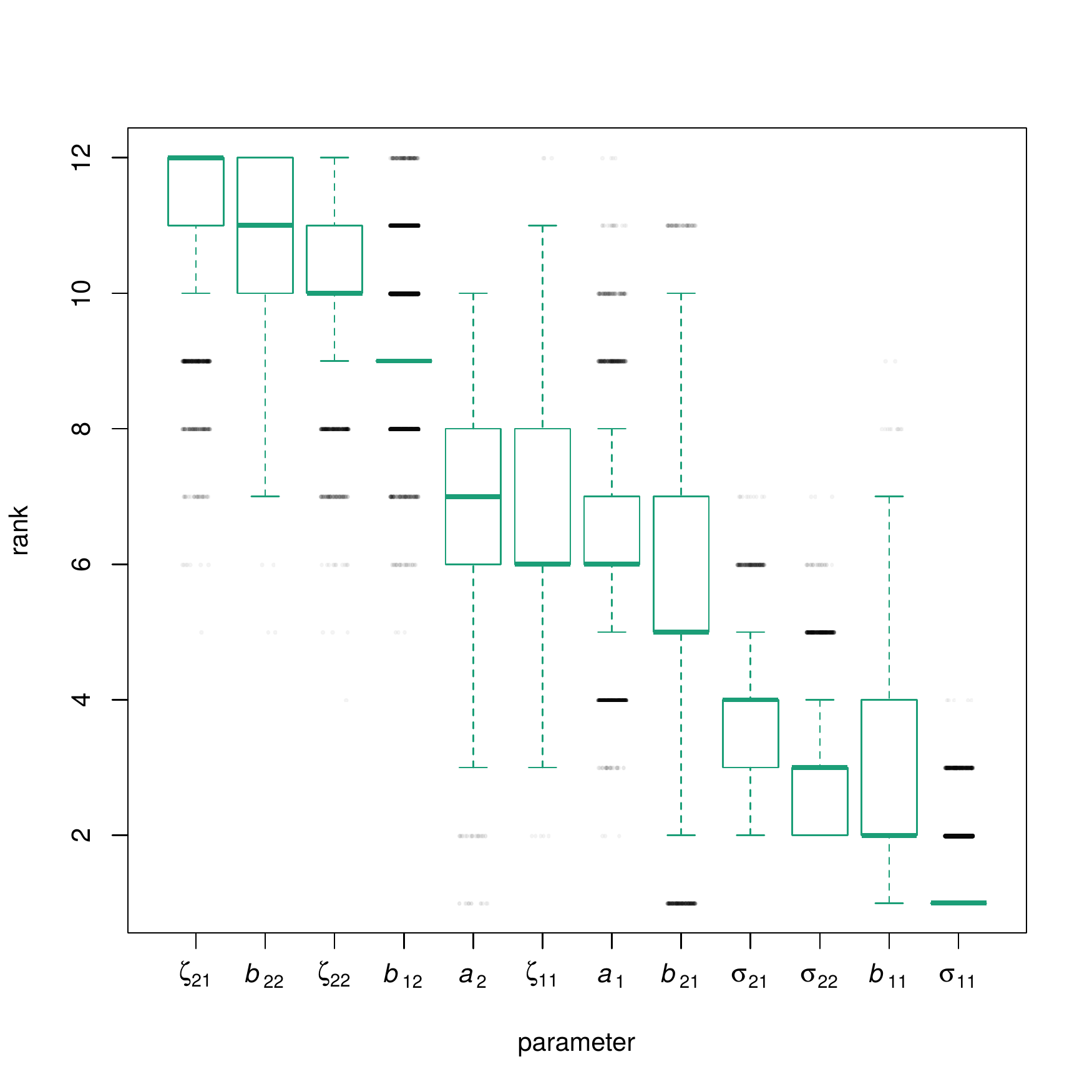}
\caption{Ranks of partial derivatives of the long-term probability of coral cover less than or equal to 0.1 with respect to elements of the $\mathbf B$ matrix, the $\mathbf a$ vector, the covariance matrix of random temporal variation $\bm \Sigma$, and the covariance matrix of among-site variability $\mathbf Z$. Parameters are ranked in descending order of median rank (higher ranks indicate larger magnitudes of partial derivative). Outliers are indicated as jittered black dots. For the covariance matrices, the elements $\sigma_{12}$ and $\zeta_{12}$ are not shown, because they are constrained to be equal to $\sigma_{21}$ and $\zeta_{21}$ respectively.}
\label{fig:rankderivs}
\end{figure}

The effects of within-site temporal variability on the probability of low coral cover were always relatively unimportant (threshold 0.1, Figure \ref{fig:rankderivs}, three of the last four positions in the ranked list; threshold 0.05, Figure \ref{fig:rankderivs0.05}, three of the last five positions; threshold 0.20, Figure \ref{fig:rankderivs0.2}, last three positions).

\bibliographystyle{hapalike}
\bibliography{kenya}

\begin{thebibliography}{}

\bibitem[Aitchison, 1986]{Aitchison86}
Aitchison, J. (1986).
\newblock {\em The statistical analysis of compositional data}.
\newblock Chapman and Hall, London.

\bibitem[Aitchison, 1989]{Aitchison89}
Aitchison, J. (1989).
\newblock Measures of location for compositional data sets.
\newblock {\em Mathematical Geology}, 21:787--790.

\bibitem[Ateweberhan and McClanahan, 2016]{Ateweberhan16}
Ateweberhan, M. and McClanahan, T.~R. (2016).
\newblock Partitioning scleractinian coral diversity across reef sites and
  regions in the {W}estern {I}ndian {O}cean.
\newblock {\em Ecosphere}, 7(5):01243.

\bibitem[Auger-M{\'e}th{\'e} et~al., 2015]{AugerMethe15}
Auger-M{\'e}th{\'e}, M., Field, C., Albertsen, C.~M., Derocher, A.~E., Lewis,
  M.~A., Jonsen, I.~D., and Mills~Flemming, J. (2015).
\newblock State-space models' dirty little secrets: even simple linear
  {G}aussian models can have estimation problems.
\newblock {\em unpublished}, arXiv:1508.04325v1.

\bibitem[Baker et~al., 2008]{Baker08}
Baker, A.~C., Glynn, P.~W., and Riegl, B. (2008).
\newblock Climate change and coral reef bleaching: an ecological assessment of
  long-term impacts, recovery trends and future outlook.
\newblock {\em Estuarine, Coastal and Shelf Science}, 80:435--471.

\bibitem[Bj{\o}rnstad and Falck, 2001]{Bjornstad01}
Bj{\o}rnstad, O.~N. and Falck, W. (2001).
\newblock Nonparametric spatial covariance functions: {E}stimation and testing.
\newblock {\em Environmental and Ecological Statistics}, 8:53--70.

\bibitem[Brook et~al., 2000]{Brook_3196}
Brook, B.~W., O'Grady, J.~J., Chapman, A.~P., Burgman, M.~A., Ak{\c{c}}akaya,
  H.~R., and Frankham, R. (2000).
\newblock Predictive accuracy of population viability analysis in conservation
  biology.
\newblock {\em Nature}, 404:385--387.

\bibitem[Bruno et~al., 2009]{Bruno09}
Bruno, J.~F., Sweatman, H., Precht, W.~F., Selig, E.~R., and Schutte, V. G.~W.
  (2009).
\newblock Assessing evidence of phase shifts from coral to macroalgal dominance
  on coral reefs.
\newblock {\em Ecology}, 90(6):1478--1484.

\bibitem[Carreiro-Silva and McClanahan, 2012]{CarreiroSilva12}
Carreiro-Silva, M. and McClanahan, T.~R. (2012).
\newblock Macrobioerosion of dead branching \emph{{P}orites}, 4 and 6 years
  after coral mass mortality.
\newblock {\em Marine Ecology Progress Series}, 458:103--122.

\bibitem[Caswell, 2001]{Cas01}
Caswell, H. (2001).
\newblock {\em Matrix population models: construction, analysis, and
  interpretation}.
\newblock Sinauer, Sunderland, MA, second edition.

\bibitem[Cinner and McClanahan, 2015]{Cinner15}
Cinner, J.~E. and McClanahan, T.~R. (2015).
\newblock A sea change on the {A}frican coast? {P}reliminary social and
  ecological outcomes of a governance transformation in {K}enyan fisheries.
\newblock {\em Global Environmental Change}, 30:133--139.

\bibitem[Connell et~al., 1997]{Con97}
Connell, J.~H., Hughes, T.~P., and Wallace, C.~C. (1997).
\newblock A 30-year study of coral abundance, recruitment, and disturbance at
  several scales in space and time.
\newblock {\em Ecological Monographs}, 67(4):461--488.

\bibitem[Cooper et~al., 2015]{Cooper15}
Cooper, J.~K., Spencer, M., and Bruno, J.~F. (2015).
\newblock Stochastic dynamics of a warmer {G}reat {B}arrier {R}eef.
\newblock {\em Ecology}, 96:1802--1811.

\bibitem[C{\^{o}}t{\'{e}} et~al., 2005]{Cote05}
C{\^{o}}t{\'{e}}, I.~M., Gill, J.~A., Gardner, T.~A., and Watkinson, A.~R.
  (2005).
\newblock Measuring coral reef decline through meta-analyses.
\newblock {\em Philosophical Transactions of the Royal Society Series B},
  360:385--395.

\bibitem[De'ath et~al., 2012]{Death12}
De'ath, G., Fabricius, K.~E., Sweatman, H., and Puotinen, M. (2012).
\newblock The 27-year decline of coral cover on the {G}reat {B}arrier {R}eef
  and its causes.
\newblock {\em Proceedings of the National Academy of Sciences of the USA},
  109:17995--17999.

\bibitem[Diamond, 1986]{Diamond86}
Diamond, J. (1986).
\newblock Overview: laboratory experiments, field experiments, and natural
  experiments.
\newblock In Diamond, J. and Case, T.~J., editors, {\em Community ecology},
  pages 3--22. Harper \& Row, New York.

\bibitem[Egozcue et~al., 2003]{Egozcue03}
Egozcue, J.~J., Pawlowsky-Glahn, V., Mateu-Figueras, G., and Barcel{\'o}-Vidal,
  C. (2003).
\newblock Isometric logratio transformations for compositional data analysis.
\newblock {\em Mathematical Geology}, 35(3):279--300.

\bibitem[Gelman et~al., 2003]{GCSR03}
Gelman, A., Carlin, J.~B., Stern, H.~S., and Rubin, D.~B. (2003).
\newblock {\em {B}ayesian {D}ata {A}nalysis}.
\newblock Chapman and Hall/CRC, Boca Raton, second edition.

\bibitem[Ginzburg et~al., 1982]{Ginzburg_1728}
Ginzburg, L.~R., Slobodkin, L.~B., Johnson, K., and Bindman, A.~G. (1982).
\newblock Quasiextinction probabilities as a measure of impact on population
  growth.
\newblock {\em Risk Analysis}, 21:171--181.

\bibitem[Gross and Edmunds, 2015]{Gross15}
Gross, K. and Edmunds, P.~J. (2015).
\newblock Stability of {C}aribbean coral communities quantified by long-term
  monitoring and autoregression models.
\newblock {\em Ecology}, 96:1812--1822.

\bibitem[Hampton et~al., 2013]{Hampton13}
Hampton, S.~E., Holmes, E.~E., Scheef, L.~P., Scheuerell, M.~D., Katz, S.~L.,
  Pendleton, D.~E., and Ward, E.~J. (2013).
\newblock Quantifying effects of abiotic and biotic drivers on community
  dynamics with multivariate autoregressive ({MAR}) models.
\newblock {\em Ecology}, 94(12):2663--2669.

\bibitem[Henze and Zirkler, 1990]{Henze90}
Henze, N. and Zirkler, B. (1990).
\newblock A class of invariant consistent tests for multivariate normality.
\newblock {\em Communications in Statistics - Theory and Methods},
  19:3595--3617.

\bibitem[Hoffman and Gelman, 2014]{Hoffman14}
Hoffman, M.~D. and Gelman, A. (2014).
\newblock The {N}o-{U}-{T}urn {S}ampler: Adaptively setting path lengths in
  {H}amiltonian {M}onte {C}arlo.
\newblock {\em Journal of Machine Learning Research}, 15:1351--1381.

\bibitem[Hyndman, 1996]{Hyndman_4081}
Hyndman, R.~J. (1996).
\newblock Computing and graphing highest density regions.
\newblock {\em The American Statistician}, 50(2):120--126.

\bibitem[Ives et~al., 2003]{Ives03}
Ives, A.~R., Dennis, B., Cottingham, K.~L., and Carpenter, S.~R. (2003).
\newblock Estimating community stability and ecological interactions from
  time-series data.
\newblock {\em Ecological Monographs}, 73(2):301--330.

\bibitem[Johnson and Wichern, 2007]{Johnson07}
Johnson, R.~A. and Wichern, D.~W. (2007).
\newblock {\em Applied multivariate statistical analysis}.
\newblock Pearson, 6th edition.

\bibitem[Kaiser, 1983]{Kaiser83}
Kaiser, L. (1983).
\newblock Unbiased estimation in line-intercept sampling.
\newblock {\em Biometrics}, 39(4):965--976.

\bibitem[Kennedy et~al., 2013]{Kennedy13}
Kennedy, E.~V., Perry, C.~T., Halloran, P.~R., Iglesias-Prieto, R.,
  Sch{\"o}nberg, C. H.~L., Wissah, M., Form, A.~U., Carricart-Ganivet, J.~P.,
  Fine, M., Eakin, C.~M., and Mumby, P.~J. (2013).
\newblock Avoiding coral reef functional collapse requires local and global
  action.
\newblock {\em Current Biology}, 23:912--918.

\bibitem[Kenward, 1979]{Kenward79}
Kenward, M.~G. (1979).
\newblock An intuitive approach to the {MANOVA} test criteria.
\newblock {\em Journal of the Royal Statistical Society Series D},
  28(3):193--198.

\bibitem[Korkmaz et~al., 2014]{Korkmaz14}
Korkmaz, S., Goksuluk, D., and Zararsiz, G. (2014).
\newblock {MVN}: An {R} package for assessing multivariate normality.
\newblock {\em The R Journal}, 6:151--162.

\bibitem[Lange et~al., 1989]{Lange89}
Lange, K.~L., Little, R. J.~A., and Taylor, J. M.~G. (1989).
\newblock Robust statistical modeling using the \emph{t} distribution.
\newblock {\em Journal of the American Statistical Association}, 84:881--896.

\bibitem[Lindegren et~al., 2009]{Lindegren09}
Lindegren, M., M{\"o}llmann, C., Nielsen, A., and Stenseth, N.~C. (2009).
\newblock Preventing the collapse of the {B}altic cod stock through an
  ecosystem-based management approach.
\newblock {\em Proceedings of the National Academy of Sciences of the USA},
  106(34):14722--14727.

\bibitem[L{\"u}tkepohl, 1993]{Lutkepohl93}
L{\"u}tkepohl, H. (1993).
\newblock {\em Introduction to multiple time series analysis}.
\newblock Springer-Verlag, Berlin, 2nd edition.

\bibitem[Magnus and Neudecker, 2007]{Magnus07}
Magnus, J.~R. and Neudecker, H. (2007).
\newblock {\em Matrix differential calculus with applications in statistics and
  econometrics}.
\newblock John Wiley \& Sons, Chichester, third edition.

\bibitem[Mardia, 1970]{Mardia70}
Mardia, K.~V. (1970).
\newblock Measures of multivariate skewnees and kurtosis with applications.
\newblock {\em Biometrika}, 57:519--530.

\bibitem[Mart{\'i}n-Fern{\'a}ndez et~al., 2003]{MartinFernandez03}
Mart{\'i}n-Fern{\'a}ndez, J.~A., Barcel{\'o}-Vidal, C., and Pawlowsky-Glahn, V.
  (2003).
\newblock Dealing with zeros and missing values in compositional data sets
  using nonparametric imputation.
\newblock {\em Mathematical Geology}, 35(3):253--278.

\bibitem[McClanahan and Arthur, 2001]{McClanahan01c}
McClanahan, T.~R. and Arthur, R. (2001).
\newblock The effect of marine reserves and habitat on populations of {E}ast
  {A}frican coral reef fishes.
\newblock {\em Ecological Applications}, 11(2):559--569.

\bibitem[McClanahan et~al., 2007]{MAM+07}
McClanahan, T.~R., Ateweberhan, M., Muhando, C.~A., Maina, J., and Mohammed,
  M.~S. (2007).
\newblock Effects of climate and seawater temperature variation on coral
  bleaching and mortality.
\newblock {\em Ecological Monographs}, 77(4):503--525.

\bibitem[McClanahan et~al., 2016]{McClanahan16}
McClanahan, T.~R., Muthiga, N.~A., and Abunge, C.~A. (2016).
\newblock Establishment of community managed fisheries' closures in {K}enya:
  early evolution of the tengefu movement.
\newblock {\em Coastal Management}, 44:1--20.

\bibitem[McClanahan et~al., 2001]{McClanahan01b}
McClanahan, T.~R., Muthiga, N.~A., and Mangi, S. (2001).
\newblock Coral and algal changes after the 1998 coral bleaching: interaction
  with reef management and herbivores on {K}enyan reefs.
\newblock {\em Coral Reefs}, 19:380--391.

\bibitem[Mellin et~al., 2014]{Mellin14}
Mellin, C., Bradshaw, C. J.~A., Fordham, D.~A., and Caley, M.~J. (2014).
\newblock Strong but opposing $\beta$-diversity-stability relationships in
  coral reef fish communities.
\newblock {\em Proceedings of the Royal Society of London Series B},
  281:20131993.

\bibitem[Mirsky, 1955]{Mirsky55}
Mirsky, L. (1955).
\newblock {\em An introduction to linear algebra}.
\newblock Oxford University Press, Oxford.

\bibitem[Mumby et~al., 2007]{MHE07}
Mumby, P.~J., Hastings, A., and Edwards, H.~J. (2007).
\newblock Thresholds and the resilience of {C}aribbean coral reefs.
\newblock {\em Nature}, 450:98--101.

\bibitem[Mutshinda et~al., 2009]{Mutshinda09}
Mutshinda, C.~M., O'Hara, R.~B., and Woiwod, I.~P. (2009).
\newblock What drives community dynamics?
\newblock {\em Proceedings of the Royal Society of London Series B},
  276:2923--2929.

\bibitem[Otto and Day, 2007]{Otto07}
Otto, S.~P. and Day, T. (2007).
\newblock {\em A biologist's guide to mathematical modeling in ecology and
  evolution}.
\newblock Princeton University Press, Princeton, New Jersey.

\bibitem[Perry et~al., 2013]{Perry13}
Perry, C.~T., Murphy, G.~N., Kench, P.~S., Smithers, S.~G., Edinger, E.~N.,
  Steneck, R.~S., and Mumby, P.~J. (2013).
\newblock Caribbean-wide decline in carbonate production threatens coral reef
  growth.
\newblock {\em Nature Communications}, 4:1402.

\bibitem[Piessens et~al., 1983]{Piessens83}
Piessens, R., de~Doncker-Kapenga, E., {\"U}berhuber, C.~W., and Kahaner, D.
  (1983).
\newblock {\em {QUADPACK}: a subroutine package for automatic integration}.
\newblock Springer-Verlag, Berlin.

\bibitem[{R Core Team}, 2015]{RCore15}
{R Core Team} (2015).
\newblock {\em R: A Language and Environment for Statistical Computing}.
\newblock R Foundation for Statistical Computing, Vienna, Austria.

\bibitem[Riley et~al., 2002]{RHB02}
Riley, K.~F., Hobson, M.~P., and Bence, S.~J. (2002).
\newblock {\em Mathematical methods for physics and engineering}.
\newblock Cambridge University Press, Cambridge, second edition.

\bibitem[Roff et~al., 2015]{Roff15}
Roff, G., Zhao, J.-X., and Mumby, P.~J. (2015).
\newblock Decadal-scale rates of reef erosion following {E}l {N}i{\~n}o-related
  mass coral mortality.
\newblock {\em Global Change Biology}, 21:4415--4424.

\bibitem[Royston, 1982]{Royston82}
Royston, J. (1982).
\newblock An extension of {S}hapiro and {W}ilk's ${W}$ test for normality to
  large samples.
\newblock {\em Applied Statistics}, 31:115--124.

\bibitem[Sandin et~al., 2008]{Sandin08}
Sandin, S.~A., Smith, J.~E., DeMartini, E.~E., Dinsdale, E.~A., Donner, S.~D.,
  Friedlander, A.~M., Konotchick, T., Malay, M., Maragos, J.~E., Obura, D.,
  Pantos, O., Paulay, G., Richie, M., Rohwer, F., Schroeder, R.~E., Walsh, S.,
  Jackson, J. B.~C., Knowlton, N., and Sala, E. (2008).
\newblock Baselines and degradation of coral reefs in the {N}orthern {L}ine
  {I}slands.
\newblock {\em PLoS ONE}, 3(2):e1548.

\bibitem[{Stan Development Team}, 2015a]{stan-software:2015}
{Stan Development Team} (2015a).
\newblock Stan: A {C}++ library for probability and sampling, version 2.7.0.

\bibitem[{Stan Development Team}, 2015b]{stan-manual:2015}
{Stan Development Team} (2015b).
\newblock {\em Stan Modeling Language Users Guide and Reference Manual, Version
  2.7.0}.

\bibitem[Vehtari et~al., 2015]{Vehtari15}
Vehtari, A., Gelman, A., and Gabry, J. (2015).
\newblock Efficient implementation of leave-one-out cross-validation and {WAIC}
  for evaluating fitted {B}ayesian models.
\newblock {\em unpublished}, arXiv:1507.04544v1.

\bibitem[Vercelloni et~al., 2014]{Vercelloni14}
Vercelloni, J., Caley, M.~J., Kayal, M., Low-Choy, S., and Mengersen, K.
  (2014).
\newblock Understanding uncertainties in non-linear population trajectories: a
  {B}ayesian semi-parametric hierarchical approach to large-scale surveys of
  coral cover.
\newblock {\em PLoS ONE}, 9(11):e110968.

\bibitem[Wilks, 1932]{Wilks32}
Wilks, S.~S. (1932).
\newblock Certain generalizations in the analysis of variance.
\newblock {\em Biometrika}, 24:471--494.

\bibitem[{\.Z}ychaluk et~al., 2012]{Zychaluk12}
{\.Z}ychaluk, K., Bruno, J.~F., Clancy, D., McClanahan, T.~R., and Spencer, M.
  (2012).
\newblock Data-driven models for regional coral-reef dynamics.
\newblock {\em Ecology Letters}, 15:151--158.

\end{thebibliography}
\clearpage

\begin{landscape}
\begin{table}[h]
\caption{Reef features. For each named reef, surveys were done at either one site, or at two sites \SIrange{20}{100}{\metre} apart. Fished reefs include community management areas with reduced harvesting intensity, and unfished reefs include those recently designated as reserves. Mean coral cover is the arithmetic mean of observed coral cover over all transects and time points.}
\label{tab:reefs}
\resizebox{\textwidth}{!}{
\begin{tabular}{ l l l l l l l l}
Reef & Sites & Location & Time points & Time range & Reef type & Management & Mean coral cover (site 1, site 2)\\
\hline
Bongoyo & 2 & 6.67 S, 39.26 E & 3 & 1995-2012 & patch & fished & 54.7, 52.1\\
Changale & 1 & 5.30 S, 39.10 E & 3 & 1995-2010 & patch & fished & 39.4\\
Changuu & 1 & 6.12 S, 39.12 E & 3 & 1997-2012 & patch & fished & 46.8\\
Chapwani & 1 & 6.07 S, 39.11 E & 3 & 1997-2012 & patch & fished & 52.5\\
Chumbe & 2 & 6.28 S, 39.17 E & 3 & 1997-2012 & patch & unfished & 70.1, 74.1\\
Diani & 2 & 4.37 S, 39.58 E & 19 & 1992-2013 & fringing & fished & 32.0, 17.5\\
Funguni & 1 & 5.27 S, 39.13 E & 3 & 1995-2010 & patch & fished & 13.7\\ 
Kanamai & 2 & 3.93 S, 39.78 E & 19 & 1991-2013 & fringing & fished & 33.0, 32.3\\
Kisite & 2 & 4.71 S, 39.37 E & 8 & 1994-2012 & patch & unfished & 33.9, 46.4\\
Makome & 1 & 5.28 S, 39.11 E & 3 & 1995-2010 & patch & fished & 32.1\\
Malindi & 2 & 3.26 S, 40.15 E & 20 & 1991-2013 & fringing & unfished & 27.9\\
Mbudya & 2 & 6.66 S, 39.25 E & 3 & 1995-2012 & patch & fished & 53.5, 68.0\\
Mombasa & 2 & 3.99 S, 39.75 E & 20 & 1991-2013 & fringing & unfished & 37.27, 29.2\\
Mradi & 1 & 3.94 S, 39.78 E & 2 & 2010-2011 & fringing & fished & 48.4\\
Nyali & 2 & 4.05 S, 39.71 E & 2 & 2006-2009 & fringing & fished & 28.1, 29.1\\
Ras Iwatine & 1 & 4.02 S, 39.73 E & 18 & 1993-2013 & fringing & fished & 10.8\\
Taa & 1 & 3.99 S, 39.77 E & 3 & 1995-2010 & patch & fished & 20.7\\
Tiwi Inside & 1 & 4.26 S, 39.61 E & 2 & 2008-2011 & fringing & fished & 36.0\\
Vipingo & 2 & 3.48 S, 39.95 E & 18 (site 1), 19 (site 2) & 1991-2013 & fringing & fished & 28.0, 28.2\\
Watamu & 1 & 3.37 S, 40.01 E & 20 & 1991-2013 & fringing & unfished & 23.2\\

\end{tabular}
}
\end{table}
\end{landscape}

\clearpage

\begin{figure}
\includegraphics[height=20cm]{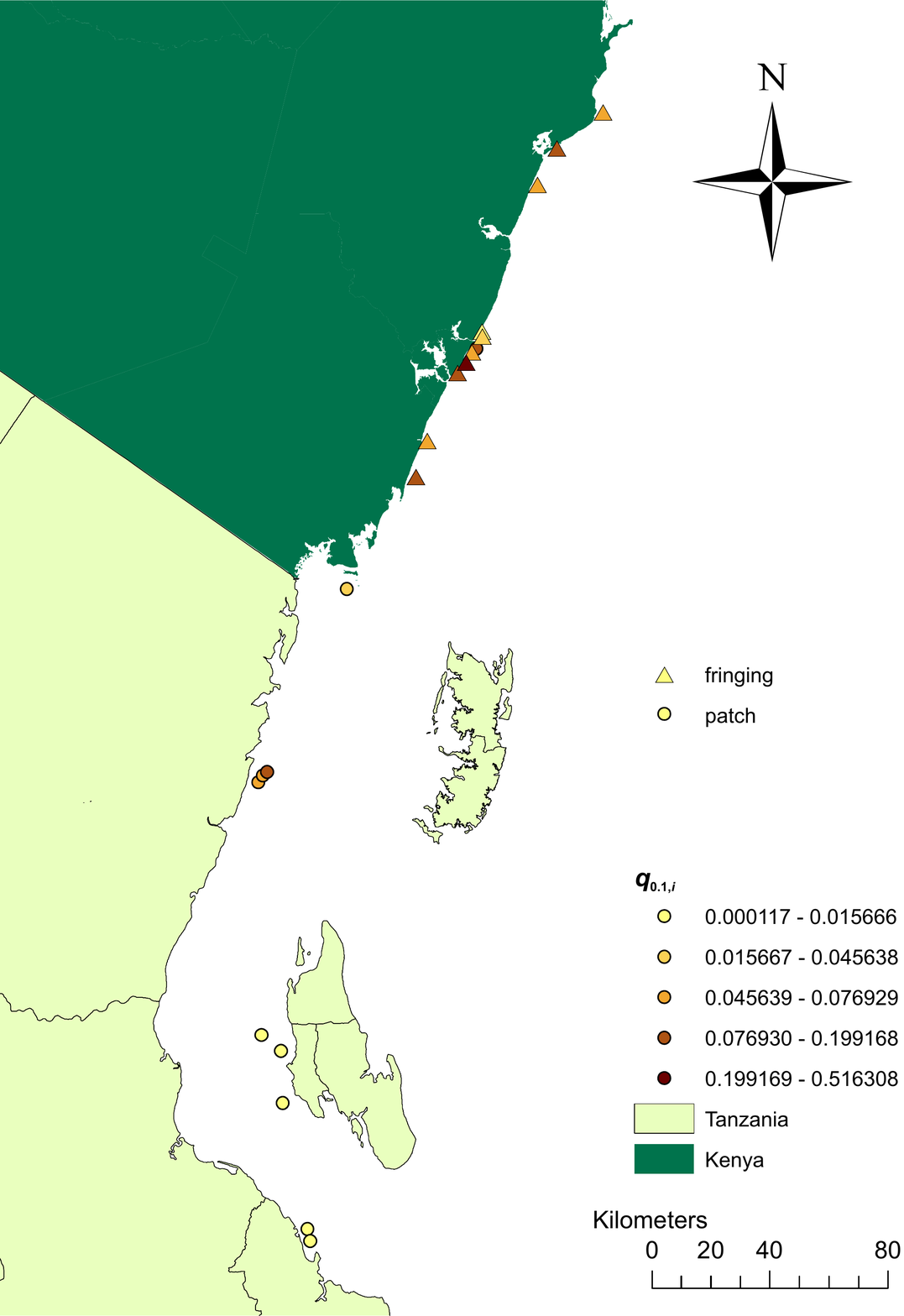}
\caption{Map of study sites, showing fringing reefs (triangles) and patch reefs (circles), shaded by the site-specific long-term probability $q_{0.1,i}$ of coral cover $\leq 0.1$ (for reefs with one site) or the mean of site-specific probabilities (for reefs with two sites).}
\label{fig:map}
\end{figure}

\clearpage

\clearpage

\clearpage

\clearpage

 \clearpage

\begin{figure}[h]
\includegraphics[height=18cm]{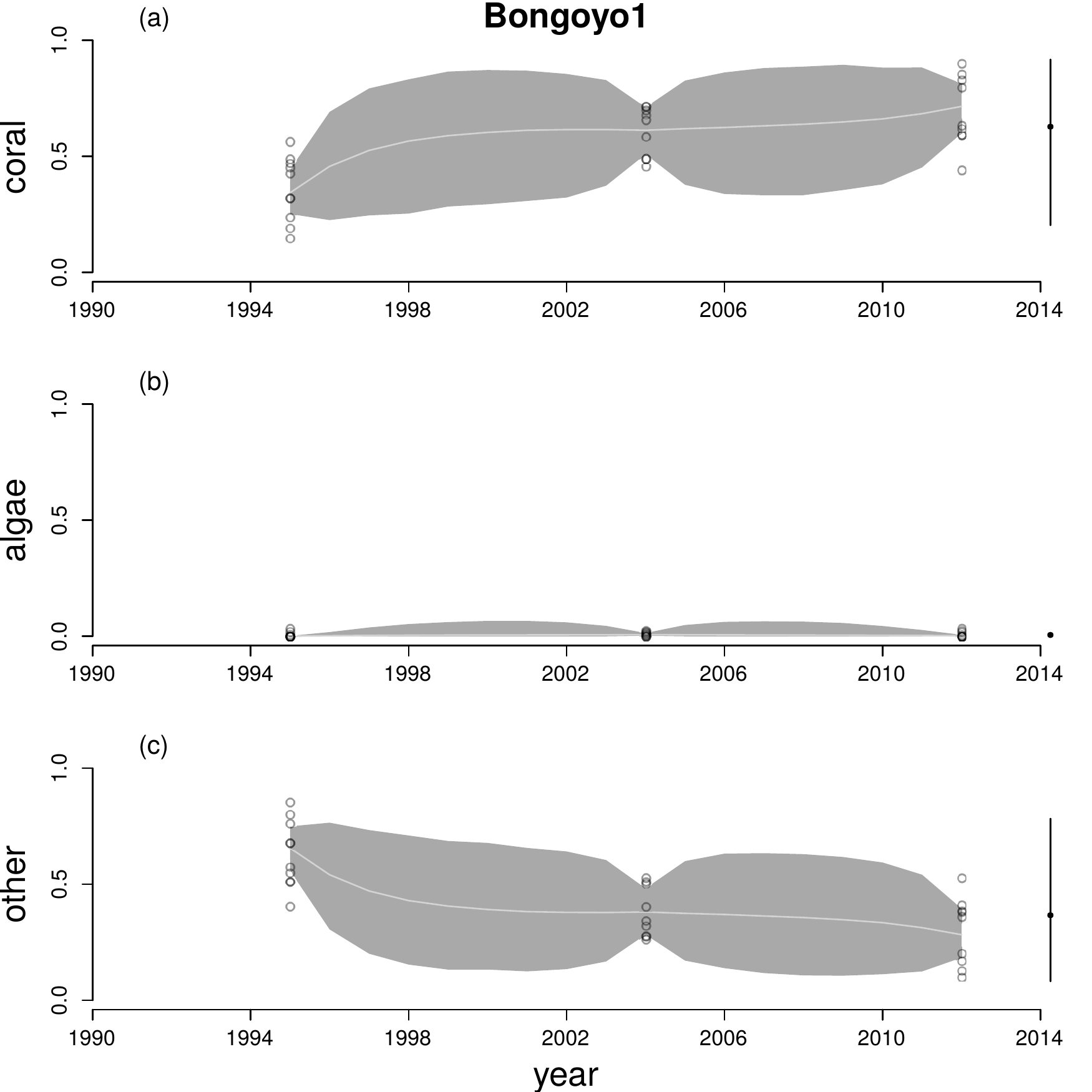}
\caption{Time series for cover of hard corals (a), macroalgae (b) and other (c) at Bongoyo1. Circles are observations from individual transects. Grey lines join back-transformed posterior mean true states from Equation \ref{eq:modelapp} and the shaded region is a 95$\%$ HPD interval. The stationary mean composition for the site is the black dot after the time series and the bar is a 95$\%$ HPD interval.}
\label{fig:Bongoyo1}
\end{figure}

\clearpage

\begin{figure}[h]
\includegraphics[height=18cm]{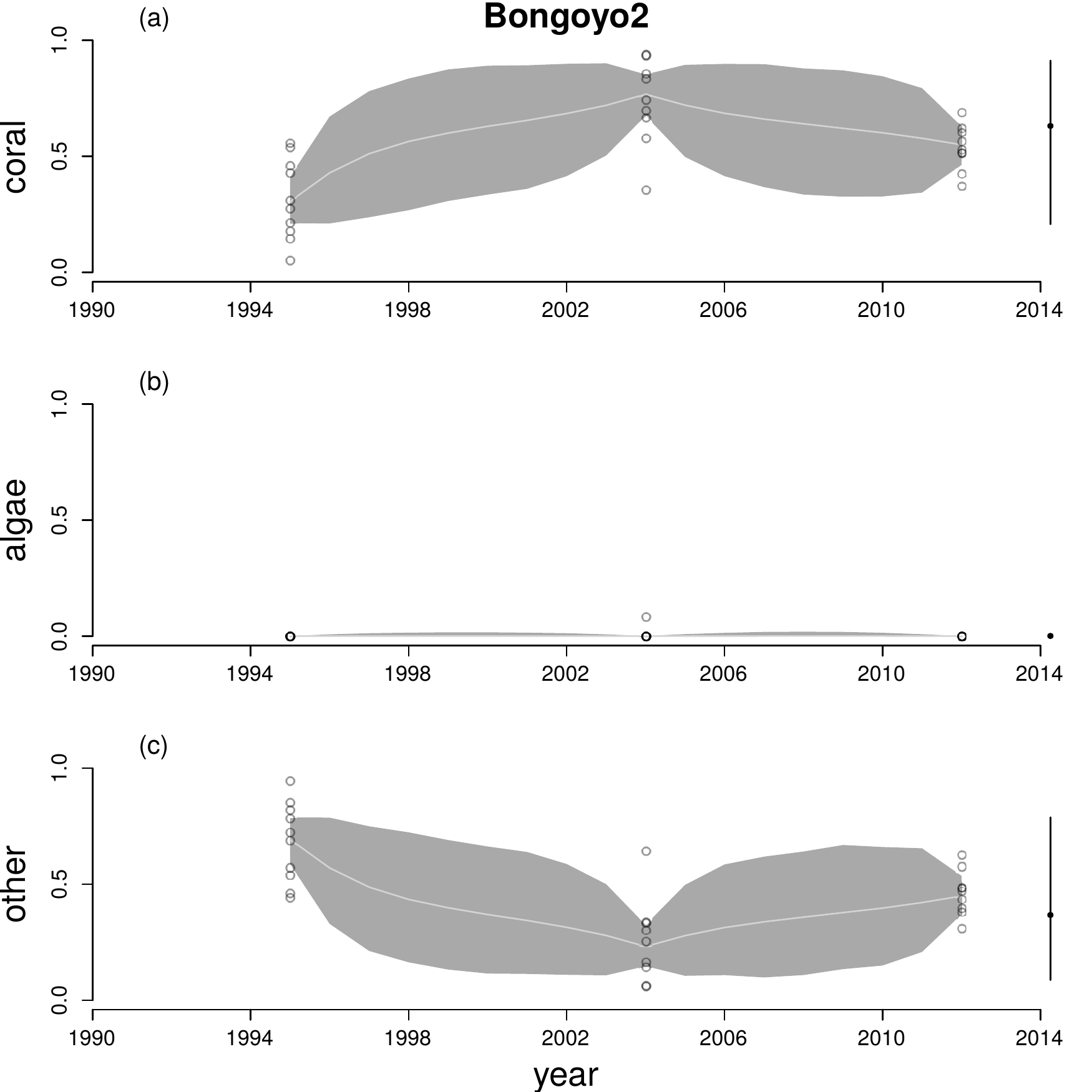}
\caption{Time series for cover of hard corals (a), macroalgae (b) and other (c) at Bongoyo2. See Figure \ref{fig:Bongoyo1} legend for explanation.}
\label{fig:Bongoyo2}
\end{figure}

\clearpage

\begin{figure}[h]
\includegraphics[height=18cm]{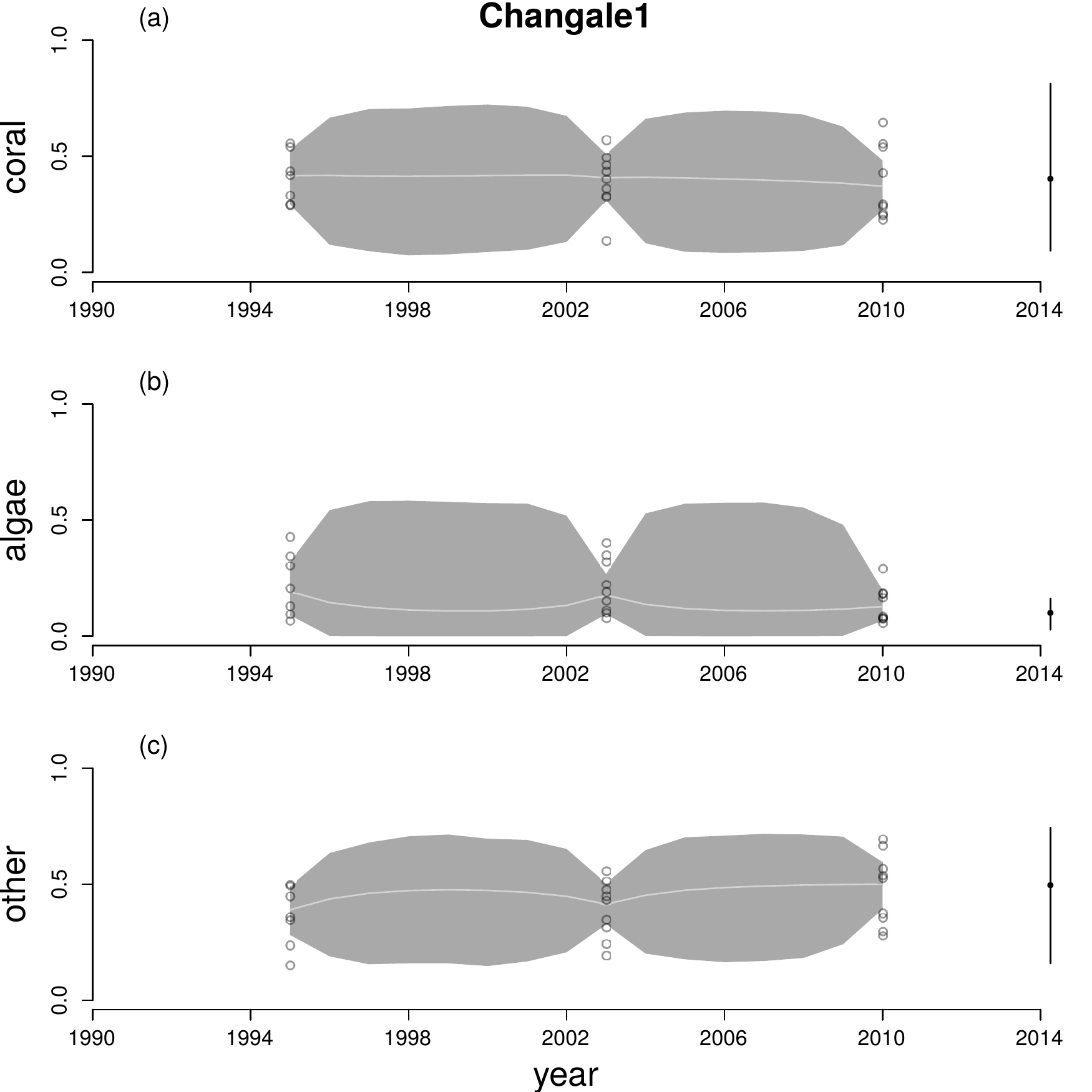}
\caption{Time series for cover of hard corals (a), macroalgae (b) and other (c) at Changale1. See Figure \ref{fig:Bongoyo1} legend for explanation.}
\label{fig:Changale1}
\end{figure}

\clearpage

\begin{figure}[h]
\includegraphics[height=18cm]{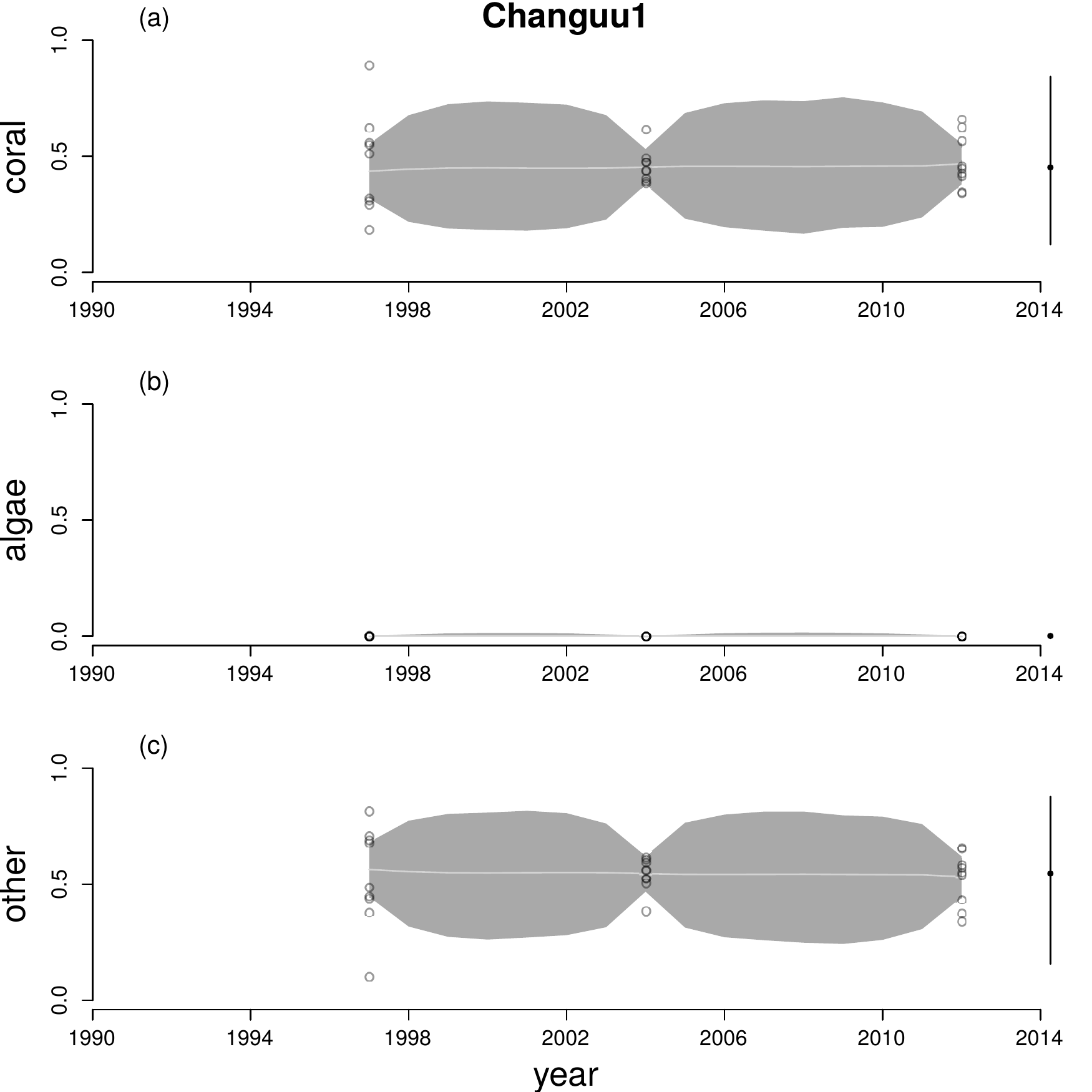}
\caption{Time series for cover of hard corals (a), macroalgae (b) and other (c) at Changuu1. See Figure \ref{fig:Bongoyo1} legend for explanation.}
\label{fig:Changuu1}
\end{figure}

\clearpage

\begin{figure}[h]
\includegraphics[height=18cm]{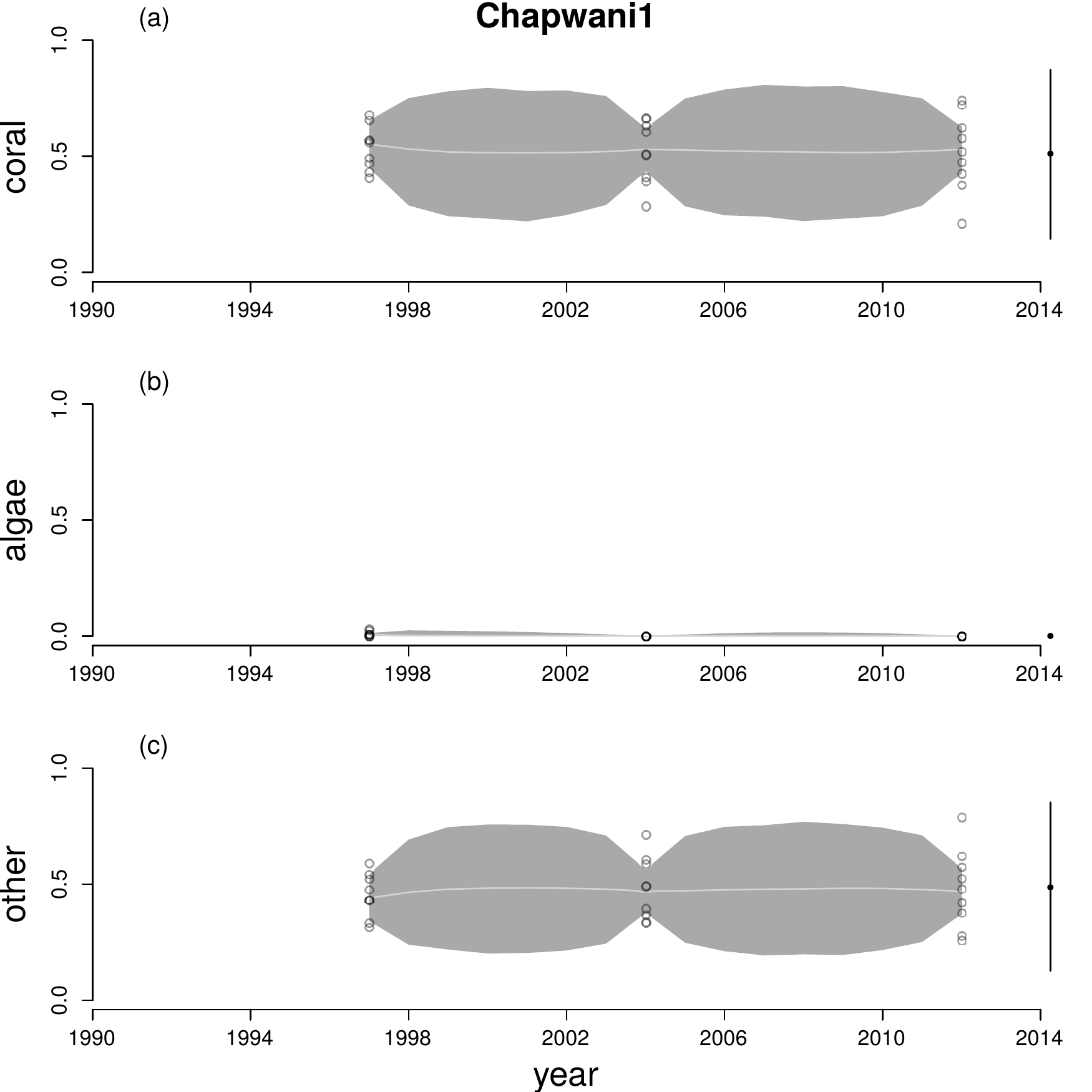}
\caption{Time series for cover of hard corals (a), macroalgae (b) and other (c) at Chapwani1. See Figure \ref{fig:Bongoyo1} legend for explanation.}
\label{fig:Chapwani1}
\end{figure}

\clearpage

\begin{figure}[h]
\includegraphics[height=18cm]{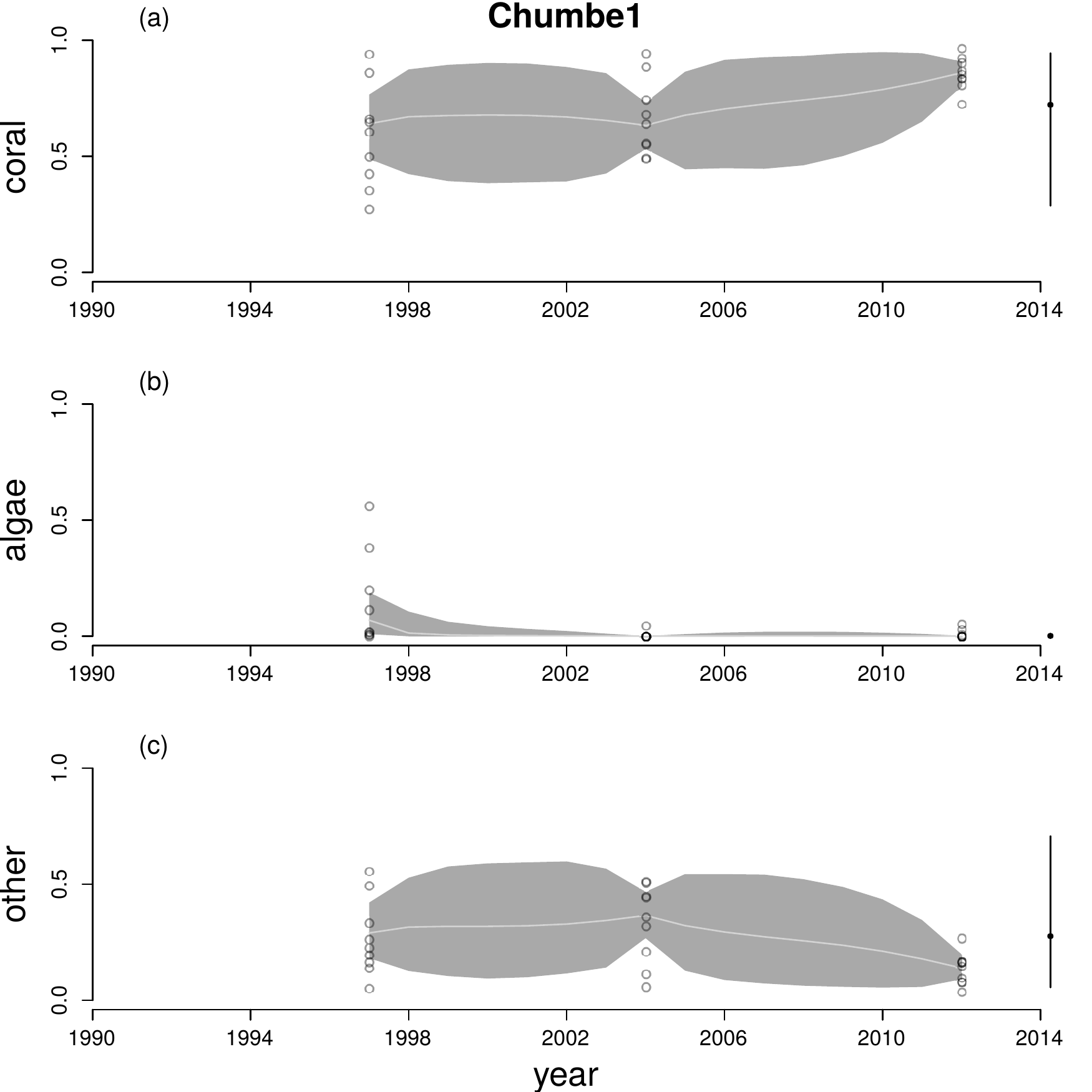}
\caption{Time series for cover of hard corals (a), macroalgae (b) and other (c) at Chumbe1. See Figure \ref{fig:Bongoyo1} legend for explanation.}
\label{fig:Chumbe1}
\end{figure}

\clearpage

\begin{figure}[h]
\includegraphics[height=18cm]{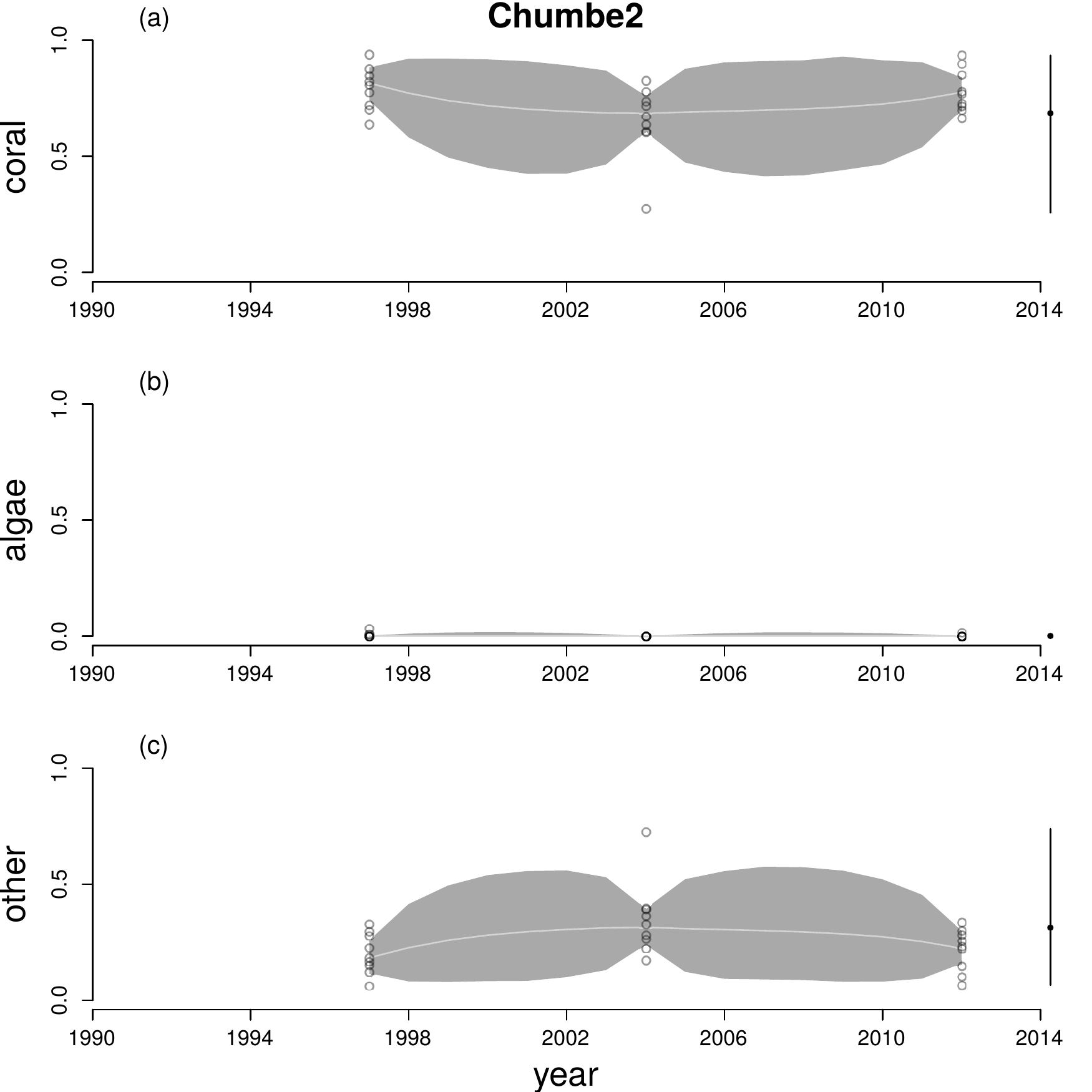}
\caption{Time series for cover of hard corals (a), macroalgae (b) and other (c) at Chumbe2. See Figure \ref{fig:Bongoyo1} legend for explanation.}
\label{fig:Chumbe2}
\end{figure}

\clearpage

\begin{figure}[h]
\includegraphics[height=18cm]{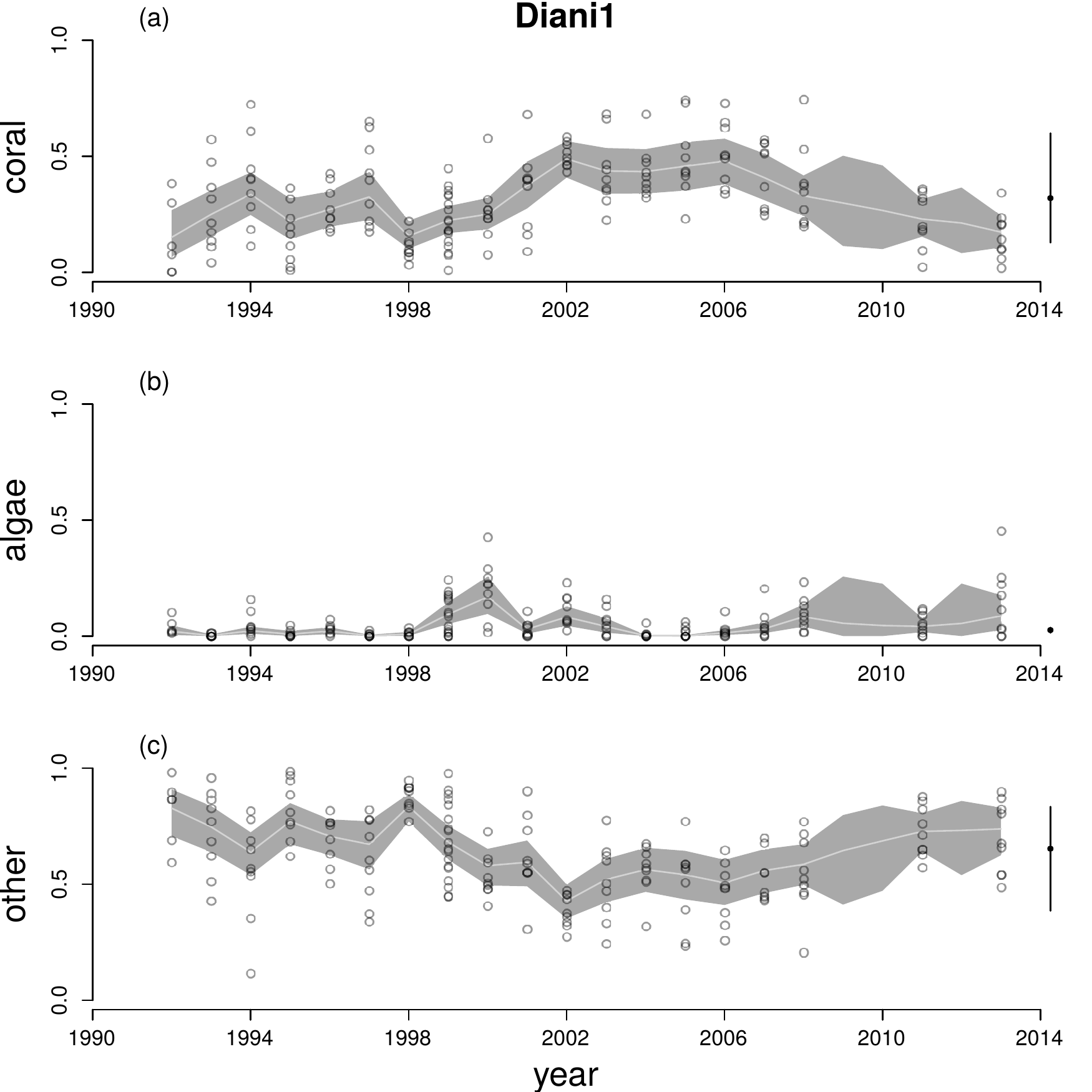}
\caption{Time series for cover of hard corals (a), macroalgae (b) and other (c) at Diani1. See Figure \ref{fig:Bongoyo1} legend for explanation.}
\label{fig:Diani1}
\end{figure}

\clearpage

\begin{figure}[h]
\includegraphics[height=18cm]{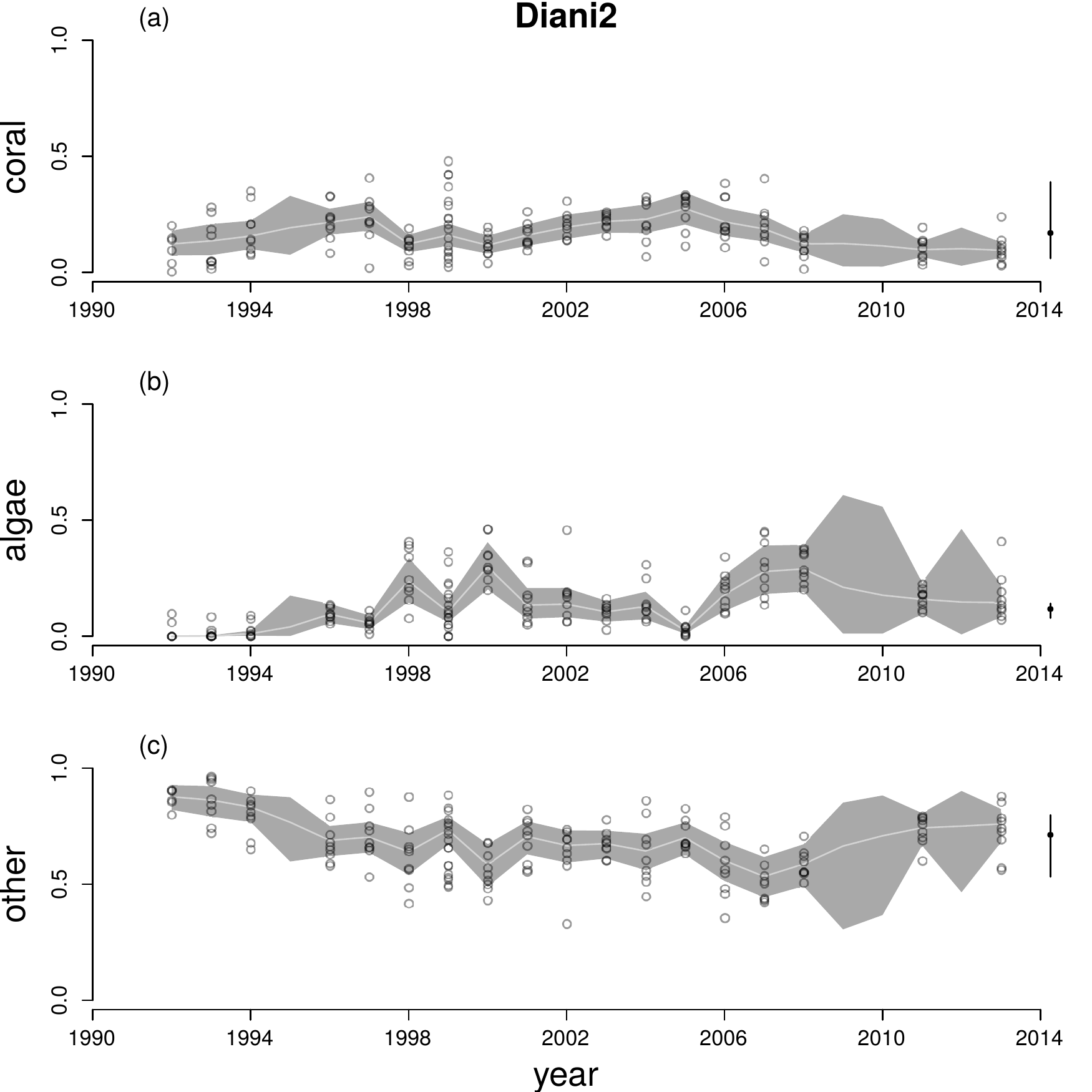}
\caption{Time series for cover of hard corals (a), macroalgae (b) and other (c) at Diani2. See Figure \ref{fig:Bongoyo1} legend for explanation.}
\label{fig:Diani2}
\end{figure}

\clearpage

\begin{figure}[h]
\includegraphics[height=18cm]{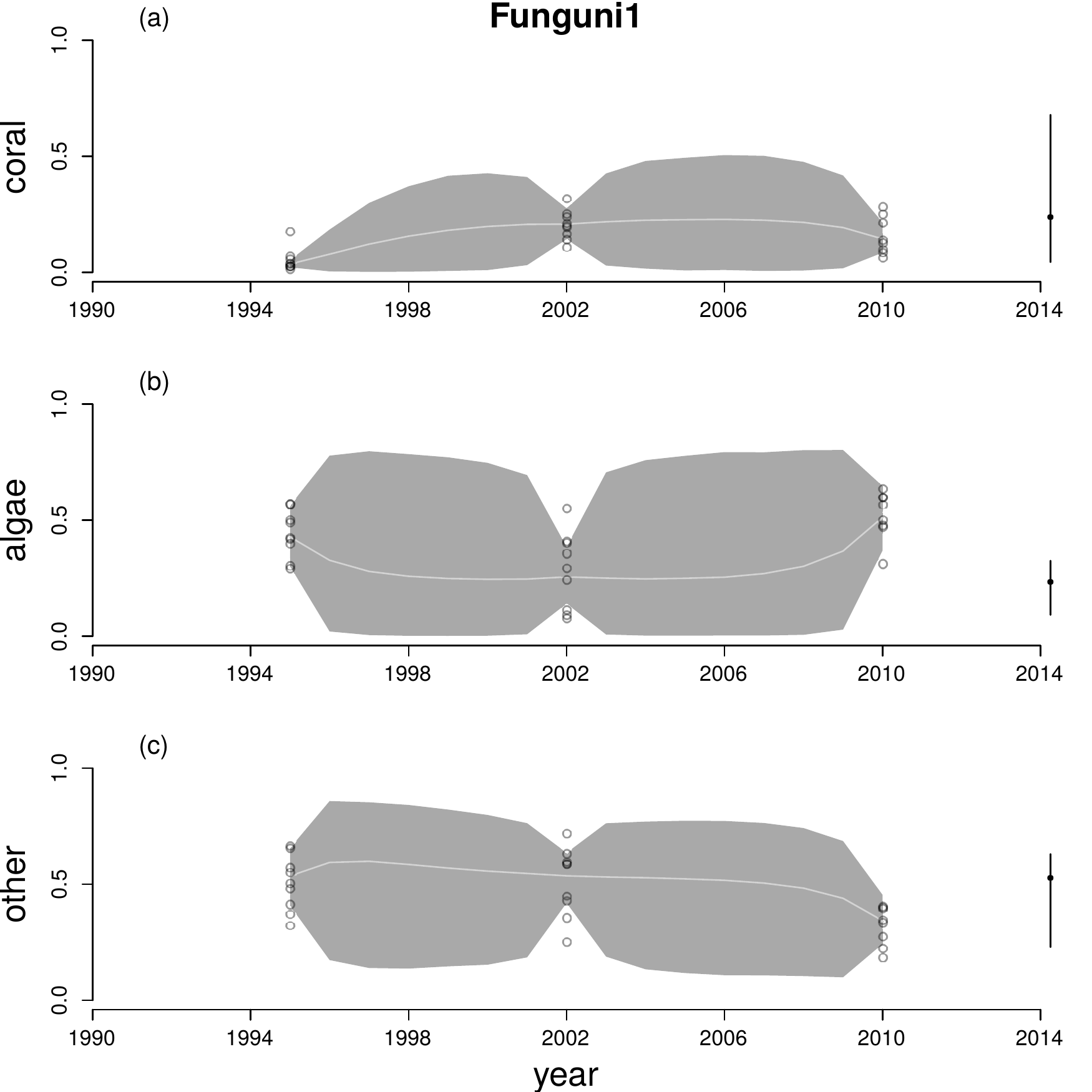}
\caption{Time series for cover of hard corals (a), macroalgae (b) and other (c) at Funguni1. See Figure \ref{fig:Bongoyo1} legend for explanation.}
\label{fig:Funguni1}
\end{figure}

\clearpage

\begin{figure}[h]
\includegraphics[height=18cm]{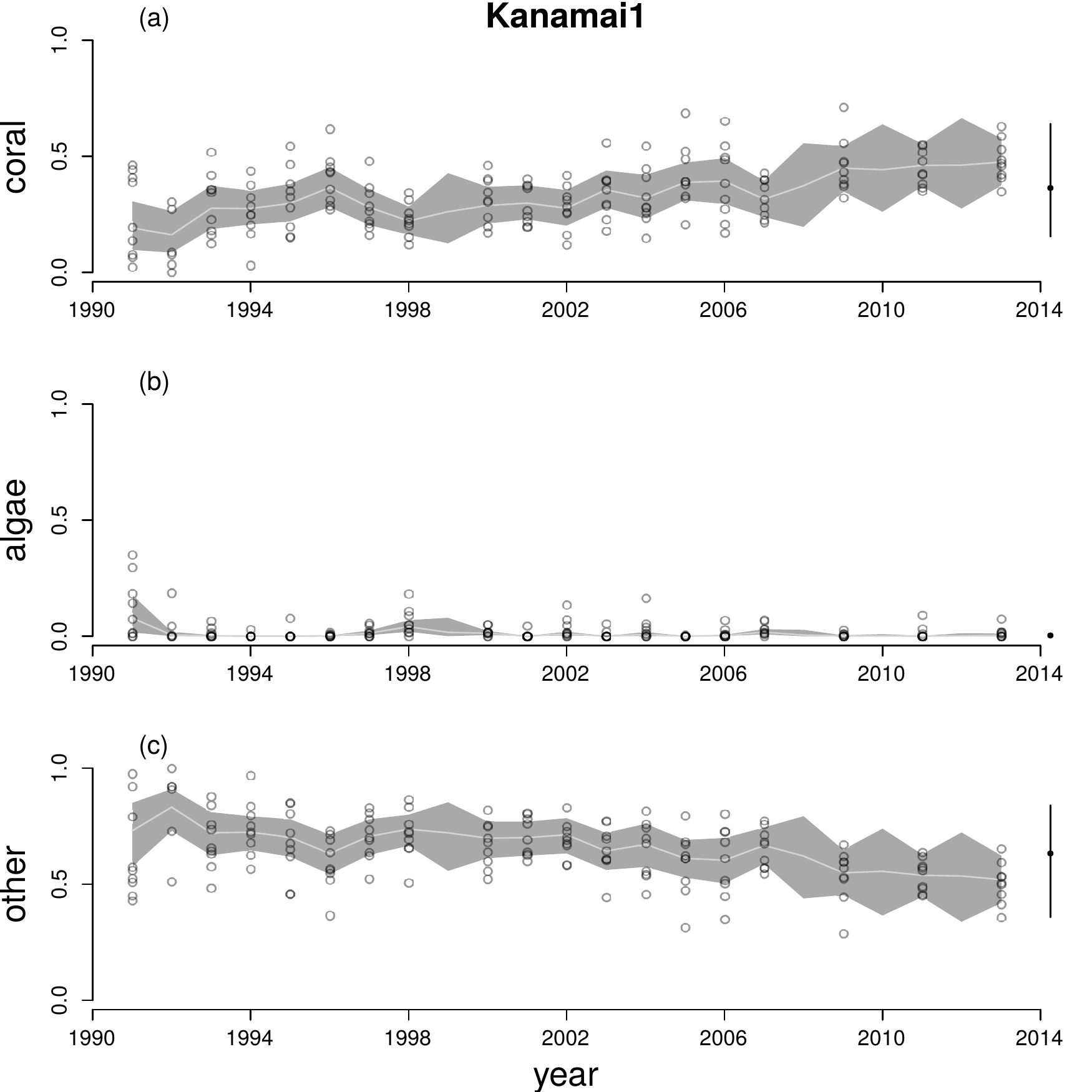}
\caption{Time series for cover of hard corals (a), macroalgae (b) and other (c) at Kanamai1. See Figure \ref{fig:Bongoyo1} legend for explanation.}
\label{fig:Kanamai1}
\end{figure}

\clearpage

\begin{figure}[h]
\includegraphics[height=18cm]{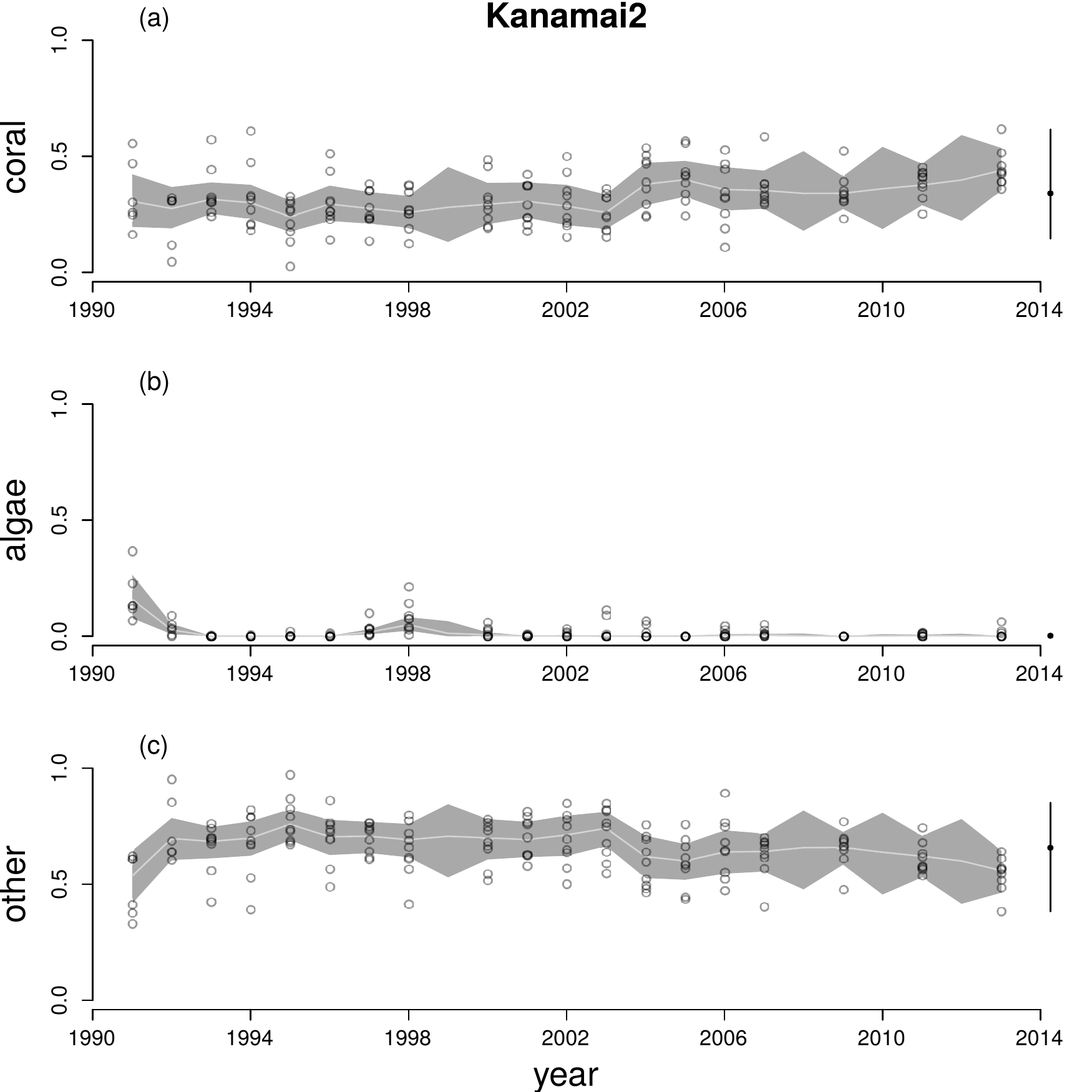}
\caption{Time series for cover of hard corals (a), macroalgae (b) and other (c) at Kanamai2. See Figure \ref{fig:Bongoyo1} legend for explanation.}
\label{fig:Kanamai2}
\end{figure}

\clearpage

\begin{figure}[h]
\includegraphics[height=18cm]{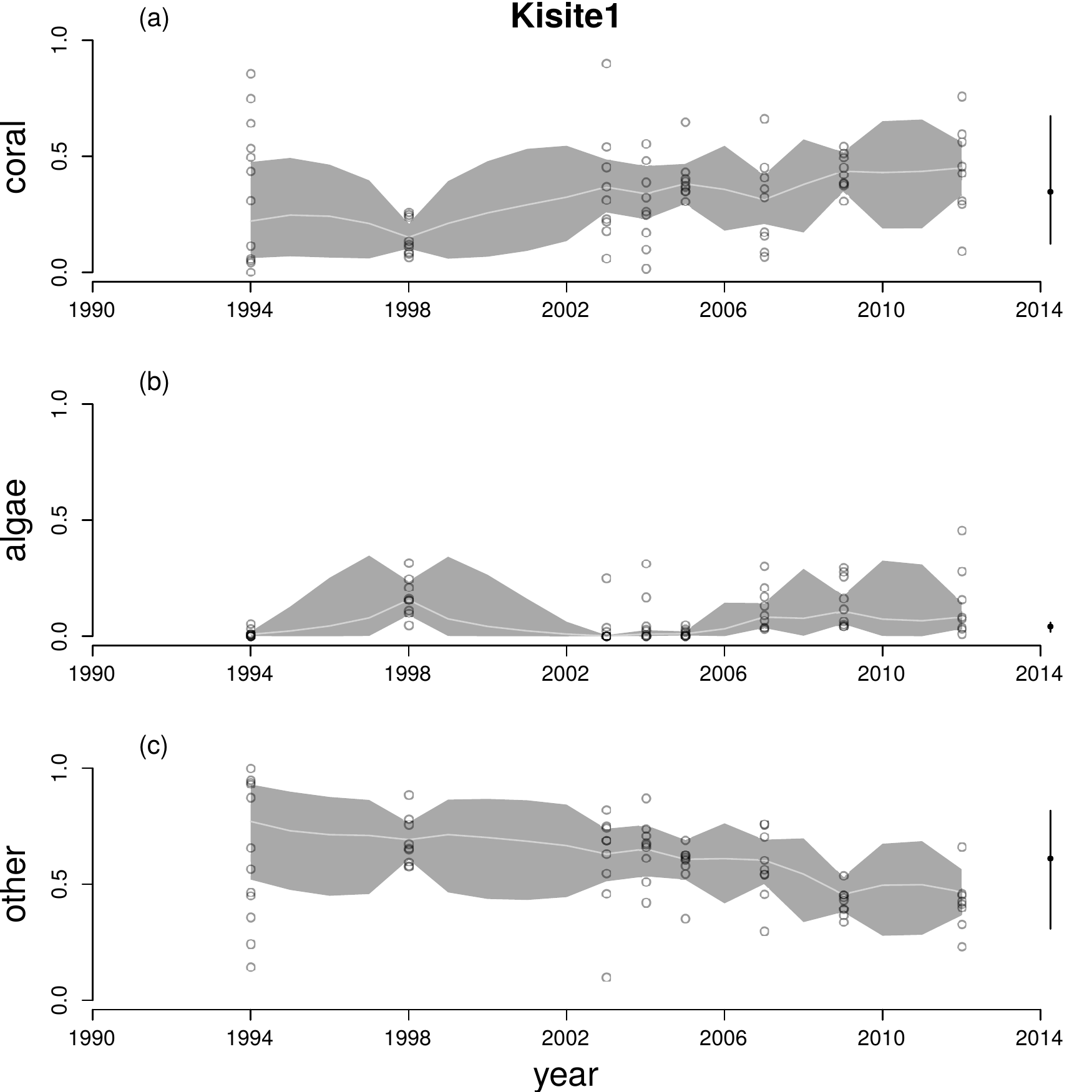}
\caption{Time series for cover of hard corals (a), macroalgae (b) and other (c) at Kisite1. See Figure \ref{fig:Bongoyo1} legend for explanation.}
\label{fig:Kisite1}
\end{figure}

\clearpage

\begin{figure}[h]
\includegraphics[height=18cm]{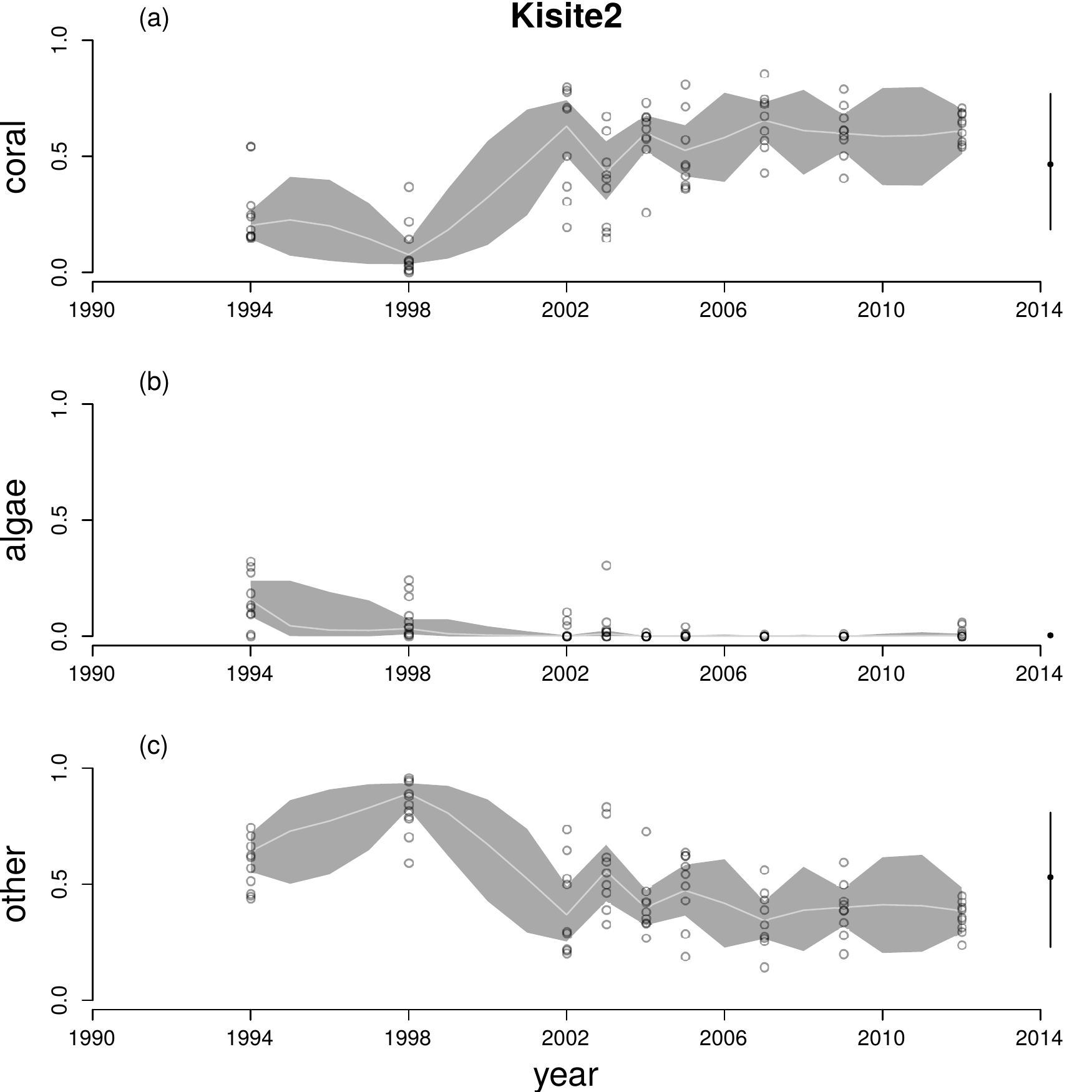}
\caption{Time series for cover of hard corals (a), macroalgae (b) and other (c) at Kisite2. See Figure \ref{fig:Bongoyo1} legend for explanation.}
\label{fig:Kisite2}
\end{figure}

\clearpage

\begin{figure}[h]
\includegraphics[height=18cm]{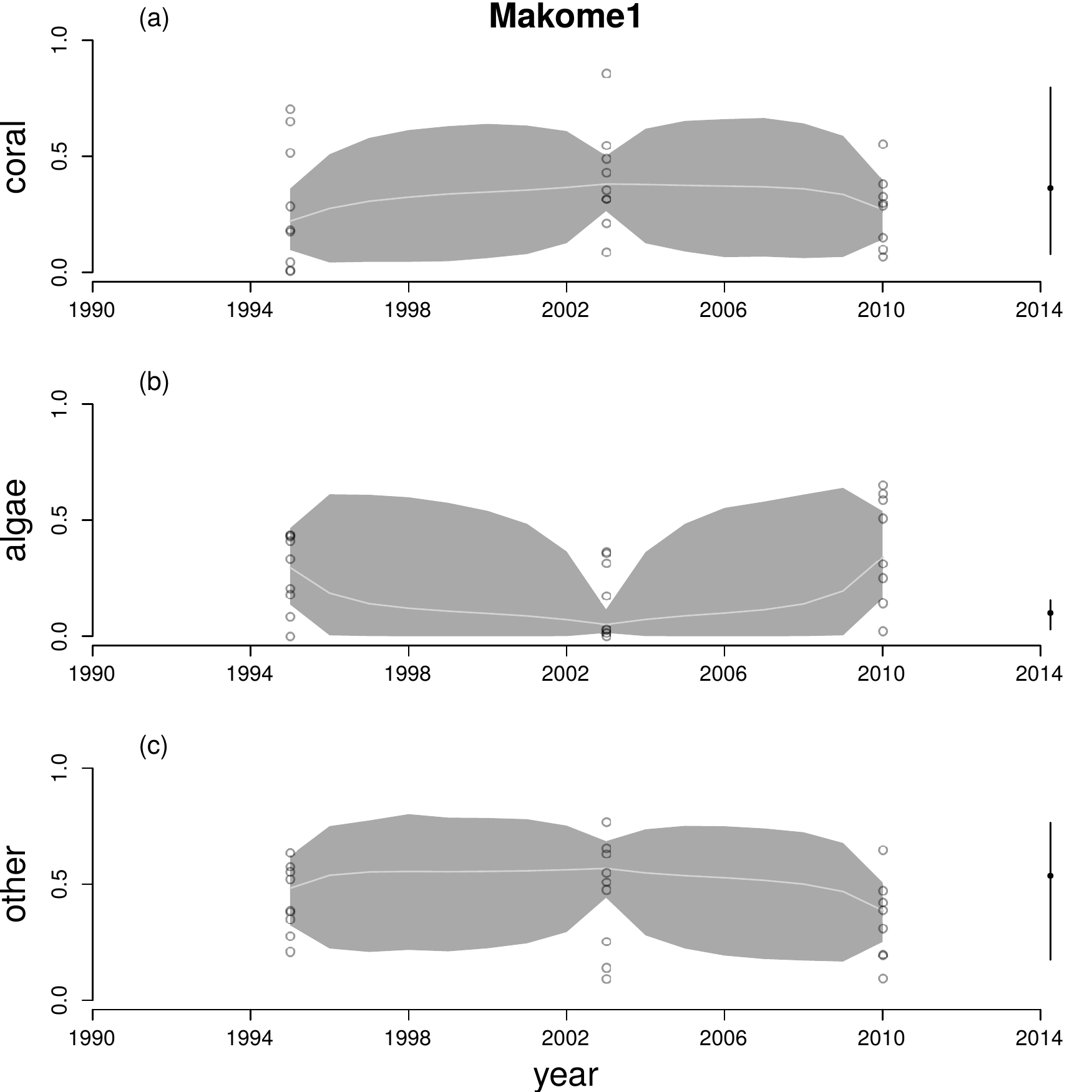}
\caption{Time series for cover of hard corals (a), macroalgae (b) and other (c) at Makome1. See Figure \ref{fig:Bongoyo1} legend for explanation.}
\label{fig:Makome1}
\end{figure}

\clearpage

\begin{figure}[h]
\includegraphics[height=18cm]{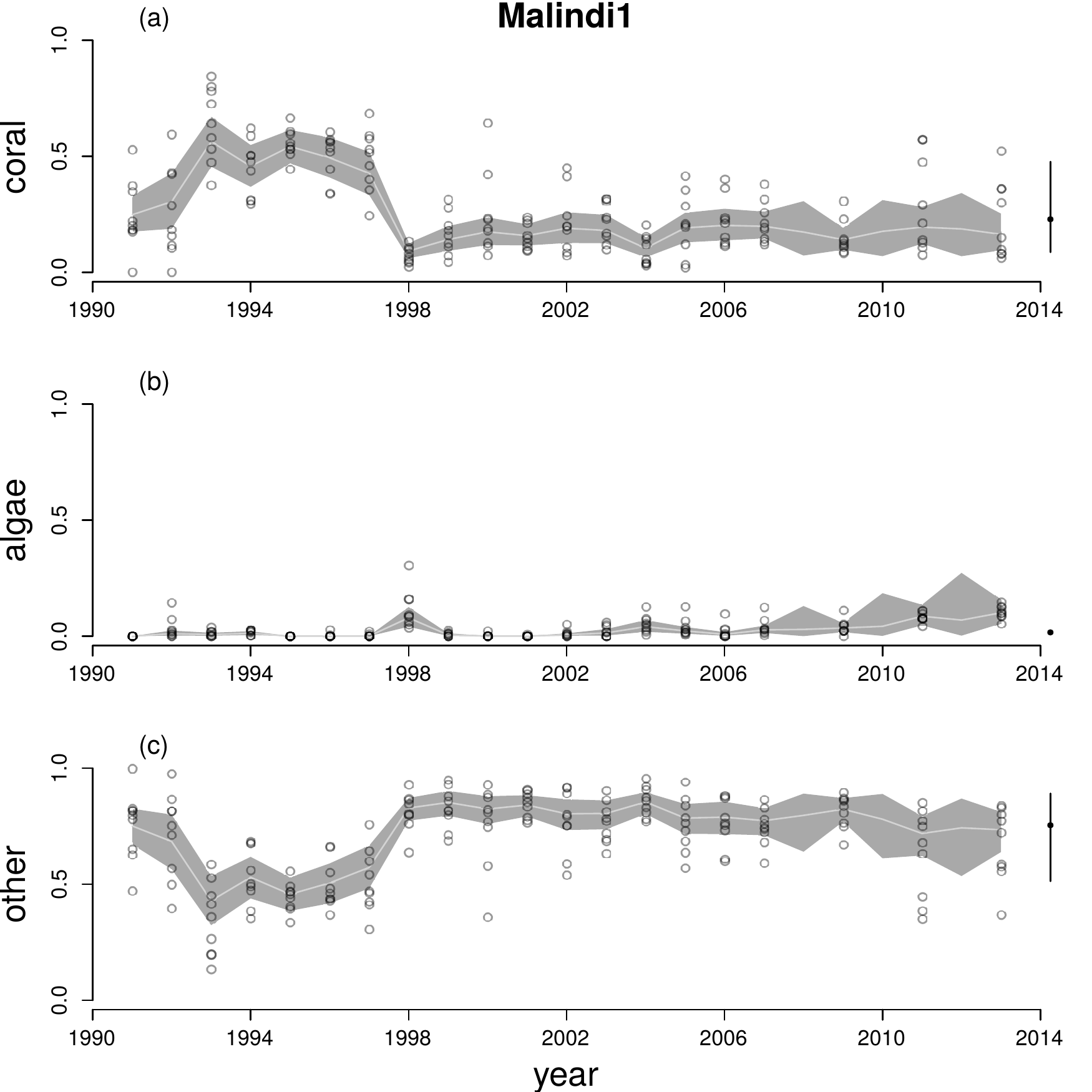}
\caption{Time series for cover of hard corals (a), macroalgae (b) and other (c) at Malindi1. See Figure \ref{fig:Bongoyo1} legend for explanation.}
\label{fig:Malindi1}
\end{figure}

\clearpage

\begin{figure}[h]
\includegraphics[height=18cm]{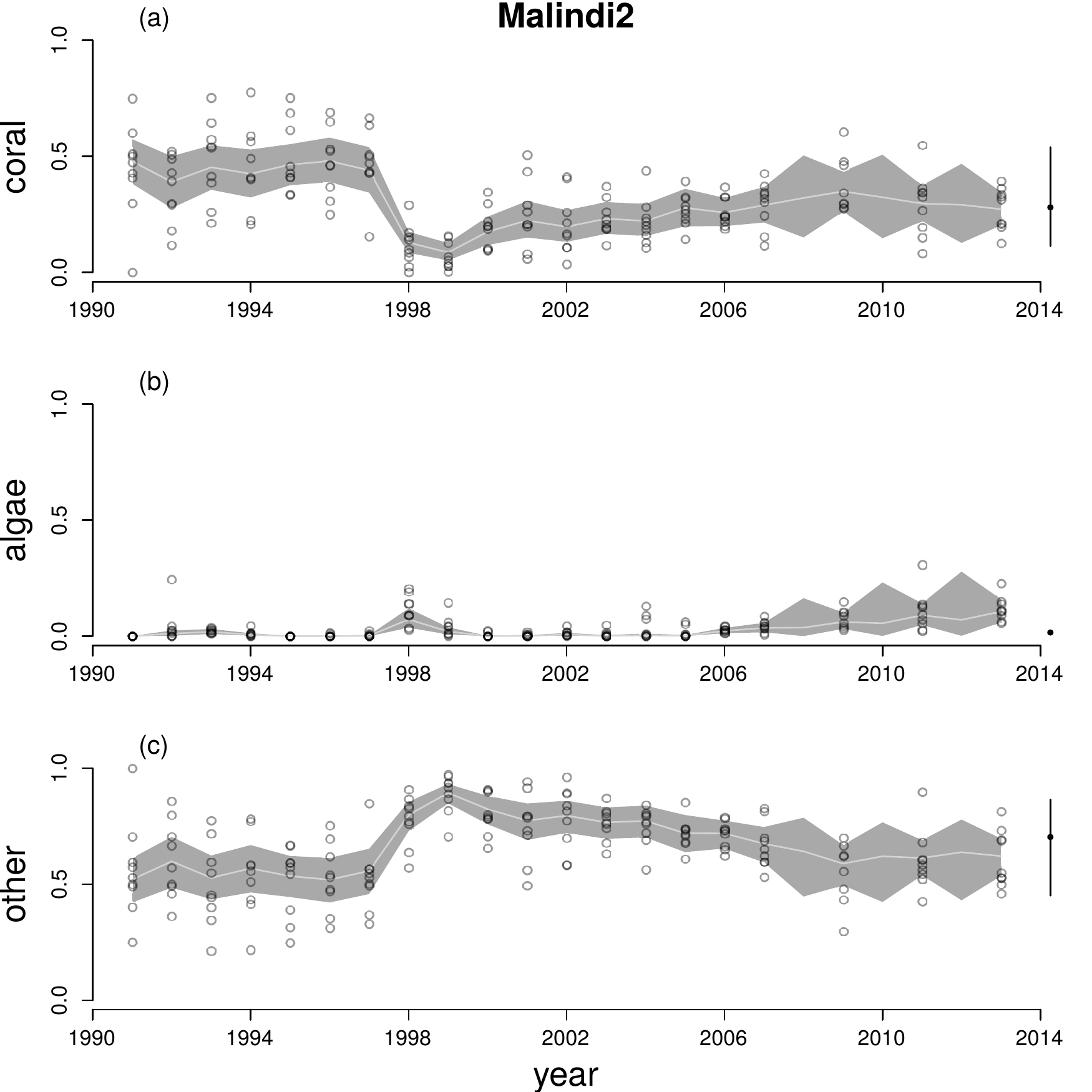}
\caption{Time series for cover of hard corals (a), macroalgae (b) and other (c) at Malindi2. See Figure \ref{fig:Bongoyo1} legend for explanation.}
\label{fig:Malindi2}
\end{figure}

\clearpage

\begin{figure}[h]
\includegraphics[height=18cm]{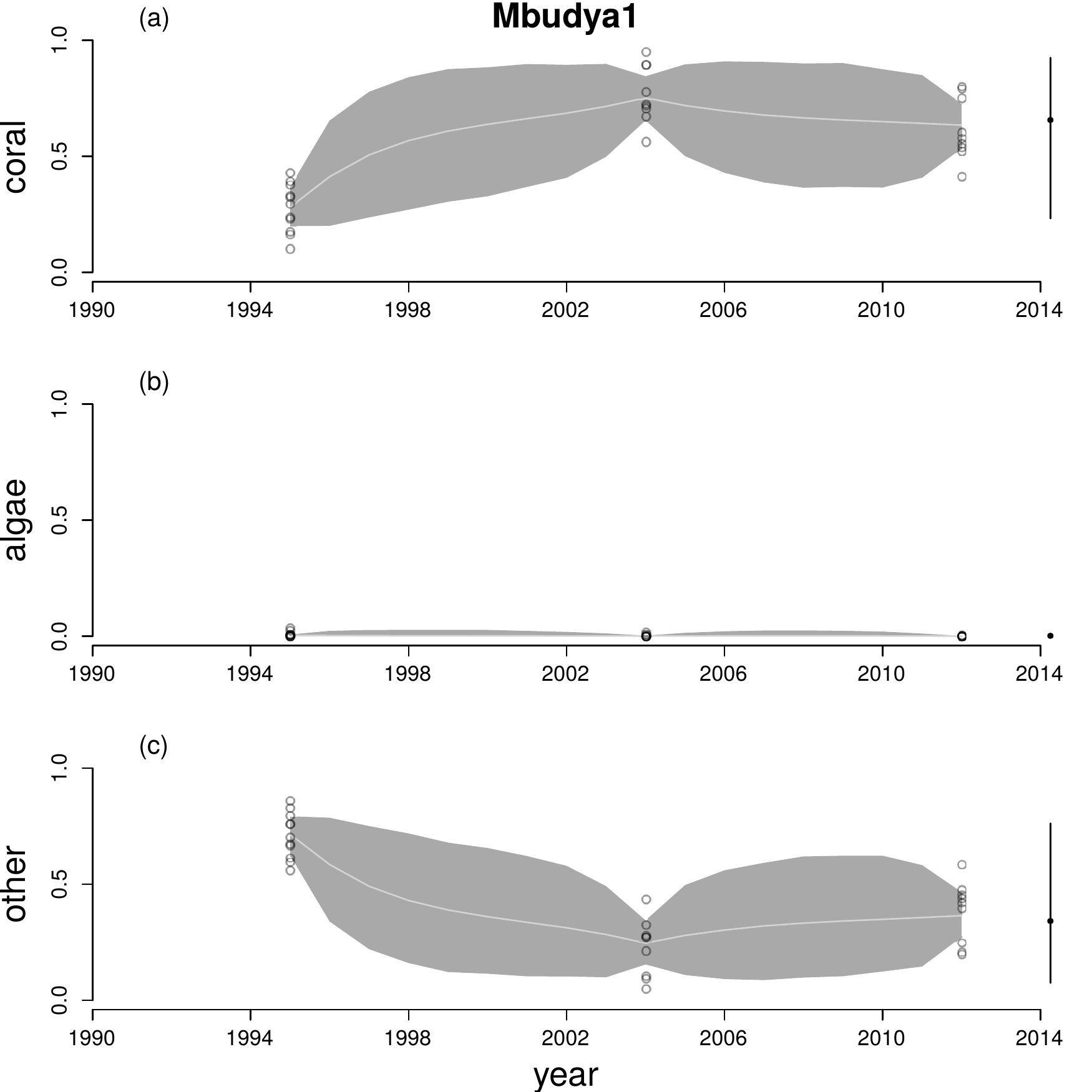}
\caption{Time series for cover of hard corals (a), macroalgae (b) and other (c) at Mbudya1. See Figure \ref{fig:Bongoyo1} legend for explanation.}
\label{fig:Mbudya1}
\end{figure}

\clearpage

\begin{figure}[h]
\includegraphics[height=18cm]{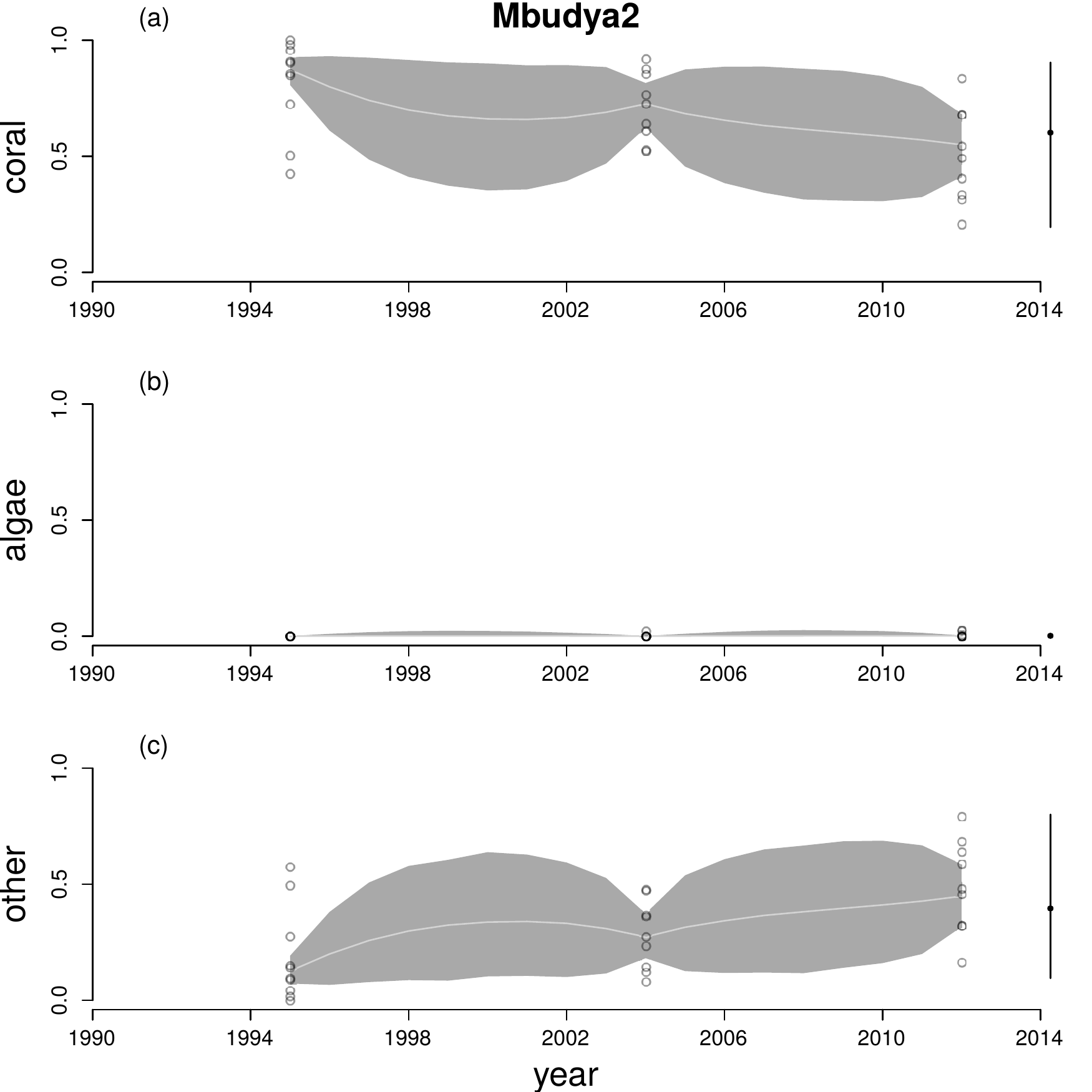}
\caption{Time series for cover of hard corals (a), macroalgae (b) and other (c) at Mbudya2. See Figure \ref{fig:Bongoyo1} legend for explanation.}
\label{fig:Mbudya2}
\end{figure}

\clearpage

\begin{figure}[h]
\includegraphics[height=18cm]{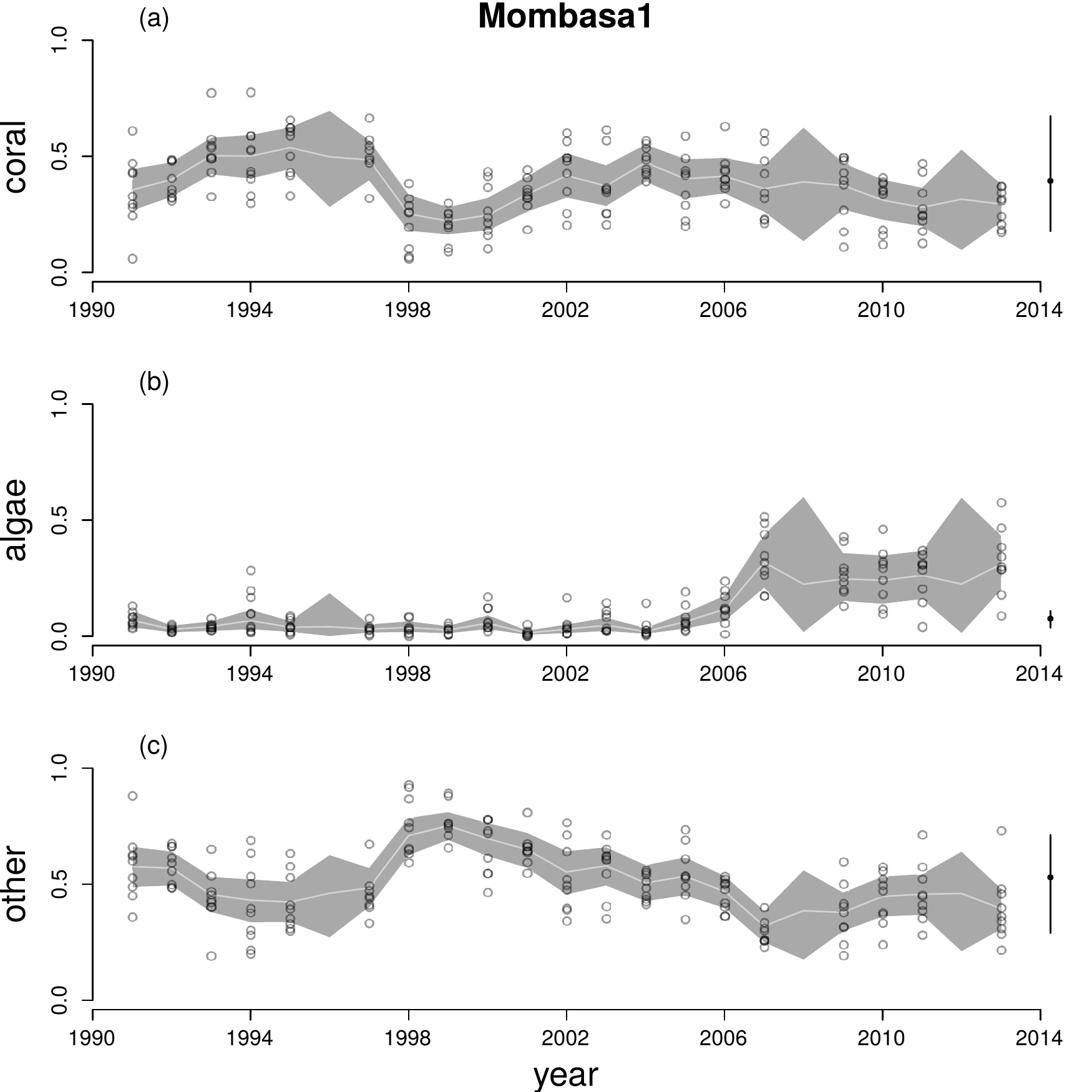}
\caption{Time series for cover of hard corals (a), macroalgae (b) and other (c) at Mombasa1. See Figure \ref{fig:Bongoyo1} legend for explanation.}
\label{fig:Mombasa1}
\end{figure}

\clearpage

\begin{figure}[h]
\includegraphics[height=18cm]{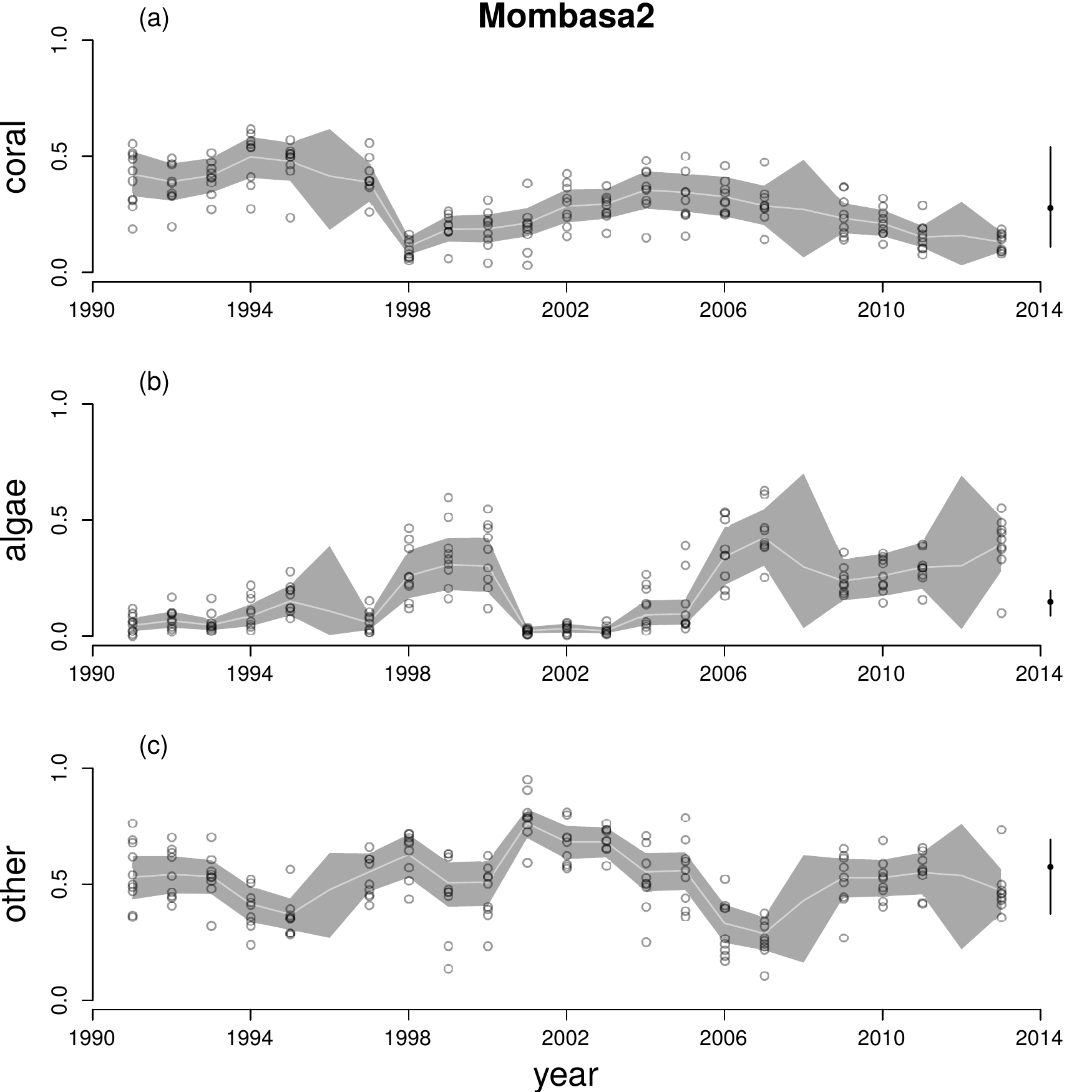}
\caption{Time series for cover of hard corals (a), macroalgae (b) and other (c) at Mombasa2. See Figure \ref{fig:Bongoyo1} legend for explanation.}
\label{fig:Mombasa2}
\end{figure}

\clearpage

\begin{figure}[h]
\includegraphics[height=18cm]{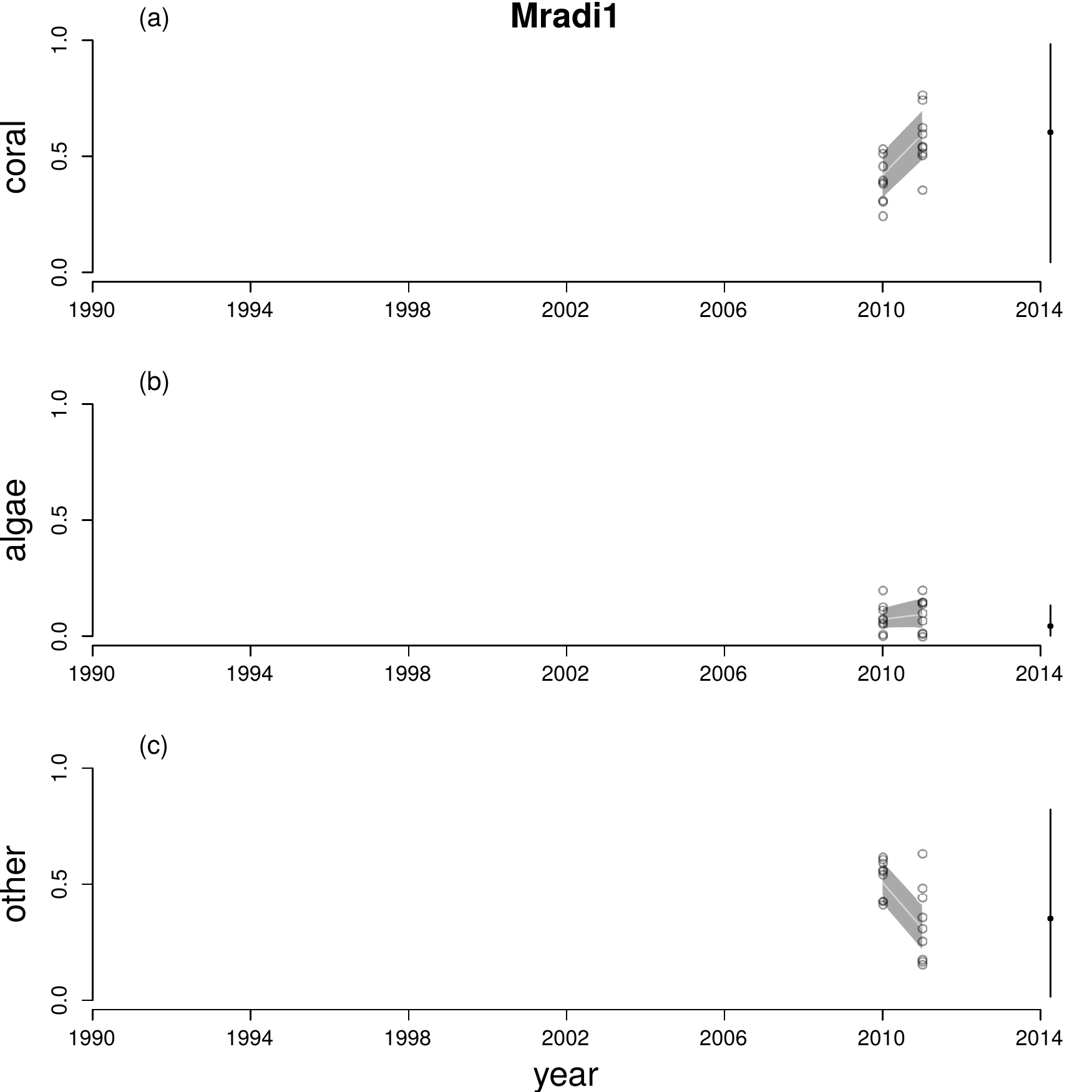}
\caption{Time series for cover of hard corals (a), macroalgae (b) and other (c) at Mradi1. See Figure \ref{fig:Bongoyo1} legend for explanation.}
\label{fig:Mradi1}
\end{figure}

\clearpage

\begin{figure}[h]
\includegraphics[height=18cm]{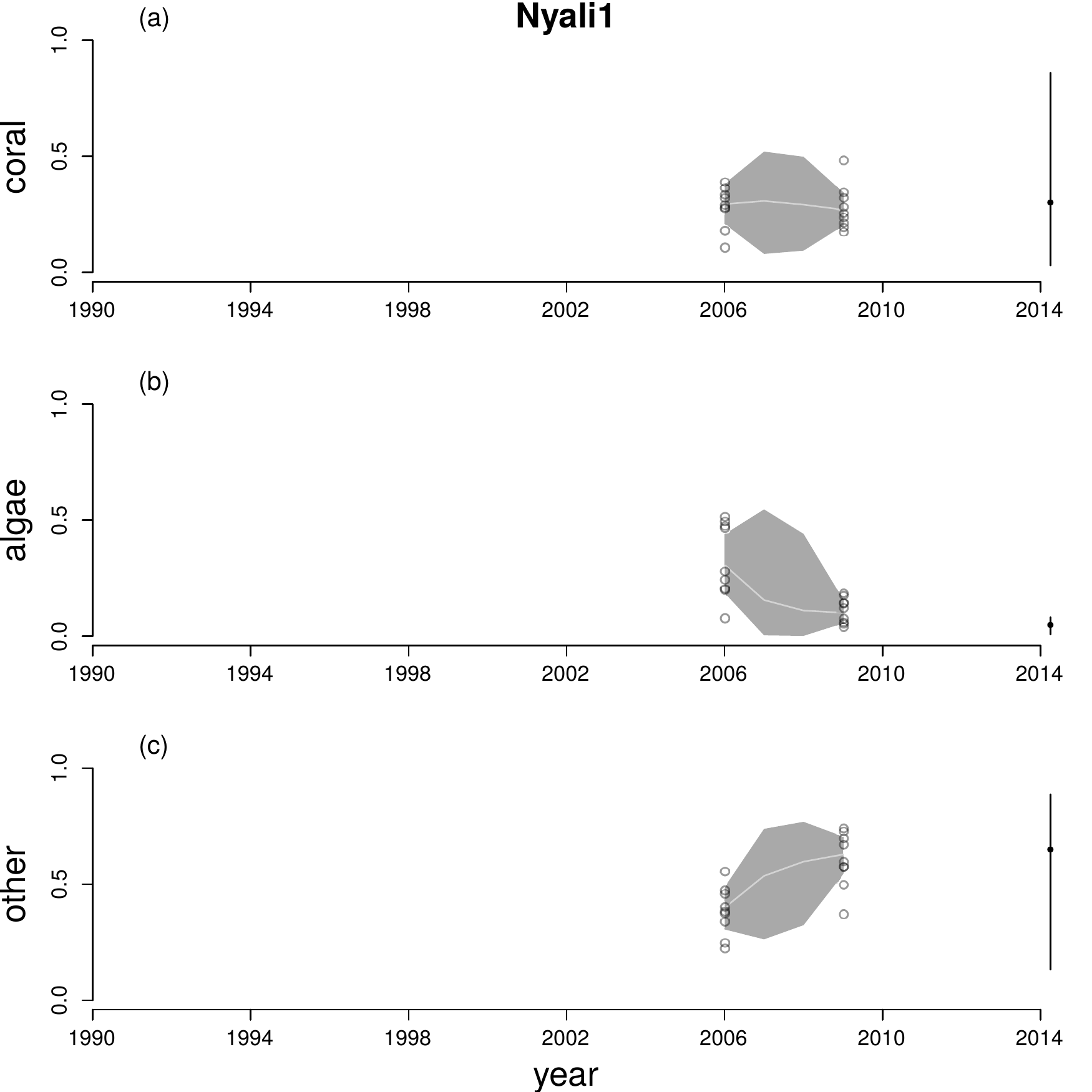}
\caption{Time series for cover of hard corals (a), macroalgae (b) and other (c) at Nyali1. See Figure \ref{fig:Bongoyo1} legend for explanation.}
\label{fig:Nyali1}
\end{figure}

\clearpage

\begin{figure}[h]
\includegraphics[height=18cm]{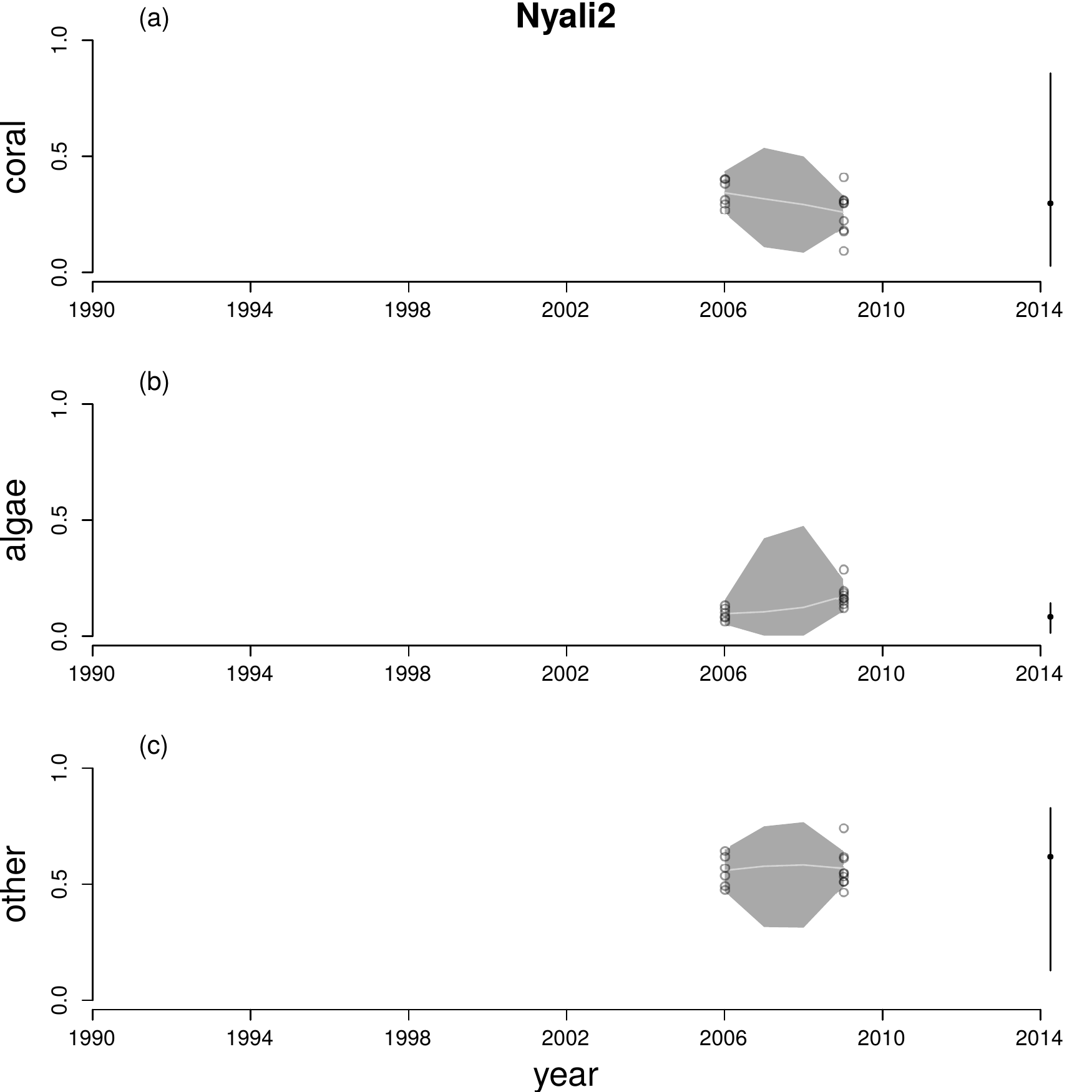}
\caption{Time series for cover of hard corals (a), macroalgae (b) and other (c) at Nyali2. See Figure \ref{fig:Bongoyo1} legend for explanation.}
\label{fig:Nyali2}
\end{figure}

\clearpage

\begin{figure}[h]
\includegraphics[height=18cm]{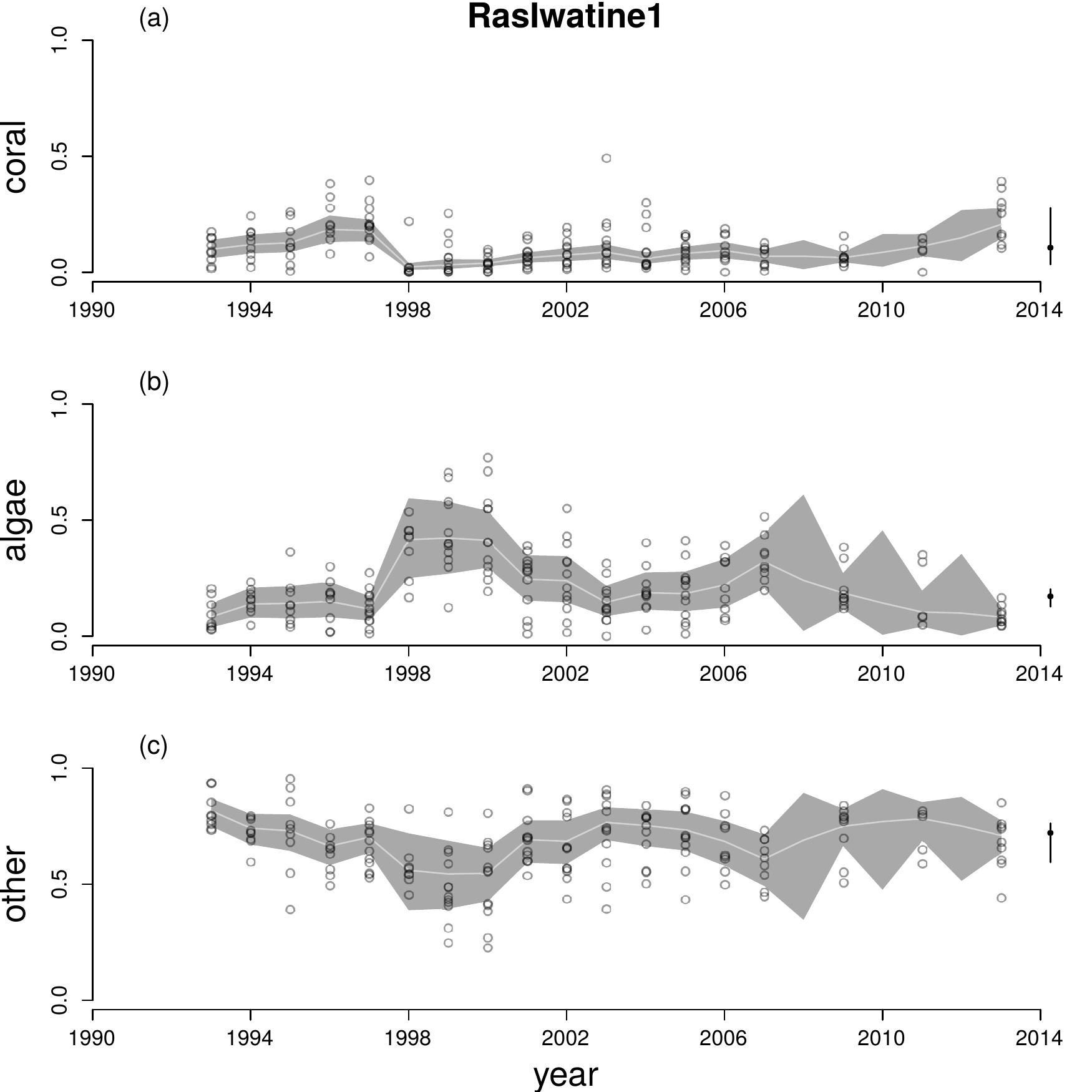}
\caption{Time series for cover of hard corals (a), macroalgae (b) and other (c) at RasIwatine1. See Figure \ref{fig:Bongoyo1} legend for explanation.}
\label{fig:RasIwatine1}
\end{figure}

\clearpage

\begin{figure}[h]
\includegraphics[height=18cm]{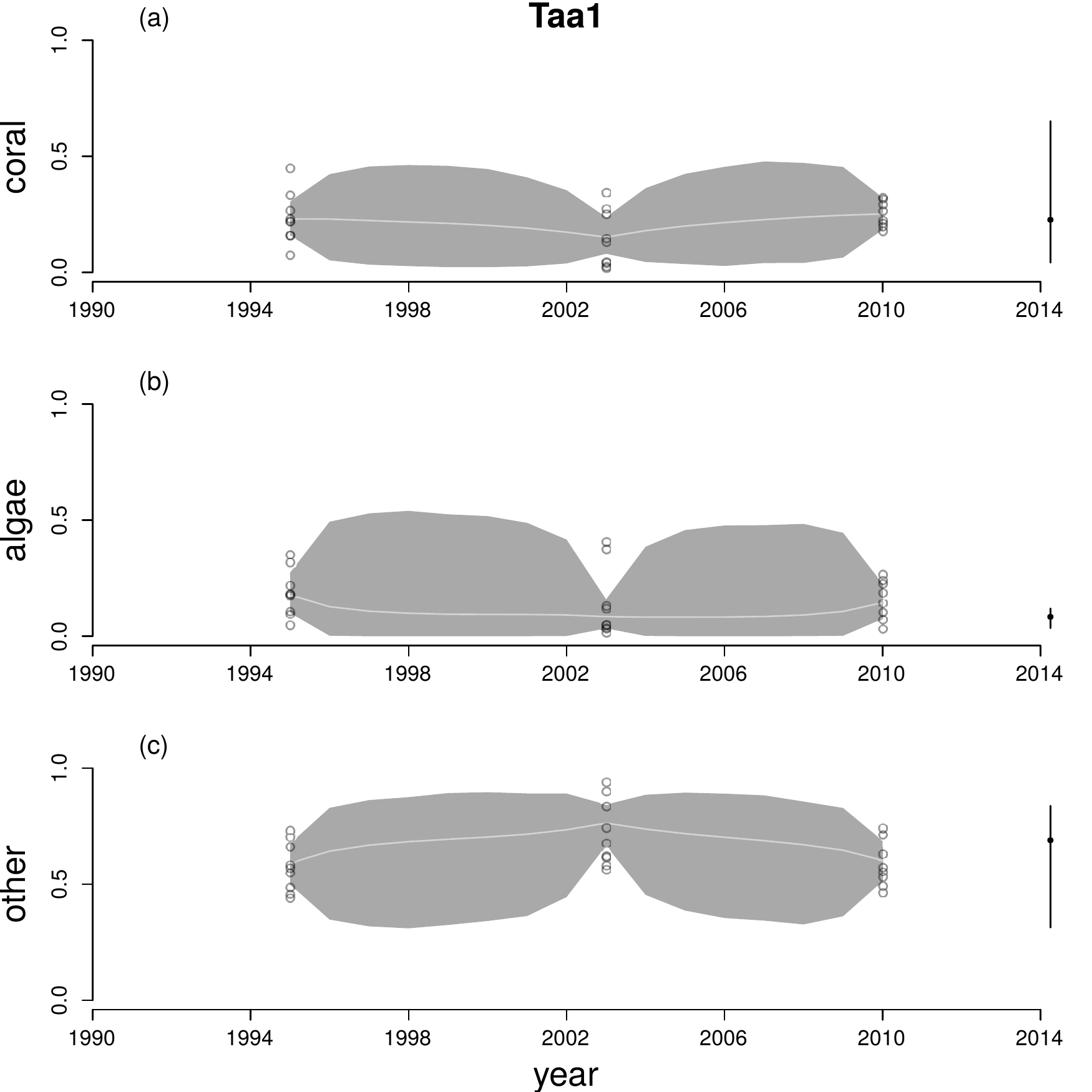}
\caption{Time series for cover of hard corals (a), macroalgae (b) and other (c) at Taa1. See Figure \ref{fig:Bongoyo1} legend for explanation.}
\label{fig:Taa1}
\end{figure}

\clearpage

\begin{figure}[h]
\includegraphics[height=18cm]{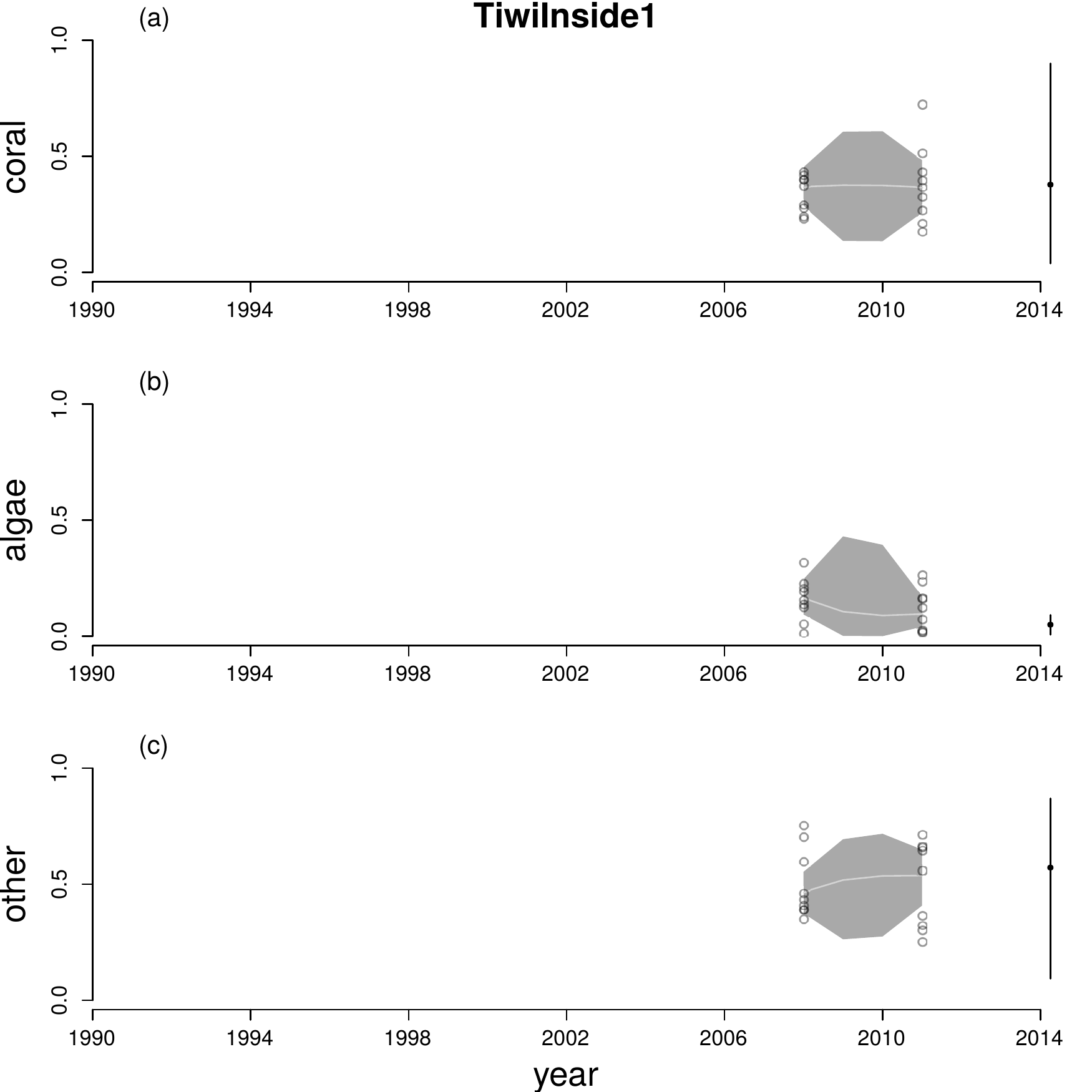}
\caption{Time series for cover of hard corals (a), macroalgae (b) and other (c) at TiwiInside1. See Figure \ref{fig:Bongoyo1} legend for explanation.}
\label{fig:TiwiInside1}
\end{figure}

\clearpage

\begin{figure}[h]
\includegraphics[height=18cm]{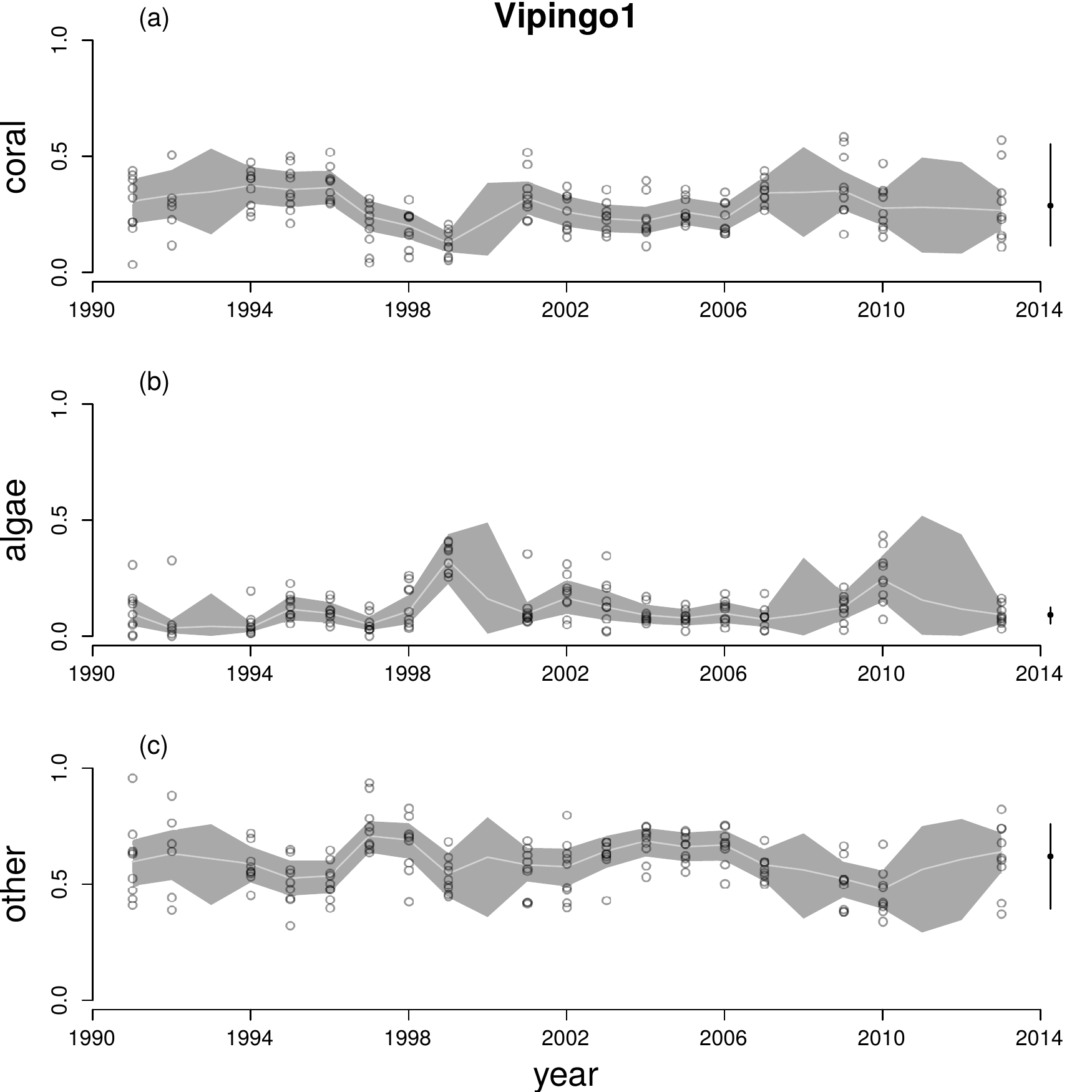}
\caption{Time series for cover of hard corals (a), macroalgae (b) and other (c) at Vipingo1. See Figure \ref{fig:Bongoyo1} legend for explanation.}
\label{fig:Vipingo1}
\end{figure}

\clearpage

\begin{figure}[h]
\includegraphics[height=18cm]{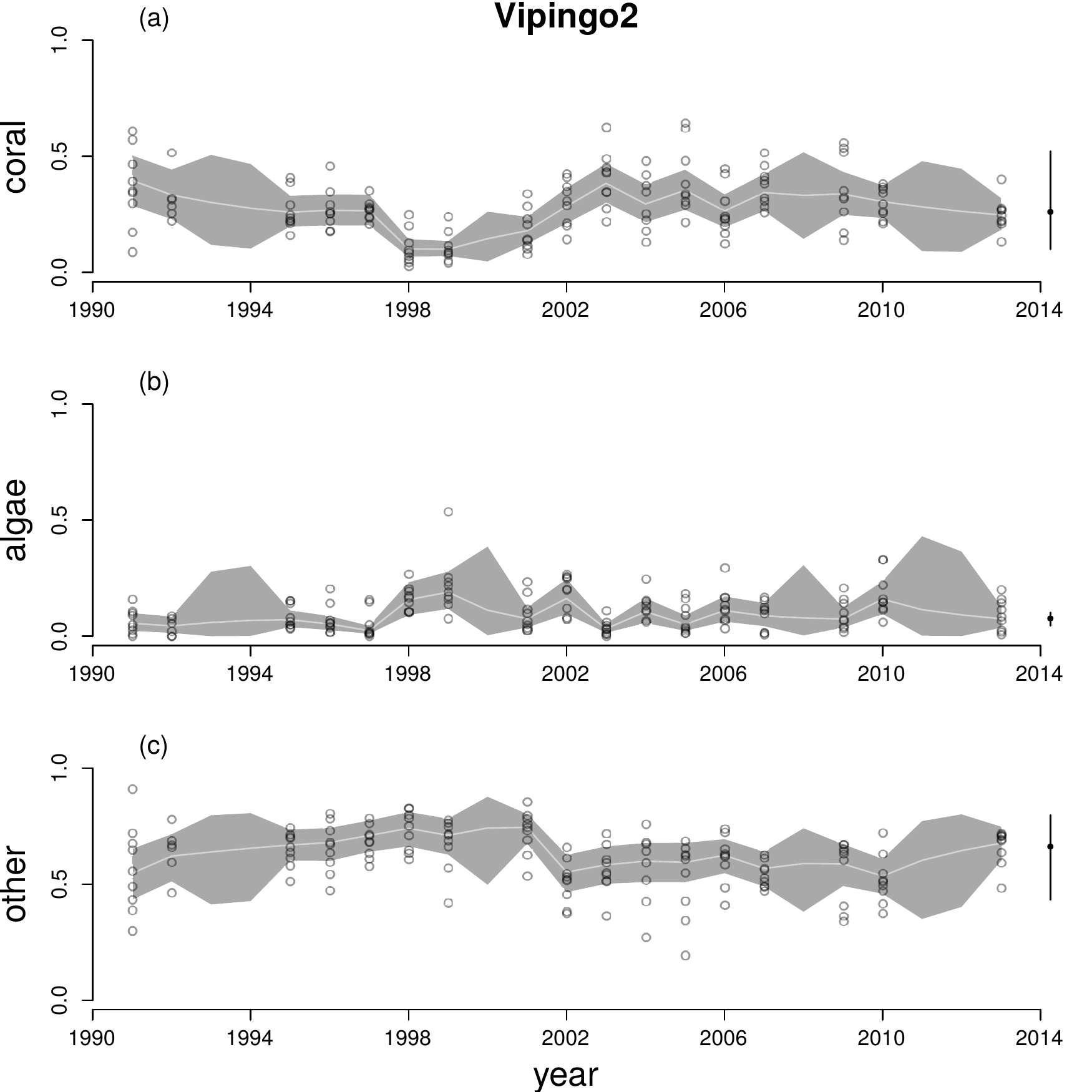}
\caption{Time series for cover of hard corals (a), macroalgae (b) and other (c) at Vipingo2. See Figure \ref{fig:Bongoyo1} legend for explanation.}
\label{fig:Vipingo2}
\end{figure}

\clearpage

\begin{figure}[h]
\includegraphics[height=18cm]{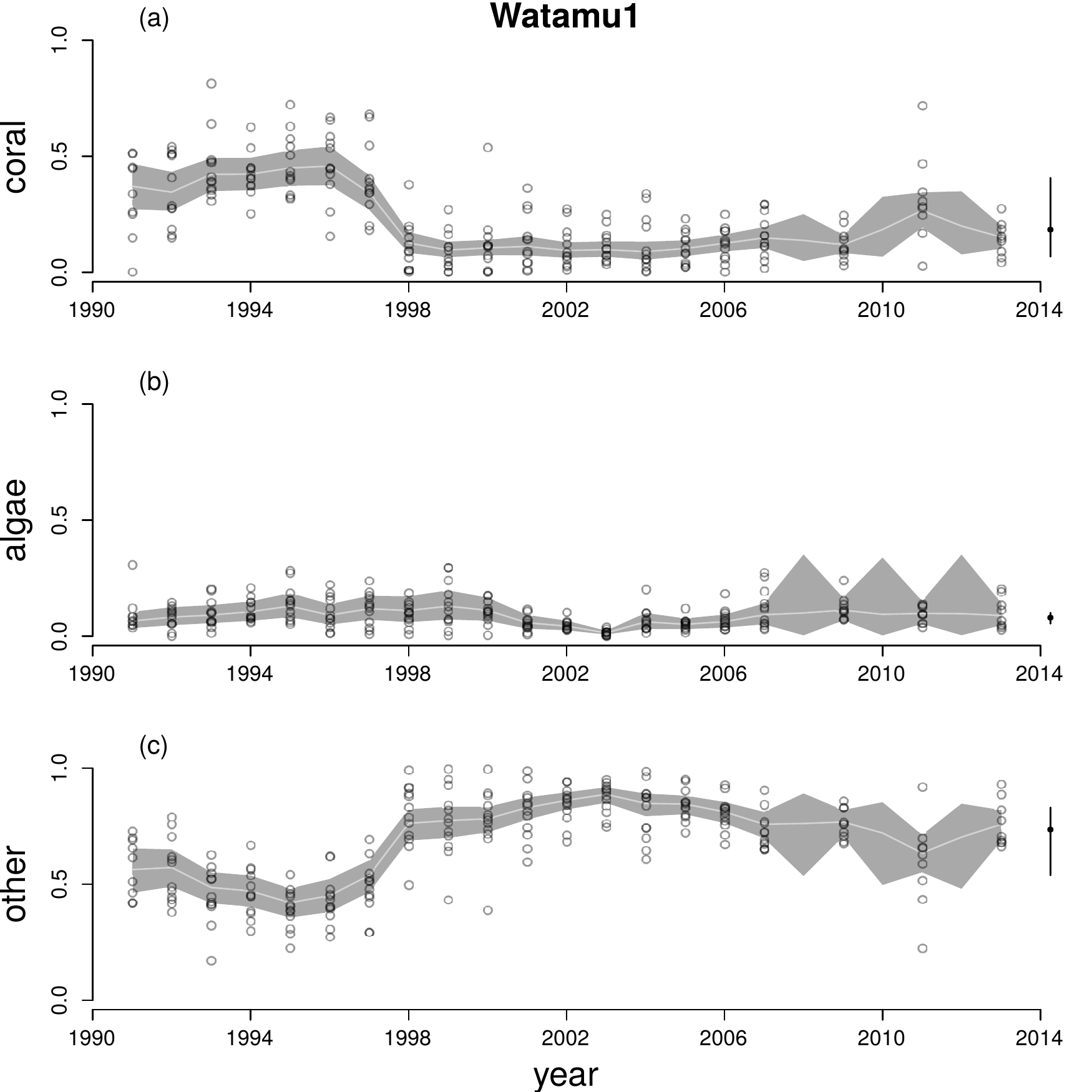}
\caption{Time series for cover of hard corals (a), macroalgae (b) and other (c) at Watamu1. See Figure \ref{fig:Bongoyo1} legend for explanation.}
\label{fig:Watamu1}
\end{figure}

\clearpage

\clearpage

\clearpage
\begin{figure}[h]
\includegraphics[height=18cm]{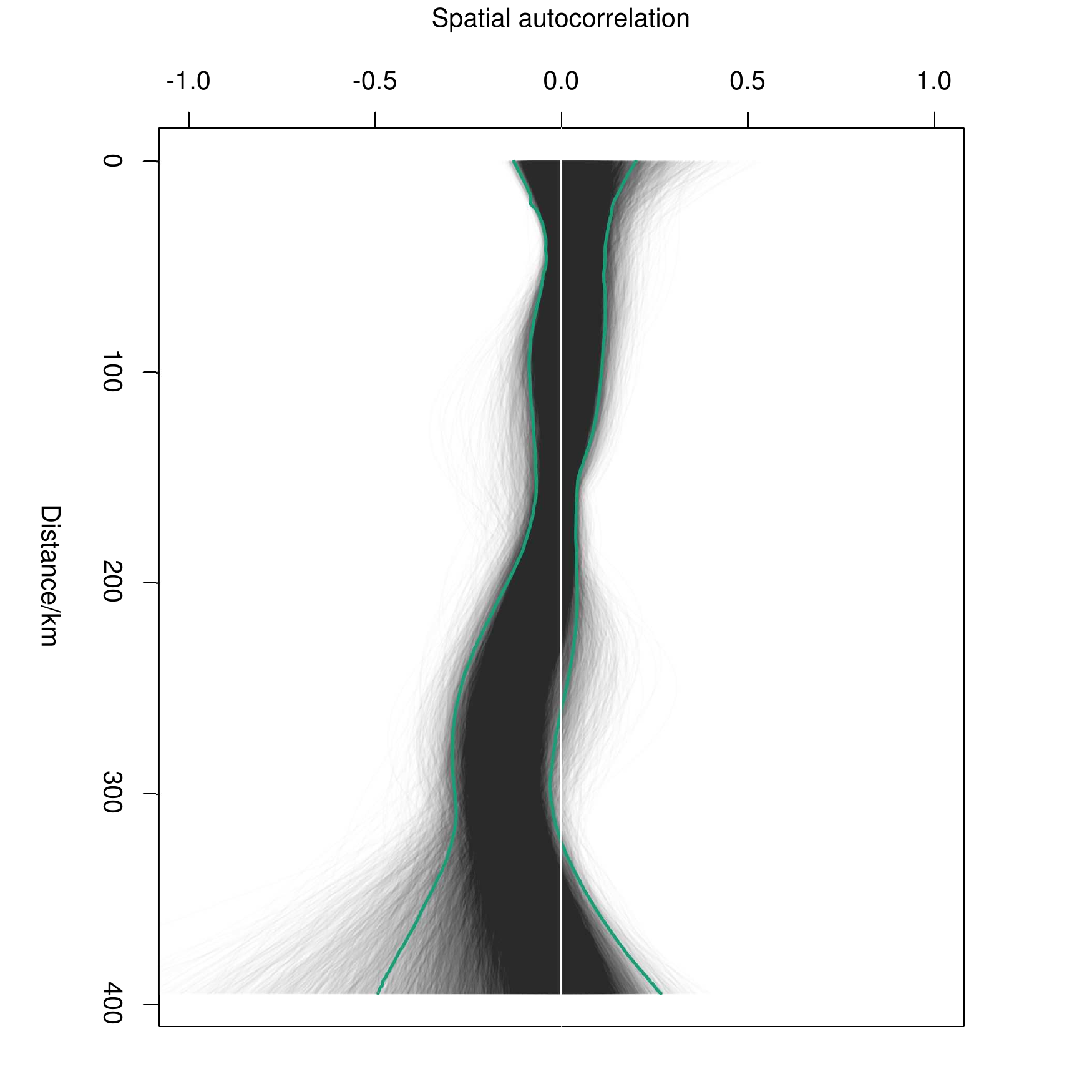}
\caption{Spline correlogram of spatial autocorrelation in $q_{0,1,i}$. Grey lines: spline correlograms from each of 20000 Monte Carlo iterations. Thick green lines: $95\%$ highest posterior density envelope. White horizontal line: zero-correlation reference line.}
\label{fig:spline}
\end{figure}

\clearpage

\clearpage

\begin{figure}[h]
\includegraphics[height=18cm]{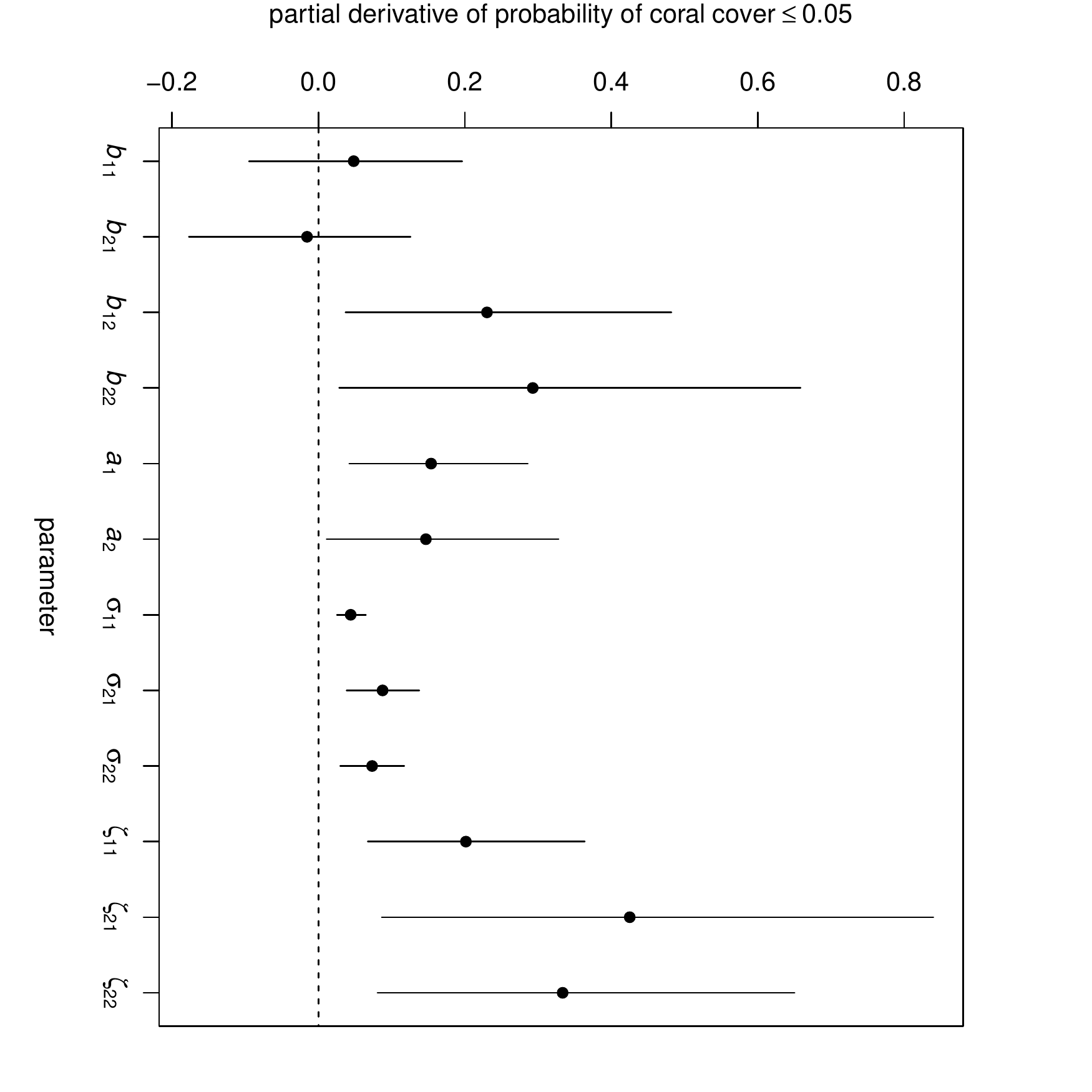}
\caption{Elements of the gradient vector of partial derivatives of the long-term probability of coral cover less than or equal to 0.05 with respect to elements of the $\mathbf B$ matrix, the $\mathbf a$ vector, the covariance matrix of random temporal variation $\bm \Sigma$, and the covariance matrix of among-site variability $\mathbf Z$. For each parameter, the dot is the posterior mean and the bar is a 95$\%$ HPD interval. For the covariance matrices, the elements $\sigma_{12}$ and $\zeta_{12}$ are not shown, because they are constrained to be equal to $\sigma_{21}$ and $\zeta_{21}$ respectively.}
\label{fig:derivs0.05}
\end{figure}

\clearpage

\clearpage

\begin{figure}[h]
\includegraphics[height=18cm]{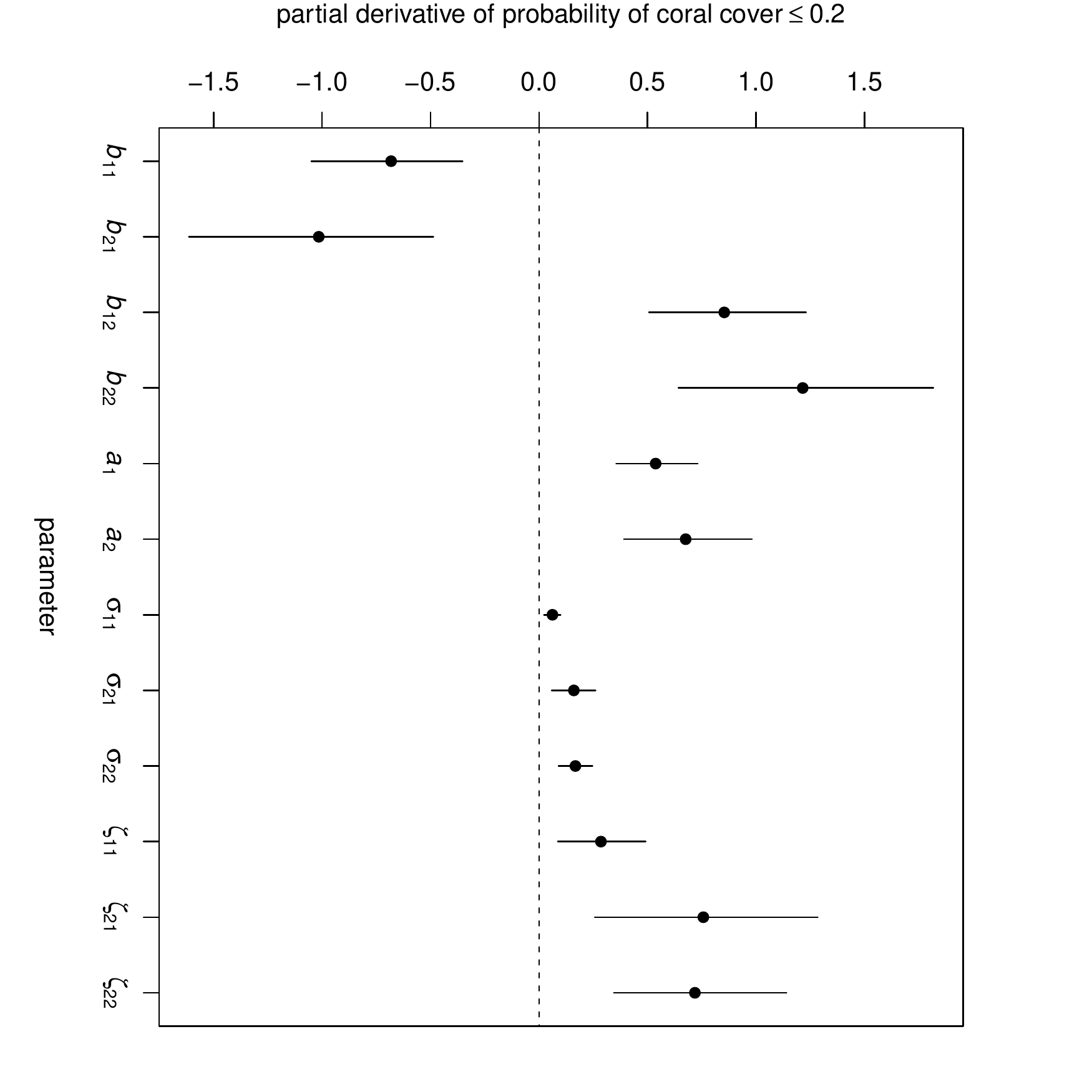}
\caption{Elements of the gradient vector of partial derivatives of the long-term probability of coral cover less than or equal to 0.2 with respect to elements of the $\mathbf B$ matrix, the $\mathbf a$ vector, the covariance matrix of random temporal variation $\bm \Sigma$, and the covariance matrix of among-site variability $\mathbf Z$. For each parameter, the dot is the posterior mean and the bar is a 95$\%$ HPD interval. For the covariance matrices, the elements $\sigma_{12}$ and $\zeta_{12}$ are not shown, because they are constrained to be equal to $\sigma_{21}$ and $\zeta_{21}$ respectively.}
\label{fig:derivs0.2}
\end{figure}

\clearpage

\end{document}